\documentclass[11pt,a4paper]{article}
\usepackage[a4paper]{geometry}
\usepackage{amssymb,latexsym,amsmath,amsfonts,amsthm,enumerate}
\usepackage{graphicx}
\usepackage{epstopdf}
\usepackage{tikz}
\usepackage{pgflibraryshapes}
\usetikzlibrary{arrows,decorations.markings}
\usepackage{subfigure}
\usepackage{overpic}
\usepackage{fullpage}
\usepackage{comment,verbatim}
\usepackage{hyperref}
\usepackage{color}
\usepackage{mathrsfs}
\usepackage[title,titletoc]{appendix}
\usepackage{ulem}
\usepackage{enumitem}
\usepackage{authblk}
\usepackage[numbers,sort&compress]{natbib}

\def\arg {\mathop{\rm arg}\nolimits}

\def\Res {\mathop{\rm Res}\nolimits}

\def\sgn {\mathop{\rm sgn}\nolimits}
\newtheorem{theorem}{Theorem}[section]

\newtheorem{proposition}[theorem]{Proposition}
\theoremstyle{definition}
\newtheorem{remark}[theorem]{Remark}

\numberwithin{equation}{section}
\newtheorem*{Notations}{Notations}
 
\numberwithin{equation}{section}
\begin{document}

\title{Large time and distance asymptotics of the one-dimensional impenetrable Bose gas and Painlev\'e IV transition}

\author[1]{Zhi-Xuan Meng}
\author[2]{Shuai-Xia Xu}
\author[1]{Yu-Qiu Zhao}
\affil[1]{Department of Mathematics, Sun Yat-sen University, Guangzhou 510275, China.}
\affil[2]{Institut Franco-Chinois de l'Energie Nucl\'{e}aire, Sun Yat-sen University,
Guangzhou 510275, China.}

\renewcommand{\thefootnote}{\fnsymbol{footnote}}
\footnotetext{Email addresses: mengzhx@mail2.sysu.edu.cn (Z.-X. Meng); xushx3@mail.sysu.edu.cn (S.-X. Xu); stszyq@mail.sysu.edu.cn (Y.-Q. Zhao)}

\date{}
\maketitle
\begin{abstract}
In the present paper, we study the time-dependent correlation function of the one-dimensional impenetrable Bose gas, which can be expressed in terms of the Fredholm determinant of a time-dependent sine kernel and the solutions of the separated NLS equations. We derive the large time and distance asymptotic expansions of  this determinant and the solutions of the separated NLS equations in both the space-like region and time-like region of the $(x,t)$-plane. Furthermore, we observe a phase transition between the asymptotic expansions in these two different regions.
The phase transition is then shown to be described by a particular solution of  the  Painlev\'e IV equation.

\end{abstract}

\tableofcontents

\noindent

\section{Introduction}

In this paper, we consider the asymptotics of the correlation functions of the one-dimensional  impenetrable Bose gas.  It is known, as shown in \cite{LL1963,L1963,LM1966},  that the one-dimensional Bose gas is exactly solvable. In the state of the thermal equilibrium at positive temperature,  the momentum distribution of the particles is given by the Fermi weight.  The thermodynamics of the model at positive temperature was developed by Yang and Yang in \cite{Yang}.   In  \cite{IIKS1990}, a completely integrable system  describing the temperature correlation functions was constructed  by developing a general  theory of integral operators. In particular, the  correlation functions were expressed as Fredholm determinants of integrable operators and the corresponding Riemann-Hilbert representations were established, which allow for the calculation of  the asymptotics of the correlation functions; see  \cite{IIK19901,IIKS1990,IIK1992, CZ2003,EFIK1997,IIK1990,IS1999,K1987,KS1990}.

In recent years, there has been  renewed interest in the  study of the Fredholm determinants and their  finite-temperature deformations, both in the physics literature and in the mathematics literature. At equal time, the correlation functions of the one-dimensional   Bose gas can be expressed in terms of the Fredholm determinants of the sine kernel  and its finite-temperature generalization. These determinants have been applied to  characterize the bulk scaling limit distributions of  particles in  noninteracting spinless fermion system, the Moshe-Neuberger-Shapiro  random matrix ensemble, and non-intersecting Brownian motions \cite{J2007,MNS1994,LW2020}.  Recently,  a completely integrable system of PDEs and  integro-differential  Painlev\'e V equation have been derived for a large class of weight functions extending the finite-temperature sine kernel  \cite{CT2024}. The asymptotics of the finite-temperature sine kernel has been derived in several different regimes in the $(x, s)$-plane, where a third-order phase transition is observed and described by an integral involving the Hastings-McLeod solution of the second Painlev\'e equation  \cite{X2024}. It is remarkable that the finite-temperature Airy-kernel determinant has been used to characterize  the solution of the Kardar-Parisi-Zhang  equation with the narrow wedge initial condition \cite{ACQ2011,CCR2020,CC2022,CCR2022}.

In the present paper, we consider the time-dependent correlation function of the one-dimensional  impenetrable Bose gas. 
At zero temperature, the correlation function can be  characterized by the determinant of the following time-dependent sine kernel \cite{IIKS1990}
\begin{equation}\label{kernel}
	K(\lambda, \mu; x,t)=\frac{f_{1}(\lambda)f_{2}(\mu )-f_{1}(\mu)f_{2}(\lambda )}{\lambda-\mu},
\end{equation}
where 
\begin{equation}
		f_{1}(\lambda)=f_{2}(\lambda)E(\lambda),~~f_{2}(\lambda )=\frac{1}{\pi}e^{it\lambda^{2}+ix\lambda}, \end{equation}
\begin{equation}\label{eq:E}
	E(\lambda)=P.V.\int_{-\infty }^{+\infty} \frac{1}{\tau-\lambda}e^{-2it\tau^{2}-2ix\tau}d\tau.
\end{equation}
Here $x$ and $t$ are the distance and time variables. Define the Fredholm determinant
\begin{equation}\label{D}
    D(x,t)=\ln\det(I+K_{x,t}), 
\end{equation}
where $K_{x,t}$ denotes the integrable operator acting on $L^2(-1,1)$ with the kernel \eqref{kernel}. 
Let  $B_{ij}$, $ (i,j=+,-)$ be the potentials
\begin{equation}\label{BC}\begin{aligned}
    B_{++}=\int_{-1}^{1}f_{1}(\mu)F_{1}(\mu)d\mu  ,\quad
    B_{+-}=\int_{-1}^{1}f_{1}(\mu)F_{2}(\mu)d\mu  ,\\
    B_{-+}=\int_{-1 }^{1}f_{2}(\mu)F_{1}(\mu)d\mu  ,\quad
	B_{--}=\int_{-1}^{1}f_{2}(\mu)F_{2}(\mu)d\mu  ,
    \end{aligned}
    \end{equation}
    where $F_{k}=(I+K_{x,t})^{-1}f_k$, $k=1,2$. 
Denote $b_{++}$ by
\begin{equation}\label{b++}
	b_{++}=B_{++}-G ,   \quad G=\int_{-\infty }^{+\infty} 
	e^{-2it\tau^{2}-2ix\tau}d\tau.
\end{equation}
Then, we have 
\begin{equation}\label{eq:partial_D}
    \partial_{xx}D(x,t)=4b_{++}B_{--}. 
\end{equation}
Furthermore, the two-point time-dependent correlation function for the one-dimensional impenetrable Bose gas can be represented by 
\begin{equation}\label{eq:correlation_function}
    	\left \langle \psi(x_{2},t_{2}) \psi^{+}(x_{1},t_{1}) \right \rangle =-\frac{1}{2\pi}e^{2it}b_{++}(x,t)\exp(D(x,t)),
\end{equation}
where the distance $x$,  time $t$ are related to $x_1,x_2$, $t_1,t_2$ and the chemical potential $h$ by
\begin{equation}
    x=\frac{1}{2}\sqrt{h}|x_{1}-x_{2}|,\quad t=\frac{1}{2}h(t_{2}-t_{1});
\end{equation}
see  \cite[Eqs.(6.1) and (6.2)]{IIKS1990} and also \cite[Chapter XIV.5]{KBI1993}. Furthermore,  it was  shown in \cite[Eq.(6.23)]{IIKS1990} and \cite[Eq.(1.18)]{IIK1992} that the potentials  $B_{ij}$ satisfy the separated NLS equations
\begin{equation}\label{1qnls}
			 \left\{\begin{array}{l}
	2i\partial_{t}b_{++}=		-\partial_{x}^2b_{++}-8b^{2}_{++}B_{--},\\
	2i\partial_{t}B_{--}=		\partial_{x}^2B_{--}+8b_{++}B^{2}_{--}.
		\end{array}\right.
	\end{equation}
The above system is the first nontrivial pair of equations of the AKNS hierarchy and it reduces to the nonlinear Schr\"odinger equation if $b_{++}=\overline{B_{--}}$; see \cite{AKNS}. It should be mentioned that  the finite-temperature unequal time correlators can be expressed in terms of the determinant of the time-dependent sine kernel multiplied by a Fermi weight. The above PDEs are also valid for the finite-temperature determinant. 
These results were obtained in \cite{IIKS1990}   by developing a general theory for the integral operators and constructing Riemann-Hilbert problems for these operators. These Riemann-Hilbert representations allow for calculations of the aymptotics of the  correlation functions.  From  these Riemann-Hilbert representations, the large time and distance asymptotics of  correlation function of impenetrable bosons at finite temperature have been derived in \cite{IIK1992}.

By a computation using \eqref{kernel}-\eqref{eq:E}, we see that the kernel \eqref{kernel} tends to  the classical sine kernel  as $t\to0$
\begin{equation}\label{eq:K0}
    K(\lambda,\mu;x,0)=-\gamma K^{(\sin)}(\lambda,\mu; x), \quad   K^{(\sin)}(\lambda,\mu; x)=\frac{\sin x (\lambda-\mu)}{\pi(\lambda-\mu)},
\end{equation}
with $\gamma=2$. This corresponds to the equal-time situation of the correlation function \eqref{eq:correlation_function};   see \cite[Eq. (1.10)]{IIKS1990}. Denote $K^{(\sin)}_{x}$ the integrable operator acting on $L^2(-1,1)$ with the sine kernel $K^{(\sin)}$. 
It is remarkable that the logarithmic derivative of the Fredholm determinant of $K^{(\sin)}_{x}$ 
\begin{equation}\label{eq:detSine}
\sigma_V(x;\gamma)=x\frac{d}{dx}   \ln\det(I-\gamma K^{(\sin)}_x) 
\end{equation}
satisfies the $\sigma$-form of the fifth Painlev\'e equation \cite[Eq. (2.27)]{JMMS1980} 
\begin{equation}\label{eq:sigmaPV}
(x\sigma_V'')^2+4(4\sigma_V-4x\sigma_V'-\sigma_V'^2)(\sigma_V-x\sigma_V')=0,\end{equation} 
with the boundary conditions 
 \begin{equation}\label{eq:PVasy_zero}
 \sigma_V(x;\gamma)=-\frac{2}{\pi}\gamma x+O(x^2), \quad x\to 0, 
\end{equation} 
and as $ x\to\infty$ \cite[Eq. (2.14)]{Dyson1976} and \cite[Eqs (1.16) and (1.21)]{MT1986}
\begin{equation}\label{eq:PVasy_infty}
 \sigma_V(x;\gamma)=\left\{\begin{array}{ll}
 4kx+O(1), & \gamma<1,\\ 
  -x^2+O(1), & \gamma=1,  \\
4kx-2x\tan( \theta(x))+O(1), &  \gamma>1,
    \end{array}\right.
\end{equation}
with $k= \frac{1}{2\pi}\ln |\gamma-1| $, $\theta(s)=2x+2k\ln x+c_0$ and $c_0=4k\ln 2-2\arg\left(\Gamma\left(ik+\frac{1}{2}\right)\right)$. 
Therefore,  the equal-time correlation function \eqref{eq:correlation_function} with $t=0$ is related to $\sigma_V(x;\gamma)$ with $\gamma=2$, which has singular asymptotic behavior as $x\to\infty$. 
Let $\gamma=1$,  integrating \eqref{eq:detSine} along $[0,x]$ leads to the famous  integral expression for the gap probability distribution of the classical sine process in random matrix theory.  In \cite{Dyson1976}, Dyson derived the large $x$ asymptotics of this determinant with $\gamma=1$ up to a conjectured constant term. The constant was  derived rigorously later in \cite{DIKZ2007, Ehrhardt2006, Krasovsky2004}.
The asymptotics of this deteminnant on the union of  several intervals, as the size of the intervals tends to infinity, have also been explored in \cite{Charlier2021, FK2024, DIZ1997}.

The present work is devoted to the studies of the time-dependent correlation function of the one-dimensional Bose gas \eqref{eq:correlation_function}, which can be expressed in terms of the  Fredholm determinant \eqref{D} and the solutions of the separated NLS equations \eqref{1qnls}. By using the Riemann-Hilbert  representation for the determinant  \eqref{D}, we derive the large time and distance asymptotic approximations of the derivatives of determinant  \eqref{D} and the solutions of the separated NLS equations in both the space-like region and time-like region of the $(x,t)$-plane. Furthermore, we observe a phase transition between the asymptotic expansions in these two different regions. The phase transition is then shown to be described by a particular solution of  the  Painlev\'e IV equation.

\subsection{Statement of results}

\begin{figure}
    \centering
    \includegraphics[width=0.7\linewidth]{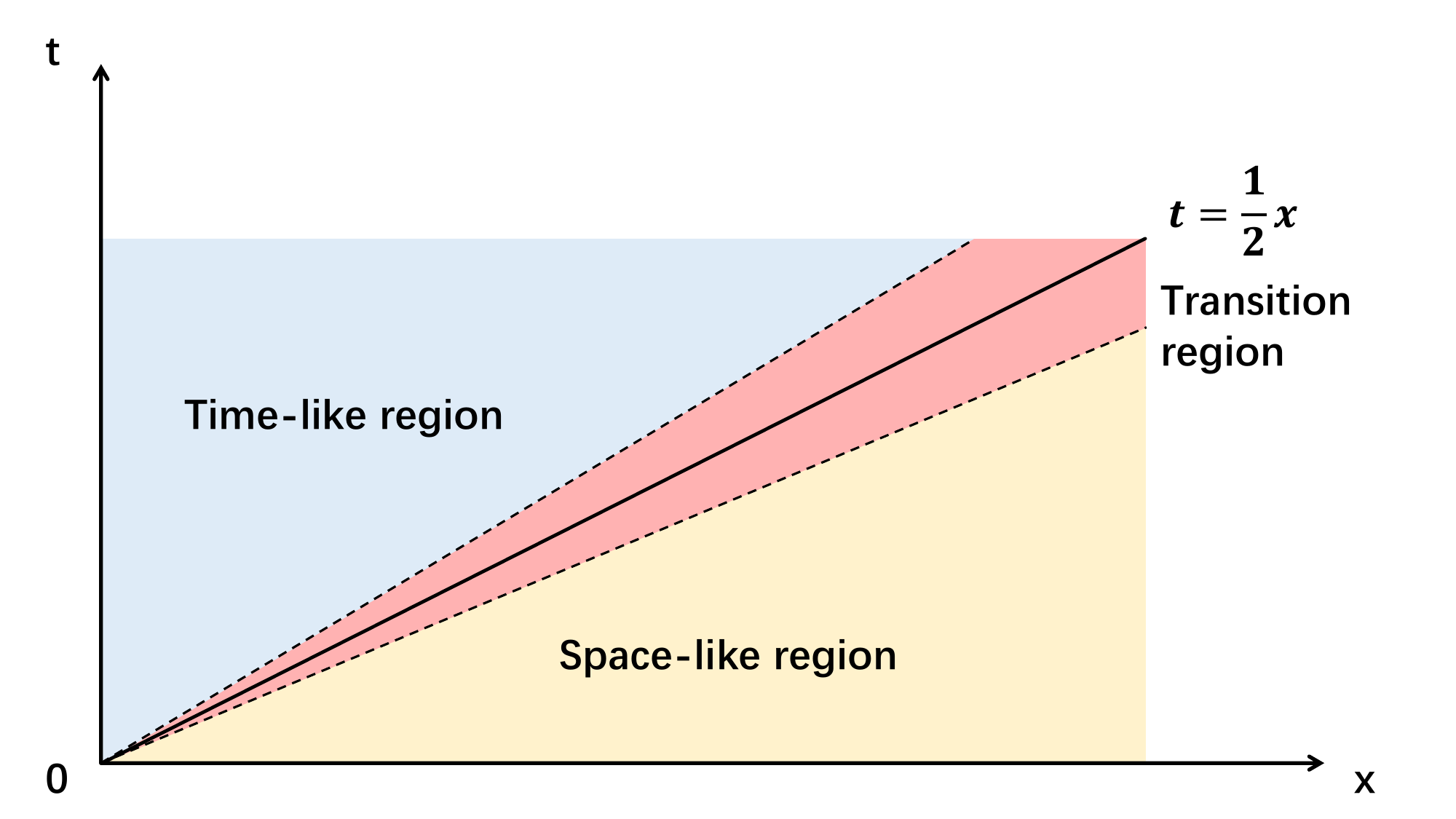}
    \caption{The space-like region, time-like region and
transition region}
    \label{space-time}
\end{figure}

To state our main results,  we define  for $x,t>0$ the space-like region, time-like region and transition region as follows:
\begin{itemize}
    \item 
    space-like region:  $\frac{x}{2t}>1+\delta$,
    \item 
    time-like region: $\frac{x}{2t}<1-\delta$,
    \item 
    transition region: $t^{\frac{1}{2}}\left|\frac{x}{2t}-1\right|\le C$,  
\end{itemize}
 with any small but fixed $\delta>0$, and any constant $C>0$; see Fig. \ref{space-time}. 
Then, we  derive the asymptotic expansions of the derivatives of determinant  \eqref{D} and the solutions of the separated NLS equations in these regions,
of which the main results are given in the following theorems. 
\begin{theorem}[Large distance asymptotics in the space-like region]\label{space} 
	Let $D(x,t)$ be the Fredholm determinant defined in \eqref{D},  we have the following asymptotic expansions  as $x\to+\infty$:
	\begin{equation}\label{th1dt}
		\partial_{t}D(x,t)=-\frac{\sqrt{2}e^{i\left(\frac{x^{2}}{2t}+2t+\frac{\pi}{4}\right)}}{\sqrt{\pi t}\cos(2x)}+O(x^{-1}),
	\end{equation}

	\begin{equation}\label{th1dx}
\partial_{x}D(x,t)=-2\tan(2x)+\frac{2\sqrt{2t}e^{i\left(\frac{x^{2}}{2t}+2t+\frac{\pi}{4}\right)}}{\sqrt{\pi }(x-2t)\cos(2x)}
\left(1-\frac{2te^{2ix}}{(x+2t)\cos(2x)}\right)+O(x^{-1}),
	\end{equation}
	where the error terms are uniform  for $(x,t)$ in  the space-like region and for x bounded away from the zeros of $\cos(2x)$. Moreover, we have the asymptotic expansions of the corresponding solutions of the separated NLS equations $b_{++}$ and  $B_{--}$, defined by \eqref{BC} and \eqref{b++}, as $x \to +\infty$:
	\begin{equation}\label{th1b}
	b_{++}(x,t)=-\frac{\pi e^{-2it}}{\cos(2x)}-\sqrt{\frac{\pi}{2t}}e^{i\left(\frac{x^{2}}{2t}-\frac{\pi}{4}\right)} \left[1-\frac{4it\tan(2x)}{x-2t}+\frac{4t^{2}e^{4ix}}{(x^{2}-4t^{2})\cos^{2}(2x)}\right]+O(x^{-1}),
	\end{equation}
\begin{equation}\label{th1B}
		B_{--}(x,t)=\frac{e^{2it}}{\pi\cos(2x)}-\frac{2\sqrt{2}t^{\frac{3}{2}}e^{i\left(\frac{x^{2}}{2t}+4t-\frac{\pi}{4}\right)}}{\pi^{\frac{3}{2}}(x^{2}-4t^{2})\cos^{2}(2x)}+O(x^{-1}),
	\end{equation}
	where the error terms are uniform  for $(x,t)$ in  the space-like region and for x bounded away from the zeros of $\cos(2x)$. 
	
\end{theorem}

\begin{theorem}[Large time asymptotics in the time-like region]\label{time}
    	Let $D(x,t)$ be the Fredholm determinant defined in \eqref{D},  we have the following asymptotic expansions  as $t\to+\infty$:
     		\begin{equation}\label{th2dt}
		\partial_{t}D(x,t)=\frac{\sqrt{2}e^{-i\left(\frac{x^{2}}{2t}+2t+\frac{\pi}{4}\right)}}{\sqrt{\pi t}\cos(2x)}+O(t^{-1}),
	\end{equation} 
			\begin{equation}\label{th2dx}
\partial_{x}D(x,t)=-2\tan(2x)-\frac{2\sqrt{2t}e^{-i\left(\frac{x^{2}}{2t}+2t+\frac{\pi}{4}\right)}}{\sqrt{\pi}(x+2t)\cos(2x)}\left(1+\frac{2te^{2ix}}{(x-2t)\cos(2x)}\right)+O(t^{-1}),
	\end{equation}
where the error terms are uniform  for $(x,t)$ in  the time-like region and for x bounded away from the zeros of $\cos(2x)$. Moreover, we have the asymptotic expansions of the corresponding solutions of the separated NLS equations $b_{++}$ and  $B_{--}$,  defined by  \eqref{BC} and \eqref{b++}, as $t \to +\infty$:
		\begin{equation}\label{th2b}
		b_{++}(x,t)=-\frac{\pi e^{-2it}}{\cos(2x)}-\frac{2\sqrt{2\pi}t^{\frac{3}{2}}e^{-i\left(\frac{x^{2}}{2t}+4t-\frac{\pi}{4}\right)} }{(x^{2}-4t^{2})\cos^{2}(2x)}+O(t^{-1}),
	\end{equation}
	\begin{equation}\label{th2B}
		B_{--}(x,t)=\frac{e^{2it}}{\pi\cos(2x)}-\frac{e^{i\left(-\frac{x^{2}}{2t}+\frac{\pi}{4}\right)} }{\pi^{\frac{3}{2}}\sqrt{2t}}\left[1+\frac{4it\tan(2x)}{x+2t}+\frac{4t^{2}e^{4ix}}{(x^{2}-4t^{2})\cos^{2}(2x)}\right]+O(t^{-1}),
	\end{equation}
where the error terms are uniform  for $(x,t)$ in  the time-like region and for x bounded away from the zeros of $\cos(2x)$.

\end{theorem}

\begin{remark}
Let $t\to 0$ in \eqref{th1dx}, we have the asymptotics as $x\to+\infty$:
\begin{equation}\label{eq: Asy_D0}
\partial_{x}D(x,0)=-2\tan(2x)+O(x^{-1}). 
	\end{equation}
	From \eqref{eq:K0} and \eqref{eq:detSine}, we have 
	\begin{equation}\label{eq:detSine_D0}
x\partial_{x}D(x,0)=x\frac{d}{dx}   \ln\det(I-\gamma K^{(\sin)}_x) =\sigma_V(x;\gamma),
\end{equation}
 where  $K^{(\sin)}_x$ is the integral operator with the sine kernel and $\sigma_V(x;\gamma)$
is the solution of the $\sigma$-form of the fifth Painlev\'e equation \eqref{eq:sigmaPV} with $\gamma=2$.  The  asymptotics \eqref{eq: Asy_D0} is consistent with the large $x$ asymptotics of $ \sigma_V(x;\gamma)$ with $\gamma=2$ as given in \eqref{eq:PVasy_infty}, which was obtained earlier by McCoy and Tang in \cite{MT1986}. 
  \end{remark}
\begin{remark}
From \eqref{th1dx}, we see that $\partial_x D$ has singular asymptotic behavior as $x\to+\infty$. This phenomenon can also be observed in the large $x$ asymptotics of the sine kernel determinant  \eqref{eq:detSine} and  $\sigma_V(x;\gamma)$  if the parameter $\gamma>1$ as shown in \eqref{eq:PVasy_infty}. Therefore, it is natural to expect that $D(x,t)$ may have poles near the zeros of $\cos(2x)$ for large $x$. To make the results valid in the above theorems, we require that $x$ is bounded away from  the zeros of $\cos(2x)$. 
It would be desirable to derive the asymptotics near the zeros of $\cos(2x)$ and the asymptotics of  $D(x,t)$ itself.  
Furthermore, similar to the sine kernel determinant, it would be interesting to study the asymptotics of the $\gamma$-parameter generalization of the determinant \eqref{D}, namely $\det(I-\gamma K_{x,t})$. We will leave these problems to further investigations. 

  \end{remark}
  
\begin{remark}
It is noted that the leading terms in both the large distance asymptotics  \eqref{th1b} and \eqref{th1B}, and the large time asymptotics \eqref{th2b} and \eqref{th2B} are given by the following special periodic solutions of the separated NLS equations  \eqref{1qnls}
\begin{equation}
\left\{\begin{array}{l}
    b_{++}(x,t)=-\frac{\pi e^{-2it}}{\cos(2x)},\\ B_{--}(x,t)=\frac{e^{2it}}{\pi\cos(2x)}.
    \end{array}\right.
\end{equation}
\end{remark}

\begin{remark}
    From Theorems \ref{space} and \ref{time}, we observe a  phase transition in the leading asymptotics of $\partial_{t}D(x,t)$ given in \eqref{th1dt} and \eqref{th2dt}. Specifically, the phase changes from $e^{i\left(\frac{x^{2}}{2t}+2t+\frac{\pi}{4}\right)}$ to $e^{-i\left(\frac{x^{2}}{2t}+2t+\frac{\pi}{4}\right)}$ as $(x,t)$ moves across the critical curve $x=2t$ as shown in Fig. \ref{space-time}. From Theorems \ref{space} and \ref{time}, similar phase transition can also be found in the large time and distance asymptotics of the other quantities, including $\partial_{x}D(x,t)$, $ b_{++}(x,t)$ and $   B_{--}(x,t)$.
   
\end{remark}

From Theorems \ref{space} and \ref{time}, we observe a phase transition near the critical curve $x=2t$ in the large time and distance  asymptotic expansions. Next, we show that the  transition can be described by a special solution of the Painlev\'e IV  equation. Let $u(s)$ be a solution of the Painlev\'e IV (PIV) equation  
\begin{equation}\label{p4}
	    \frac{\mathrm{d}^{2}u}{\mathrm{d}s^{2}}=\frac{1}{2u}\left(\frac{\mathrm{d}u}{\mathrm{d}s}\right)^{2}+\frac{3}{2}u^{3}+4su^{2}+2(s^{2}+1-2\Theta_{\infty})u-\frac{8\Theta^{2}}{u},
	\end{equation}
	and we define
     \begin{equation}\label{y1}
	\frac{1}{y}\frac{\mathrm{d}y}{\mathrm{d}s}=-u-2s,
\end{equation}
\begin{equation}\label{K}
	z=\frac{1}{4}\left(-\frac{\mathrm{d}u}{\mathrm{d}s}+u^{2}+2su+4\Theta\right),
\end{equation}
and the associate Hamiltonian 
 \begin{equation}\label{H1}
	H=\frac{z}{u}(z-2\Theta)-\left(\frac{u}{2}+s\right)(z-\Theta-\Theta_{\infty}).
\end{equation}
The following solution of the PIV equation  with  the parameters $\Theta=0$ and $\Theta_{\infty}=\frac{1}{2}$ plays a central role in the asymptotics in the transition region.

\begin{proposition}[Large $s$ asymptotics of PIV]\label{thPIV}
  For the parameters $\Theta=0$ and $\Theta_{\infty}=\frac{1}{2}$, there exists a unique solution of PIV equation \eqref{p4} corresponding  to the special Stokes multipliers
   \begin{equation}
      \{s_{1}=s_{2}=2i,\quad s_{3}=s_{4}=0\}.
   \end{equation}
This solution has the following asymptotic expansions  
   \begin{equation}\label{th1u}
	u(s)=-2s-\frac{2ie^{-s^{2}}}{\sqrt{\pi}}+O(s^{-1}),\quad e^{\frac{\pi i}{4}}s\to+\infty,
\end{equation}
   \begin{equation}\label{th2u}
	u(s)=-2s-\frac{2e^{s^{2}}}{\sqrt{\pi}}+O(s^{-1}),\quad e^{\frac{\pi i}{4}}s\to-\infty.
\end{equation}
Moreover, we have the following asymptotic expansions of  $y(s)$ and  the associate Hamiltonian $H(s)$ defined in \eqref{y1} and \eqref{H1}, respectively: 
\begin{equation}\label{th1y}
	y(s)=2-\frac{2ie^{-s^{2}}}{\sqrt{\pi}s}+O(s^{-2}),\quad e^{\frac{\pi i}{4}}s\to+\infty,
\end{equation}
\begin{equation}\label{th1H}
	H(s)=O(s^{-1}),\quad e^{\frac{\pi i}{4}}s\to+\infty,
\end{equation}
\begin{equation}\label{th2y}
y(s)=2+\frac{2e^{s^{2}}}{\sqrt{\pi}s}+O(s^{-2}),\quad e^{\frac{\pi i}{4}}s\to-\infty,
\end{equation}
\begin{equation}\label{th2H}
	H(s)=-\frac{e^{s^{2}}}{\sqrt{\pi}}+O(s^{-1}),\quad e^{\frac{\pi i}{4}}s\to-\infty.
\end{equation}
\end{proposition}

\begin{theorem}[Asymptotics in the transition region]\label{transition}
     Let $D(x,t)$ be the Fredholm determinant defined in \eqref{D},  we have the following asymptotic expansions  as $x,t\to+\infty$:
     		\begin{equation}\label{th3dt}
\partial_{t}D(x,t)=\frac{2\sqrt{2}ie^{\frac{\pi i}{4}}}{\sqrt{t}}\left[H(s)-\frac{y(s)}{2}\left(\frac{u(s)}{2}+s\right)\frac{e^{4ix}}{1+\frac{y(s)}{2}e^{4ix}}\right]+O(t^{-1}),
	\end{equation}

    \begin{equation}\label{th3dx}
\partial_{x}D(x,t)=2i\frac{\frac{y(s)}{2}e^{4ix}-1}{1+\frac{y(s)}{2}e^{4ix}}+\sqrt{\frac{2}{t}}e^{-\frac{\pi}{4}i}\frac{H(s)-\frac{y^{2}(s)}{4}\left(\frac{u(s)}{2}+s-H(s)\right)e^{8ix}}{\left(1+\frac{y(s)}{2}e^{4ix}\right)^{2}}+O(t^{-1}).
\end{equation}
Moreover, we have the asymptotic expansions of the corresponding solutions of the separated NLS equations $b_{++}$ and  $B_{--}$,  defined by  \eqref{BC} and \eqref{b++}, as $t \to +\infty$:		
	\begin{equation}\label{th3b}\begin{aligned}
		b_{++}(x,t)=&-\frac{\pi y(s)e^{2i(x-t)}}{1+\frac{y(s)}{2}e^{4ix}}+\frac{\pi y(s)e^{2i(x-t)+\frac{\pi i}{4}}}{2\sqrt{2t}\left(1+\frac{y(s)}{2}e^{4ix}\right)^{2}} \left[y(s)H(s)e^{4ix}-(1+y(s)e^{4ix})\left(\frac{u(s)}{2}+s\right)\right]\\&+O(t^{-1}),
	\end{aligned}\end{equation}
	\begin{equation}\label{th3B}
		B_{--}(x,t)=\frac{2e^{2i(x+t)}}{\pi\left(1+\frac{y(s)}{2}e^{4ix}\right)}+\frac{e^{2i(x+t)+\frac{i\pi}{4}}}{\pi\sqrt{2t}\left(1+\frac{y(s)}{2}e^{4ix}\right)^{2}}\left[2H(s)+\frac{y(s)}{2}\left(\frac{u(s)}{2}+s\right)e^{4ix}\right]+O(t^{-1}).
	\end{equation} 
Here the variable $s=e^{-\frac{\pi i}{4}}\sqrt{2t}\left(\frac{x}{2t}-1\right)$ and the error terms are uniform  for $(x,t)$ in  the transition region such that $s$ is bounded.  The function  $u(s)$ is the solution of the PIV equation in \eqref{p4} and $y(s)$ and $H(s)$ are defined by \eqref{y1} and \eqref{H1}  with the properties specified in Proposition \ref{thPIV}. \end{theorem}

\begin{remark}
    As $t^{\frac{1}{2}}\left(\frac{x}{2t}-1\right)\to+\infty$, from \eqref{th1u}, \eqref{th1y} and \eqref{th1H}, we see that the  Painlev\'e IV asymptotics \eqref{th3dt} degenerates to \eqref{th1dt}. On the other hand, as $t^{\frac{1}{2}}\left(\frac{x}{2t}-1\right)\to-\infty$,  from \eqref{th2u}, \eqref{th2y} and \eqref{th2H}, the asymptotics \eqref{th3dt} is reduced  to \eqref{th2dt}.  Therefore, the Painlev\'e IV asymptotics describes the phase transition between the asymptotics of $\partial_{t}D(x,t)$ in the space-like and time-like regions as given in  \eqref{th1dt} and  \eqref{th2dt}.   Similarly, 
    the Painlev\'e IV asymptotics shown in Theorem \ref{transition} also describe the  phase transition in the asymptotics of the quantities 
$\partial_{x}D(x,t)$, $ b_{++}(x,t)$ and $ B_{--}(x,t)$. 
\end{remark}

\begin{Notations}
In this paper, we will  frequently use the following notations.
   \begin{itemize}
       \item If $A$ is a matrix, then $(A)_{ij}$ denotes its $(i,j)$-th entry and $A^{T}$ represents  its transpose.
       \item We define $U(a,\delta)$ as the open disc centered at $a$ with radius $\delta>0$:
\begin{equation}
U(a,\delta):=\{z\in\mathbb{C}:|z-a|<\delta\},\end{equation}
and $\partial U(a,\delta)$ as its boundary with the clockwise orientation. 
       \item The Pauli matrices are defined as follows:
       \begin{equation}
    \sigma_{1}=\begin{pmatrix}
        0&1\\1&0
    \end{pmatrix},\quad\sigma_{3}=\begin{pmatrix}
        1&0\\0&-1
    \end{pmatrix},\quad\sigma_{+}=\begin{pmatrix}
        0&1\\0&0
    \end{pmatrix}, \quad \sigma_{-}=\begin{pmatrix}
        0&0\\1&0
    \end{pmatrix}.
\end{equation}
\item We will carry out  Deift-Zhou nonlinear steepest descent analysis for the Riemann-Hilbert problems several times. Each time, we will use the same notations such as $T$, $P^{(\infty)}$, $P^{(\pm1)}$ and $R$. These notations will have different meaning in each context, and we expect this will not cause confusion.
   \end{itemize}
\end{Notations}

The rest of the present paper is arranged as follows. In Section \ref{m&L}, we introduce  the Riemann-Hilbert (RH) problem for the Fredholm determinant \eqref{D}, which was introduced in \cite{IIK1992}. The RH problem for the classical PIV equation is also given in this section. In Sections \ref{sec.s} and \ref{sec.t}, we perform the Deift-Zhou nonlinear steepest descent analysis \cite{DZ1993,DMVZ19991,DMVZ19992}  of the RH problem for the determinant in the space-like region and the time-like region, respectively. The  proofs of Theorems \ref{space} and \ref{time} are given at the end of  Sections \ref{sec.s} and \ref{sec.t}, respectively. In Section \ref{sec.3}, we  consider the asymptotics in the transition region and prove Theorem \ref{transition}. Finally, in Section \ref{sec.pIV}, we derive the asymptotics of a special solution of the PIV equation  and prove Proposition \ref{thPIV} by using the associate RH problem.

\section{RH problem for the determinant and the PIV equation}\label{m&L}
 In this section, we intruduce the RH problem representation for the determinant in \eqref{D} as constructed in  \cite{IIKS1990,IIK1992} and then present the RH problem for the PIV equation \eqref{p4}.

\subsection{RH problem for the determinant }\label{mrhp}

 The kernel in \eqref{kernel} can be expressed in the following form 
\begin{equation}
	K(\lambda,\mu)=\frac{\overrightarrow{f}^{T}(\lambda)\overrightarrow{g}(\mu)}{\lambda-\mu},
\end{equation}
where
\begin{equation}
	\overrightarrow{f}(\lambda)=\begin{pmatrix}
	f_{1}(\lambda)	\\ f_{2}(\lambda)
	\end{pmatrix} ,\quad \overrightarrow{g}(\lambda)=\begin{pmatrix}
	f_{2}(\lambda)	\\ -f_{1}(\lambda)
	\end{pmatrix}.
\end{equation}
Then the kernel of the resolvent operator $\left(I+K_{x,t}\right)^{-1}K_{x,t}$ can be expressed as \cite{IIKS1990}:
\begin{equation}
R(\lambda,\mu)=\frac{\overrightarrow{F}^{T}(\lambda)\overrightarrow{G}(\mu)}{\lambda-\mu},
\end{equation}
with
\begin{equation}
	\overrightarrow{F}(\lambda)=\left(I+K_{x,t}\right)^{-1}\overrightarrow{f}(\lambda),\quad \overrightarrow{G}(\lambda)=\left(I+K_{x,t}^{T}\right)^{-1}\overrightarrow{g}(\lambda).
\end{equation}
Here $K_{x,t}$ denotes the integrable operator acting on $L^2(-1,1)$ with the kernel \eqref{kernel} and
  $K^{T}_{x,t}$ represents the  real adjoint of the operator $K_{x,t}$ with the kernel
      \begin{equation}
           K^{T}(\lambda,\mu)=K(\mu, \lambda).
     \end{equation}
Furthermore, $\overrightarrow{F}$ and $\overrightarrow{G}$ can be expressed as
\begin{equation}
	\overrightarrow{F}(\lambda)=X(\lambda)\overrightarrow{f}(\lambda),\quad
	\overrightarrow{G}(\lambda)=X^{-T}(\lambda)\overrightarrow{g}(\lambda),
\end{equation}
where $X$ satisfies the following RH problem; see \cite{IIK1992, IIKS1990}.

\subsubsection*{RH problem for $X$}
\begin{description}
\item(1)
$X(\lambda)$ is analytic for $\lambda\in\mathbb{C}\setminus[-1,1]$. 
\item(2)
$X(\lambda)$ has continuous boundary values $X_{\pm}(\lambda)$ as $\lambda$ approaches the real axis from the positive and negative sides, respectively. And they satisfy the relation
\begin{equation}
    X_{+}(\lambda)= X_{-}(\lambda)\left(I+2{\pi}i\overrightarrow{f}(\lambda)\overrightarrow{g}^{T}(\lambda)\right), \quad\lambda\in (-1,1).
\end{equation} 

\item(3)
$X(\lambda )= I+O\left(\frac{1}{\lambda}\right)$, as $\lambda\to\infty$.
\end{description}

Then the solution to the above RH problem is expressible in terms of the functions $\overrightarrow{F}$ and $\overrightarrow{g}$ by using the Cauchy integral 
\begin{equation} 
	X(\lambda)=I+\int_{-1}^{1}\frac{\overrightarrow{F}(\mu)\overrightarrow{g}^{T}(\mu)}{\mu-\lambda}d\mu.
\end{equation}
The behavior of $X$ at infinity can be expressed as
\begin{equation}
	X(\lambda )=I+\frac{X_{1}}{\lambda}+\frac{X_{2}}{\lambda^{2}}+O\left(\frac{1}{\lambda^{3}}\right). 
\end{equation}
We denote $X_{1}$ and $X_{2}$  by
\begin{equation}\label{eq:X12}
	X_{1}=
	\begin{pmatrix}
		-B_{-+}&B_{++}  \\
		-B_{--} &B_{+-}
	\end{pmatrix}, \quad
	X_{2}=
	\begin{pmatrix}
		-C_{-+}&C_{++}  \\
		-C_{--} &C_{+-}
	\end{pmatrix},
\end{equation} 
then the potentials  $B_{ij}$ ($i,j=+,-$) are given by \eqref{BC}. 

In order to simplify the jump condition, we introduce the transformation 
\begin{equation}
	Y(\lambda)=X(\lambda)
	\begin{pmatrix}
		1&\int_{-\infty }^{+\infty}\frac{e^{-2i\theta(\tau)}}{\tau-\lambda}d\tau  \\
		0&1
	\end{pmatrix},
\end{equation}
where the phase function
\begin{equation} 
	\theta(\lambda)=t{\lambda}^2+x\lambda .
\end{equation}
Then $Y$ satisfies the following RH problem.

\subsubsection*{RH problem for $Y$}
\begin{description}
\item(1)
$Y(\lambda)$ is analytic for $\lambda\in\mathbb{C}\setminus\mathbb{R}$.

\item(2)
$Y_{+}(\lambda)= Y_{-}(\lambda)J_{Y}(\lambda), \lambda\in \mathbb{R}$,
\begin{equation}
    J_{Y}(\lambda)=\left\{\begin{array}{ll}
\begin{pmatrix}
		-1 & 0 \\
		\frac{2i}{\pi}e^{2i\theta(\lambda)} & -1
	\end{pmatrix}, &\quad\lambda \in (-1,1), \\
 \begin{pmatrix}
		1 & 2\pi ie^{-2i\theta(\lambda)} \\
		0& 1
	\end{pmatrix}, 
	&\quad\lambda \in (-\infty,-1)\cup(1,+\infty).
\end{array}\right.
\end{equation}

\item(3)
$Y(\lambda )=I+\frac{Y_{1}}{\lambda}+\frac{Y_{2}}{\lambda^{2}}+O\left(\frac{1}{\lambda^{3}}\right)$, as $\lambda\to\infty$, where
\begin{equation}\label{Y}
    Y_{1}=\begin{pmatrix}
        -B_{-+}& b_{++}\\ -B_{--} &B_{+-}
    \end{pmatrix},\quad
    Y_{2}=\begin{pmatrix}
        -C_{-+} & C_{++}+B_{-+}G-\int_{-\infty}^{+\infty}\tau e^{-2i\theta(\tau)}d\tau\\-C_{--}& C_{+-}+B_{--}G,
    \end{pmatrix},
\end{equation}
with $b_{++}$ and $G$ defined in \eqref{b++}, and $B_{ij}, C_{ij}$ given in \eqref{eq:X12}.
\item(4)
 $Y(\lambda)=O(\ln|\lambda\mp1|)$, as $\lambda\to\pm1$.

\end{description}

Let 
\begin{equation}\label{Psi}
	\Psi(\lambda)=Y(\lambda)e^{-i\theta(\lambda)\sigma_{3} }.
\end{equation}
We have the following Lax pair  \cite{AKNS, ZS1972}:
	\begin{equation}\label{lax pair}
		\left\{\begin{array}{l}
		\Psi_{x}(\lambda;x,t)=L_{1}(\lambda;x,t)\Psi(\lambda;x,t),\\
		\Psi_{t}(\lambda;x,t)=L_{2}(\lambda;x,t)\Psi(\lambda;x,t),
			\end{array}\right.
	\end{equation}
	where 
	\begin{equation} \label{eqL2}   
		L_{1}(x,t)=-i\lambda\sigma_{3}+i[\sigma_{3},Y_{1}],
	\end{equation}
\begin{equation}\label{eqL3}
	L_{2}(x,t)=-i\lambda^{2}\sigma_{3}+i\lambda[\sigma_{3},Y_{1}]+i[\sigma_{3},Y_{2}]-i[\sigma_{3},Y_{1}]Y_{1},
	\end{equation}
	with $Y_{1}$ and $Y_{2}$ given in \eqref{Y}. The zero-curvature equation 
 \begin{equation}
 \frac{\partial L_{2}}{\partial x}-\frac{\partial L_{1}}{\partial t}-[L_{1},L_{2}]=0,\end{equation}
gives us  the separated NLS equations \eqref{1qnls}.

According to \cite{IIKS1990} and \cite[Chapter XIV.5]{KBI1993}, we have
    \begin{equation}
        \partial_{x}D=-2iB_{+-},\quad \partial_{t}D=-2iGB_{--}-2i(C_{+-}+C_{-+}),
    \end{equation}
    where $G$, $B_{ij}$ and $C_{ij}$ appeared in \eqref{Y}. 
Therefore, the derivatives of  the determinant \eqref{D} and  the associate solutions of the separated NLS equations $b_{++}$ and  $B_{--}$,  defined by  \eqref{BC} and \eqref{b++} can be expressed in terms of the elements of the solution to the RH problem. 
\begin{proposition} \label{expression}
The derivatives of  $D$ in \eqref{D} and the associate solutions of the separated NLS equations $b_{++}$ and  $B_{--}$,  defined by  \eqref{BC} and \eqref{b++},  can be expressed in terms of the elements of  $Y_{1}$ and $Y_{2}$ in \eqref{Y}

\begin{equation}
    \partial_{t}D=2i\left((Y_{2})_{11}-(Y_{2})_{22}\right),
\end{equation}
\begin{equation}
    \partial_{x}D=2i(Y_{1})_{11}=-2i(Y_{1})_{22},
\end{equation}
    \begin{equation}
	b_{++}= (Y_{1})_{12},
\end{equation}
\begin{equation}
	B_{--}=-\left(Y_{1}\right)_{21}.
\end{equation}
\end{proposition}

\subsection{RH problem for the PIV equation}\label{PainleveIV}
In this section, we recall the RH problem for the PIV equation \eqref{p4} constructed in \cite[Chapter 5.1]{FIK2006}. A particular solution of the PIV equation with special parameters plays important roles in the description of the asymptotics of the logarithmic derivative of the Fredholm determinant and the corresponding solutions of the NLS equations in \eqref{1qnls} in the transition region.

\subsubsection*{RH problem for $\Psi$ }
\begin{description}
\item(1)
$\Psi(\xi,s)$ ($\Psi(\xi)$ for short) is analytic for $\xi\in\mathbb{C}\setminus{\cup^{4}_{i=1}\Sigma_{i}}$, where $\Sigma_{i}$, $i=1,\dots,4$, are shown in Fig. \ref{PIV1}. 
\begin{figure}
    \centering
    \includegraphics[width=0.7\linewidth]{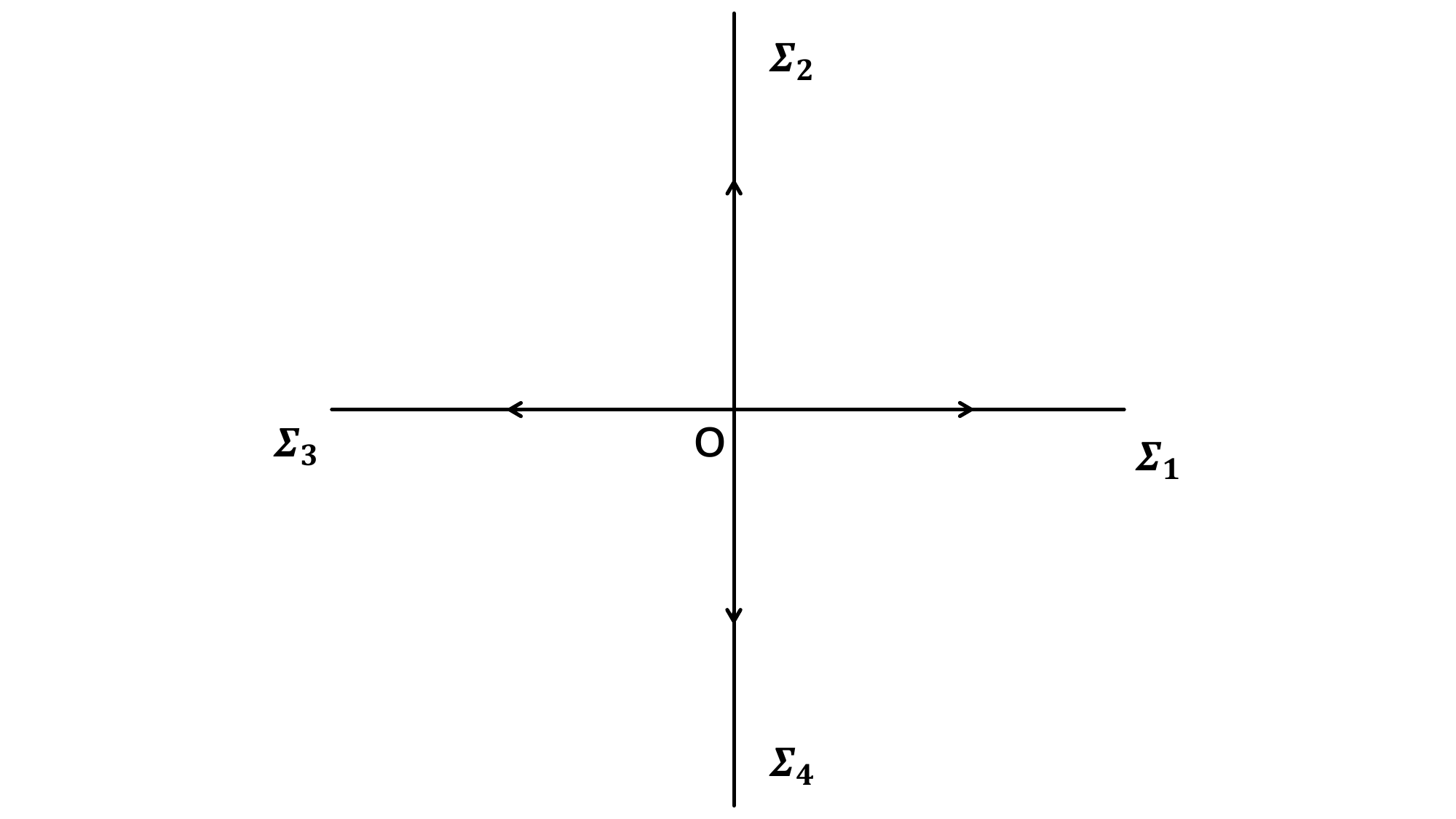}
    \caption{The jump contour of the RH problem for $\Psi$}
    \label{PIV1}
\end{figure}

\item(2)
$\Psi_{+}(\xi)=\Psi_{-}(\xi)S_{i}$, $\xi\in\Sigma_{i}$, $i=1,2,3$; $\Psi_{+}(\xi)=\Psi_{-}(\xi)S_{4}e^{2\pi i\Theta_{\infty}\sigma_{3}}$ for $\xi\in \Sigma_{4}$, where
\begin{align*}
S_{1}=\begin{pmatrix}
    1&0\\s_{1}&1
\end{pmatrix},\quad
S_{2}=\begin{pmatrix}
    1&s_{2}\\0&1
\end{pmatrix},\quad
S_{3}=\begin{pmatrix}
    1&0\\s_{3}&1
\end{pmatrix},\quad
S_{4}=\begin{pmatrix}
    1&s_{4}\\0&1
\end{pmatrix}.
\end{align*}
The Stokes multipliers $s_{i}$, $i=1,\dots,4$, satisfy
 \begin{equation}
(1+s_{2}s_{3})e^{2\pi i\Theta_{\infty}}+[s_{1}s_{4}+(1+s_{3}s_{4})(1+s_{1}s_{2})]e^{-2\pi i\Theta_{\infty}}=2\cos 2\pi\Theta.
\end{equation}
\item(3)
As $\xi\to\infty$,
\begin{equation}\label{Psiinfty}
	\Psi(\xi)=\left(I+\frac{\Psi_{1}(s)}{\xi}+\frac{\Psi_{2}(s)}{\xi^{2}}+O(\xi^{-3})\right)e^{(\frac{\xi^{2}}{2}+s\xi)\sigma_{3}} \xi^{-\Theta_{\infty}\sigma_{3}},
\end{equation}
where the branch for  $\xi^{\Theta_{\infty}}$ is chosen such that $\arg \xi\in(-\frac{\pi}{2},\frac{3\pi}{2})$.

\item(4)
As $\xi\to 0$, for $\Theta\ne 0$, $\Psi(\xi)=\Psi^{(0)}(\xi)\xi^{\Theta\sigma_{3}}$,  where $\Psi^{(0)}(\xi)$ is analytic near $\xi=0$. For $\Theta=0$, we have $\Psi(\xi)=O(\ln |\xi|)$, as $\xi\to0$.

\end{description}

Then, the Painlev\'e IV tanscendent $u$, and the quantities $y$ and $H$   can be expressed in terms of the elements of $\Psi_{1}$ and $\Psi_{2}$ 
as follows:

\begin{equation}\label{y(s)}
    y(s)=-2(\Psi_{1})_{12}(s),
\end{equation}
\begin{equation}\label{H(s)}
    H(s)=(\Psi_{1})_{22}(s),
\end{equation}
\begin{equation}\label{1u(s)}
	u(s)=-2s-\frac{d}{ds}\ln(\Psi_{1})_{12}(s),
\end{equation}
or
\begin{equation}\label{u(s)}
    u(s)=2(\Psi_{1})_{22}(s)-2s-2\frac{(\Psi_{2})_{12}(s)}{(\Psi_{1})_{12}(s)}.
\end{equation}
In our situation,  we take the parameters $\Theta=0$ and $\Theta_{\infty}=\frac{1}{2}$, and the Stokes multipliers $s_{1}=s_{2}=2i$ and $s_{3}=s_{4}=0$.

\section{Asymptotic analysis in the space-like region }\label{sec.s}
In this section, we derive the large $x$ asymptotics of the derivatives of $D$ defined in \eqref{D} and the corresponding solutions of the separated NLS equations  in the space-like region $\frac{x}{2t}>1+\delta$, for any small but fixed $\delta>0$, by performing Deift-Zhou nonlinear steepest descent analysis \cite{DZ1993,DMVZ19991,DMVZ19992}  of the RH problem for $Y$.

\subsection{Deformation of the jump contour}\label{1.1}
Define 
\begin{figure}
    \centering
    \includegraphics[width=0.7\linewidth]{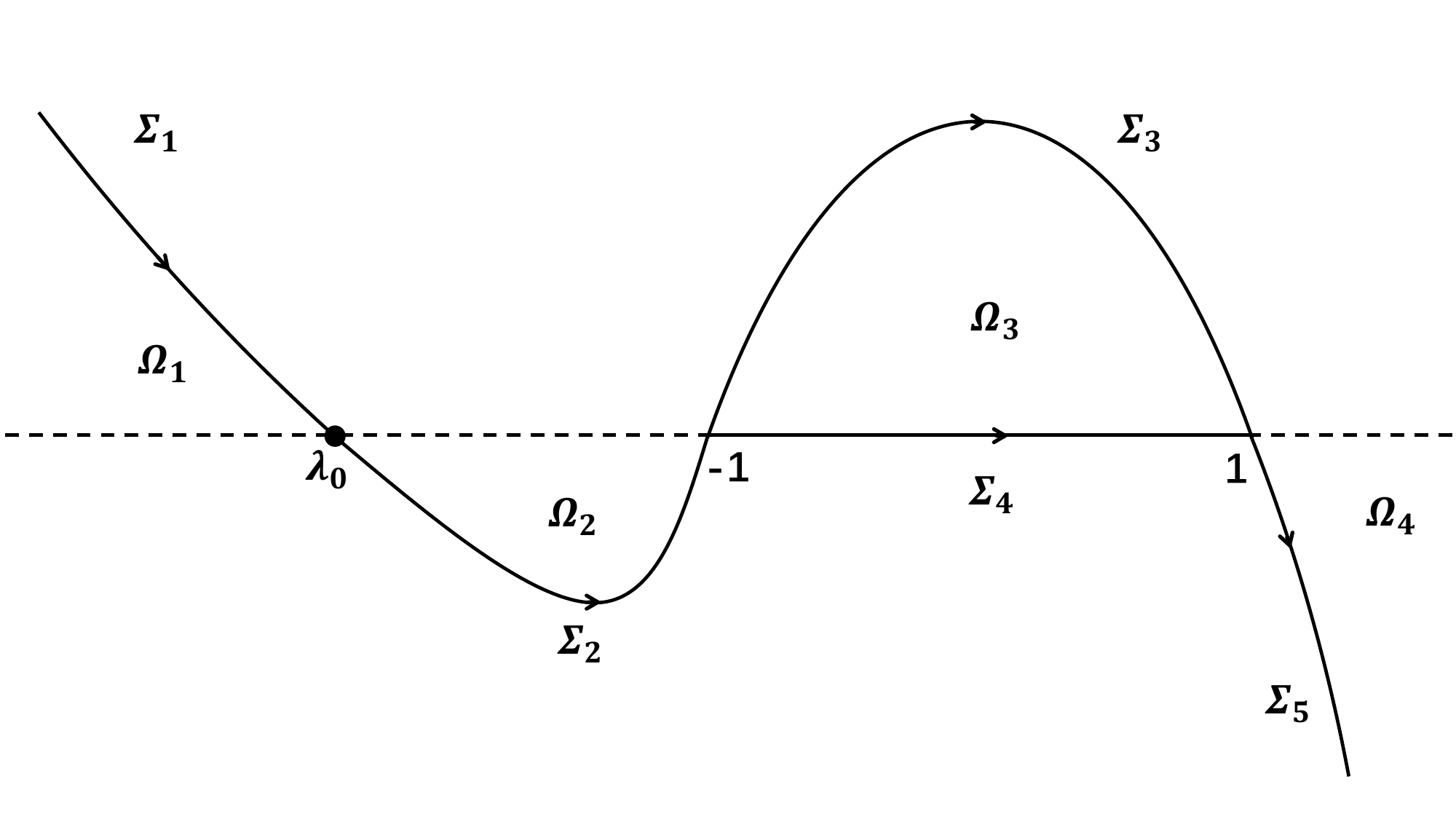}
    \caption{Deformation of the jump contour}
    \label{f1}
\end{figure}
\begin{equation}
	T(\lambda)=
	\left\{\begin{array}{ll}
		 Y(\lambda)\begin{pmatrix}
		 	1& -2\pi ie^{-2i\theta(\lambda)}\\
		 	0&1
		 \end{pmatrix},&\quad \lambda\in \Omega_{1},\\
		 Y(\lambda)\begin{pmatrix}
		 	1& 2\pi ie^{-2i\theta(\lambda)}\\
		 	0&1
		 \end{pmatrix},&\quad \lambda\in \Omega_{2}\cup\Omega_{4},\\
		 Y(\lambda)\begin{pmatrix}
		 	1& 0\\
		 	\frac{2i}{\pi}e^{2i\theta(\lambda)}&1
		 \end{pmatrix}, &\quad \lambda\in\Omega_{3},\\
          Y(\lambda),&\quad \lambda\in\mathbb{C}\setminus\bigcup_{i=1}^{4}\Omega_{i},
	\end{array}\right.
\end{equation}
where the regions $\Omega_{i}$, $i=1,\dots,4$, are illustrated in Fig. \ref{f1}. Then $T$ solves the following RH problem.

\subsubsection*{RH problem for $T$}
\begin{description}
	\item(1)
    $T(\lambda)$ is analytic for $\lambda\in\mathbb{C}\setminus \Sigma$, where $\Sigma$ is shown in Fig \ref{f1}.
    
\item(2)
$T_{+}(\lambda)=T_{+}(\lambda)J_{T}(\lambda)$ , $\lambda\in\Sigma$, where
\begin{equation}\label{1JT}
	J_{T}(\lambda)=\left\{\begin{array}{ll}
	 \begin{pmatrix}
	 1	& 2\pi ie^{-2i\theta(\lambda)}\\
	 0	&1
	 \end{pmatrix}, &\quad \lambda\in\Sigma_{1}\cup\Sigma_{2}\cup\Sigma_{5},\\
	\begin{pmatrix}
		1& 0\\
	-\frac{2i}{\pi}e^{2i\theta(\lambda)}	&1
	\end{pmatrix}, &\quad\lambda\in\Sigma_{3}	,\\
	-I, &\quad\lambda\in\Sigma_{4}
		 .
	\end{array}\right.
\end{equation}

\item(3)
As $\lambda\to\infty$, $T(\lambda )=I+O\left(\frac{1}{\lambda}\right)$.

\item(4)
As $\lambda\to\pm1$, $T(\lambda)=O(\ln(\lambda\mp1))$.
\end{description}

\subsection{Global parametrix}\label{1.2}
From \eqref{1JT}, we have $J_{T}(\lambda)\to I$, as $x\to+\infty$ for $ \lambda\in\Sigma\setminus\Sigma_{4}$. As $x\to+\infty$, it is expected that $T$ can be approximated by a solution to the following RH problem with the remaining jump matrix along the interval $(-1,1)$.

\subsubsection*{RH problem for $P^{(\infty)}$}
\begin{description}
	\item(1)
    $P^{(\infty)}(\lambda)$ is analytic for $\lambda\in\mathbb{C}\setminus [-1,1]$.

\item(2)
$P^{(\infty)}_{+}(\lambda)=-P^{(\infty)}_{-}(\lambda)$ , $\lambda\in(-1,1)$.

\item(3)
$P^{(\infty)}(\lambda )=I+O\left(\frac{1}{\lambda}\right)$, as $\lambda\to\infty$.
\end{description}

The solution to the RH problem for $P^{(\infty)}$ can be constructed as follows:
\begin{equation}\label{N}
	P^{(\infty)}(\lambda)=\left(\frac{\lambda-1}{\lambda+1}\right)^{\frac{1}{2}\sigma_{3}},
\end{equation}
where $\left(\frac{\lambda-1}{\lambda+1}\right)^{\frac{1}{2}}$ takes the branch cut along $[-1,1]$ and behaves like $1$ as $\lambda\to\infty$.

\subsection{Local parametrices near $\lambda=\pm 1$}\label{1.3}
In this subsection, we seek two parametrices $P^{(\pm 1)}$ that satisfy the same jump conditions as $T$ on $\Sigma$ in the neighborhoods $U(\pm 1,\delta)$, for some $\delta>0$.

\subsubsection{Local parametrix near $\lambda=-1$}\label{1.3.1}
In this section, we construct the solution to the following RH problem for $P^{(-1)}$.
\subsubsection*{RH problem for $P^{(-1)}$}
\begin{description}
	\item(1)
    $P^{(-1)}(\lambda)$ is analytic for $\lambda\in U(-1,\delta)\setminus \Sigma$.

\item(2)
$P^{(-1)}(\lambda)$ has the same jumps as $T(\lambda)$ on $U(-1,\delta)\cap\Sigma$.

\item(3)
On the boundary $\partial U(-1,\delta)$, $P^{(-1)}(\lambda)$ satisfies 
\begin{equation}\label{1P-1N-1}
	P^{(-1)}(\lambda)\{P^{(\infty)}(\lambda)\}^{-1}=\begin{pmatrix}  1& \frac{\lambda-1}{\lambda+1}\pi e^{2i(x-t)} \\ 0&1 \end{pmatrix}+O(x^{-1}), \quad x \to +\infty.
\end{equation}

\end{description}

We define the following conformal mapping
\begin{equation}\label{1cm-1}
	\xi(\lambda)=2\left[\theta(\lambda)-\theta(-1)\right]=2(x-2t)(\lambda+1)+2t(\lambda+1)^{2}.
\end{equation}
As $\lambda\to -1$, we have
\begin{equation}\label{1Cm-1}
	\xi(\lambda)\sim2(x-2t)(\lambda+1).
\end{equation}

Let $\Phi^{(CHF)}$ be the confluent hypergeometric parametrix with the parameter $\beta=\frac{1}{2}$, as given in Appendix \ref{CHF}. The solution to the above RH problem can be constructed as follows:
\begin{equation}\label{1P-1}
	P^{(-1)}(\lambda)=E^{(-1)}(\lambda)P^{(-1)}_{0}(\xi(\lambda))e^{i(t\lambda^{2}+x\lambda)\sigma_{3}},\quad \lambda\in U(-1,\delta),
\end{equation}
where
\begin{equation}\label{1P0-1}
	P^{(-1)}_{0}(\xi)=
	\left\{\begin{array}{ll}
	\Phi^{(CHF)}(\xi)(e^{\pi i}\pi)^{-\frac{1}{2}\sigma_{3}}, &\quad \arg\xi\in(0,\frac{\pi}{3}) \cup (\frac{2\pi}{3},\pi),  \\
	\Phi^{(CHF)}(\xi)\begin{pmatrix} 1 & 0\\ -i&1 \end{pmatrix}(e^{\pi i}\pi)^{-\frac{1}{2}\sigma_{3}}, &\quad \arg\xi\in(\frac{\pi}{3},\frac{\pi}{2}),  \\
	\Phi^{(CHF)}(\xi)\begin{pmatrix} 1 & 0\\ i&1 \end{pmatrix}(e^{\pi i}\pi)^{-\frac{1}{2}\sigma_{3}}, &\quad \arg\xi\in(\frac{\pi}{2},\frac{2\pi}{3}),  \\
	\Phi^{(CHF)}(\xi)\begin{pmatrix} 0 & i \\ i&0 \end{pmatrix}(e^{\pi i}\pi)^{-\frac{1}{2}\sigma_{3}}, &\quad \arg\xi\in(\pi,\frac{4\pi}{3}) \cup (-\frac{\pi}{3}, 0),  \\
	\Phi^{(CHF)}(\xi)\begin{pmatrix} 0 & i \\ i&1 \end{pmatrix}(e^{\pi i}\pi)^{-\frac{1}{2}\sigma_{3}}, &\quad \arg\xi\in(\frac{4\pi}{3},\frac{3\pi}{2}), \\
	\Phi^{(CHF)}(\xi)\begin{pmatrix}  0& i \\ i&-1 \end{pmatrix}(e^{\pi i}\pi)^{-\frac{1}{2}\sigma_{3}}, &\quad \arg\xi\in(-\frac{\pi}{2},-\frac{\pi}{3}),
    \end{array}\right.
\end{equation}
and
\begin{equation}\label{1E-1}
	E^{(-1)}(\lambda)=P^{(\infty)}(\lambda)\xi(\lambda)^{\frac{1}{2}\sigma_{3}}e^{i(x-t)\sigma_{3}}(e^{\pi i}\pi)^{\frac{1}{2}\sigma_{3}}.
\end{equation}
It follows from \eqref{N} and \eqref{1Cm-1} that $E^{(-1)}(\lambda)$ is analytic for $\lambda\in U(-1,\delta)$. From \eqref{N}, \eqref{1P-1}-\eqref{1E-1} and \eqref{CHFat00}, we have \eqref{1P-1N-1}.

\subsubsection{Local parametrix near $\lambda=1$}\label{1.3.2}
In this section, we seek the solution to the following RH problem for $P^{(1)}$.

\subsubsection*{RH problem for $P^{(1)}$}
\begin{description}
	\item(1)
    $P^{(1)}(\lambda)$ is analytic for $\lambda\in U(1,\delta)\setminus \Sigma$.

\item(2)
$P^{(1)}(\lambda)$ has the same jumps as $T(\lambda)$ on $U(1,\delta)\cap\Sigma$.

\item(3)
On the boundary $\partial U(1,\delta)$, $P^{(1)}(\lambda)$ satisfies 
\begin{equation}\label{1P1N-1}
	P^{(1)}(\lambda)\{P^{(\infty)}(\lambda)\}^{-1}=\begin{pmatrix}  1&  0\\-\frac{\lambda+1}{(\lambda-1)\pi} e^{2i(x+t)} &1 \end{pmatrix}+O(x^{-1}), \quad x \to +\infty.
\end{equation}

\end{description}

We define the following conformal mapping
\begin{equation}\label{1cm1}
	\xi(\lambda)=2[\theta(\lambda)-\theta(1)]=2(x+2t)(\lambda-1)+2t(\lambda-1)^{2}.
\end{equation}
As $\lambda\to 1$, we have
\begin{equation}\label{1Cm1}
	\xi(\lambda)\sim2(x+2t)(\lambda-1).
\end{equation}

Let $\Phi^{(CHF)}$ be the confluent hypergeometric parametrix with the parameter $\beta=-\frac{1}{2}$, as given in Appendix \ref{CHF}. The solution to the above RH problem can be constructed as follows:
\begin{equation}\label{1P1}
	P^{(1)}(\lambda)=E^{(1)}(\lambda)P^{(1)}_{0}(\xi(\lambda))e^{i(t\lambda^{2}+x\lambda)\sigma_{3}}, \quad \lambda\in U(1,\delta),
\end{equation}
where
\begin{equation}\label{1P01}
	P^{(1)}_{0}(\xi)=
	\left\{\begin{array}{ll}
		\Phi^{(CHF)}(\xi)(e^{\pi i}\pi)^{-\frac{1}{2}\sigma_{3}}, &\quad \arg\xi\in(0,\frac{\pi}{3}) \cup (\frac{2\pi}{3},\pi),  \\
		\Phi^{(CHF)}(\xi)\begin{pmatrix} 1 &0 \\ i&1 \end{pmatrix}(e^{\pi i}\pi)^{-\frac{1}{2}\sigma_{3}}, &\quad \arg\xi\in(\frac{\pi}{3},\frac{\pi}{2}),  \\
		\Phi^{(CHF)}(\xi)\begin{pmatrix} 1 &0 \\ -i&1 \end{pmatrix}(e^{\pi i}\pi)^{-\frac{1}{2}\sigma_{3}}, &\quad \arg\xi\in(\frac{\pi}{2},\frac{2\pi}{3}),  \\
		\Phi^{(CHF)}(\xi)\begin{pmatrix} 0 & i \\ i& 0\end{pmatrix}(e^{\pi i}\pi)^{-\frac{1}{2}\sigma_{3}}, &\quad \arg\xi\in(\pi,\frac{4\pi}{3}) \cup (-\frac{\pi}{3}, 0),  \\
		\Phi^{(CHF)}(\xi)\begin{pmatrix} 0& i \\ i&-1 \end{pmatrix}(e^{\pi i}\pi)^{-\frac{1}{2}\sigma_{3}}, &\quad \arg\xi\in(\frac{4\pi}{3},\frac{3\pi}{2}), \\
		\Phi^{(CHF)}(\xi)\begin{pmatrix} 0 & i \\ i& 1 \end{pmatrix}(e^{\pi i}\pi)^{-\frac{1}{2}\sigma_{3}}, &\quad \arg\xi\in(-\frac{\pi}{2},-\frac{\pi}{3}),
	\end{array}\right.
\end{equation}
and
\begin{equation}\label{1E1}
	E^{(1)}(\lambda)=P^{(\infty)}(\lambda)\xi^{-\frac{1}{2}\sigma_{3}}e^{-i(x+t)\sigma_{3}}(e^{\pi i}\pi)^{\frac{1}{2}\sigma_{3}}.
\end{equation}
It follows from \eqref{N} and \eqref{1Cm1} that $E^{(1)}(\lambda)$ is analytic for $\lambda\in U(1,\delta)$. From \eqref{N}, \eqref{1P1}-\eqref{1E1} and \eqref{CHFat00}, we have \eqref{1P1N-1}.

\subsection{Local parametrix near the stationary point }\label{1.4}
In this subsection, we seek a parametrix $P^{(0)}$ that satisfies the same jump conditions as  $T$ on $\Sigma$ within the neighborhood $U(\lambda_{0},\delta)$, for some $\delta>0$, where $\lambda_{0}=-\frac{x}{2t}$.

\subsubsection*{RH problem for $P^{(0)}$}
\begin{description}
	\item(1)
    $P^{(0)}(\lambda)$ is analytic for $\lambda\in U(\lambda_{0},\delta)\setminus \Sigma$.

\item(2)
$P^{(0)}(\lambda)$ has the same jumps as $T(\lambda)$ on $U(\lambda_{0},\delta)\cap\Sigma$.

\item(3)
On the boundary $\partial U(\lambda_{0},\delta)$, $P^{(0)}(\lambda)$ satisfies 
\begin{equation}\label{1P0N-1}
	P^{(0)}(\lambda)\{P^{(\infty)}(\lambda)\}^{-1}=I+O(x^{-\frac{1}{2}}), \quad x \to +\infty.
\end{equation}

\end{description}

The solution to the above RH problem can be constructed by using the Cauchy integral
\begin{equation}\label{1P0}
	P^{(0)}(\lambda)=P^{(\infty)}(\lambda)\left(I+\int_{\Gamma_{0}} \frac{e^{-2i(ts^{2}+xs)}}{s-\lambda}ds\begin{pmatrix}
	0	& 1\\
	0	& 0
	\end{pmatrix}\right),
\end{equation}
where $P^{(\infty)}$ is defined in \eqref{N}. The integral contour is defined as $\Gamma_{0}=U(\lambda_{0},\delta)\cap(\Sigma_{1}\cup\Sigma_{2})$, where $\Sigma_{1}$ and $\Sigma_{2}$ are shown in Fig. \ref{f1}.
As $x\to+\infty$, we have the asymptotics of the integral by using the steepest descent method
\begin{equation}\begin{aligned}\label{1p0n-1}
    \int_{\Gamma_{0}} \frac{e^{-2i(ts^{2}+xs)}}{s-\lambda}ds&= e^{i\frac{x^{2}}{2t}}\int_{\Gamma_{0}}\frac{e^{-2it(s+\frac{x}{2t})^{2}}}{(s+\frac{x}{2t})-(\lambda+\frac{x}{2t})}ds \\
    &= e^{i\frac{x^{2}}{2t}}\int_{\Gamma_{1}}\frac{\frac{x}{2t}e^{-2i\frac{x^{2}}{4t}(u+1)^{2}}}{\frac{x}{2t}(u+1)-(\lambda+\frac{x}{2t})}du\\
  &=-\sqrt{\frac{\pi}{2t}}e^{i\left(\frac{x^{2}}{2t}-\frac{\pi}{4}\right)}(\lambda-\lambda_{0})^{-1}+O(x^{-\frac{3}{2}}),
\end{aligned}\end{equation}
where integral contour $\Gamma_{1}=\frac{2t}{x}\Gamma_{0}$. From \eqref{N}, \eqref{1P0} and \eqref{1p0n-1}, we have \eqref{1P0N-1}.

\subsection{RH problem for $M$}\label{1.5}
As $x\to+\infty$, from \eqref{1P-1N-1} and \eqref{1P1N-1}, we see that $P^{(-1)}\left\{P^{(\infty)}\right\}^{-1}$ and $P^{(1)}\left\{P^{(\infty)}\right\}^{-1}$ do not tend to the identity matrix on $\partial U(-1,\delta)$ and $\partial U(1,\delta)$, respectively. To resolve this issue, we construct a matrix-valued function $M(\lambda)$, which solves the remaining jumps on $\partial U(-1,\delta)$ and $ \partial U(1,\delta)$.

\subsubsection*{RH problem for $M$}
\begin{description}
    \item (1)
    $M(\lambda)$ is analytic for $\lambda\in\mathbb{C}\setminus\left(\partial U(-1,\delta)\cup \partial  U(1,\delta)\right)$.
 
    \item(2) On the boundaries $\partial U(-1,\delta)$ and $\partial U(1,\delta)$, we have
    \begin{equation}
        M_{+}(\lambda)=M_{-}(\lambda)\begin{pmatrix}  1& \frac{\lambda-1}{\lambda+1}\pi e^{2i(x-t)} \\ 0&1 \end{pmatrix}, \quad \lambda\in\partial U(-1,\delta),
        \end{equation}
        \begin{equation}
        M_{+}(\lambda)=M_{-}(\lambda)\begin{pmatrix}  1&  0\\-\frac{\lambda+1}{\pi(\lambda-1)} e^{2i(x+t)} &1 \end{pmatrix}, \quad\lambda\in\partial U(1,\delta).
        \end{equation}
 
    \item(3)
    As $\lambda\to\infty$, $M(\lambda)=I+O\left(\frac{1}{\lambda}\right)$.
\end{description}

Let
\begin{equation}\label{A}
	A(\lambda)=I+\frac{B}{\lambda+1}+\frac{C}{\lambda-1},
\end{equation}
then we seek a solution to the above RH problem of the following form:
\begin{equation}\label{M}
    M(\lambda)=\left\{\begin{array}{ll}
A(\lambda)\begin{pmatrix}  1& -\frac{\lambda-1}{\lambda+1}\pi e^{2i(x-t)} \\ 0&1 \end{pmatrix}, &\quad \lambda\in U(-1,\delta),\\
A(\lambda)\begin{pmatrix}  1&  0\\\frac{\lambda+1}{\pi(\lambda-1)} e^{2i(x+t)} &1 \end{pmatrix}, &\quad  \lambda\in U(1,\delta),\\
A(\lambda), &\quad \lambda\in \mathbb{C}\setminus(U(-1,\delta)\cup U(1,\delta)).
    \end{array}\right.
\end{equation}
By the condition that $M$ is analytic near $\lambda=\pm1$, we determine the coefficients in \eqref{A}
\begin{equation}
	B=\begin{pmatrix}
	0	& -\frac{2\pi e^{2i(x-t)}}{1+e^{4ix}}\\
	0	&-\frac{2e^{4ix}}{1+e^{4ix}}
	\end{pmatrix},\quad C=\begin{pmatrix}
\frac{2e^{4ix}}{1+e^{4ix}}	& 0 \\
-\frac{\frac{2}{\pi} e^{2i(x+t)}}{1+e^{4ix}}	& 0
	\end{pmatrix}.
\end{equation}
From \eqref{N}, \eqref{1P-1}, \eqref{1P1}, \eqref{1P0} and \eqref{M}, we have as $x\to+\infty$,
\begin{equation}\label{1P1n-1}
    M_{-}(\lambda)P^{(1)}(\lambda)\{P^{(\infty)}(\lambda)\}^{-1}M_{+}^{-1}(\lambda)=I+O\left(x^{-1}\right),\quad \lambda\in\partial U(1,\delta),
\end{equation}
\begin{equation}\label{1P-1n-1}
    M_{-}(\lambda)P^{(-1)}(\lambda)\{P^{(\infty)}(\lambda)\}^{-1}M_{+}^{-1}(\lambda)=I+O\left(x^{-1}\right),\quad \lambda\in\partial U(-1,\delta),
\end{equation}
\begin{equation}\label{1P0n-1}
   M(\lambda) P^{(0)}(\lambda)\{P^{(\infty)}(\lambda)\}^{-1}M^{-1}(\lambda)=I+O\left(x^{-\frac{1}{2}}\right),\quad \lambda\in\partial U(\lambda_{0},\delta).
\end{equation}

\subsection{Final transformation}\label{1.6}
The final transformation is defined as
\begin{equation}
   R(\lambda)=
	\left\{\begin{array}{ll}
		T(\lambda)\left\{M(\lambda)P^{(\infty)}(\lambda)\right\}^{-1}, &\quad \lambda\in\mathbb{C}\setminus \left(U(\lambda_{0},\delta)\cup U(1,\delta)\cup U(-1,\delta)\right)  ,\\
		T(\lambda)\left\{M(\lambda)P^{(0)}(\lambda)\right\}^{-1}, &\quad\lambda\in U(\lambda_{0},\delta)\setminus\Sigma,\\
		T(\lambda)\left\{M(\lambda)P^{(1)}(\lambda)\right\}^{-1}, &\quad\lambda\in U(1,\delta)\setminus\Sigma,\\
		T(\lambda)\left\{M(\lambda)P^{(-1)}(\lambda)\right\}^{-1}, &\quad\lambda\in U(-1,\delta)\setminus\Sigma.
	\end{array}\right.
\end{equation}
Then $R$ fulfills the following RH problem.

\subsubsection*{RH problem for $R$}
\begin{description}
	\item(1)
    $R(\lambda)$ is analytic for $\lambda\in\mathbb{C}\setminus \Sigma$, where the contour is shown in Fig. \ref{f1R}.
\begin{figure}
    \centering
    \includegraphics[width=0.7\linewidth]{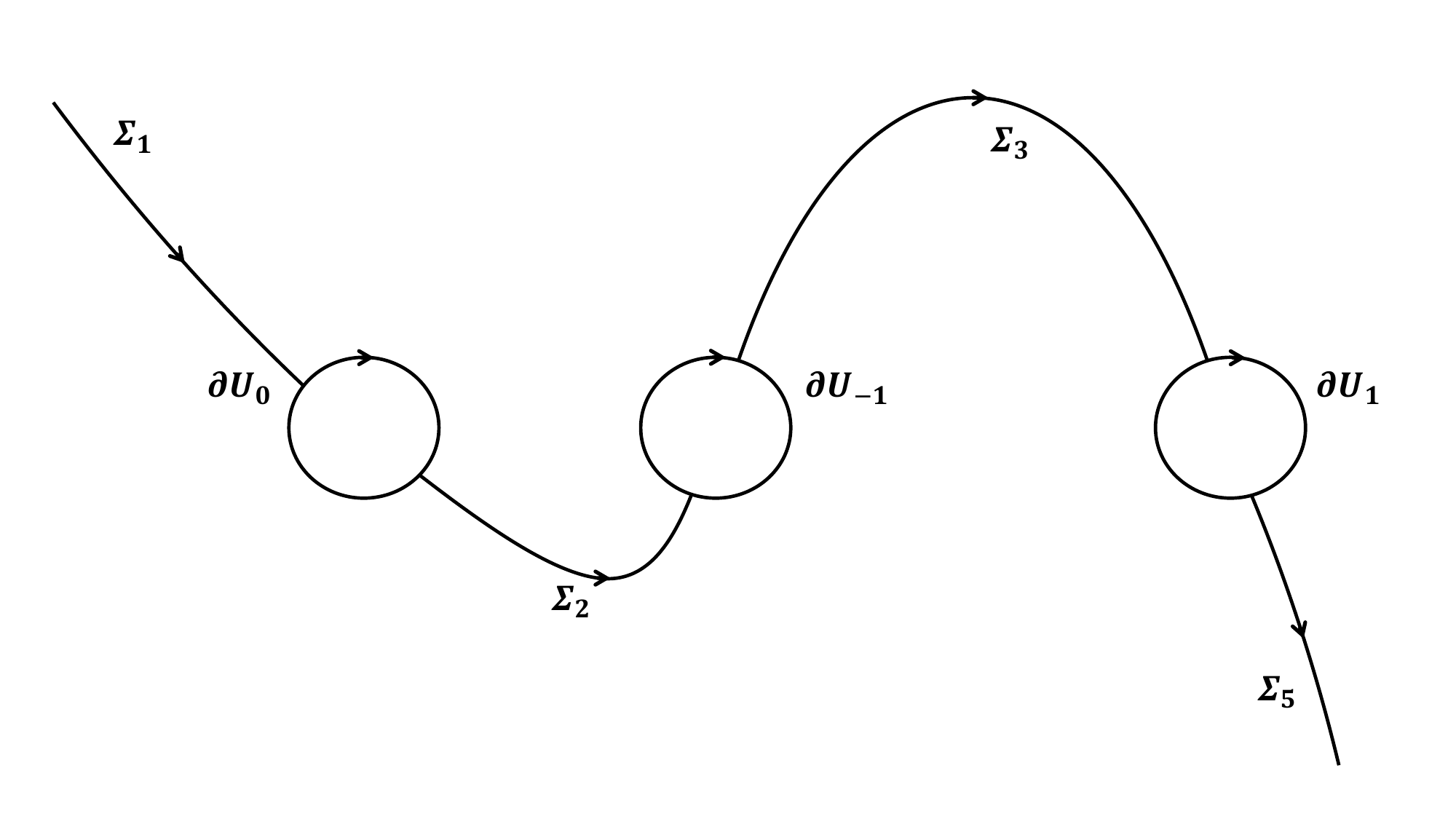}
    \caption{The jump contour of the RH problem for $R$}
    \label{f1R}
\end{figure}

\item(2)
$R_{+}(\lambda)= R_{-}(\lambda)J_{R}(\lambda)$, $ \lambda\in  \Sigma $, where
\begin{equation}
	J_{R}(\lambda)=\left\{\begin{array}{ll}
		M(\lambda)P^{(0)}(\lambda)\left\{P^{(\infty)}(\lambda)\right\}^{-1}M^{-1}(\lambda), &\quad \lambda\in \partial U(\lambda_{0},\delta),\\
		M_{-}(\lambda)P^{(1)}(\lambda)\left\{P^{(\infty)}(\lambda)\right\}^{-1}M_{+}^{-1}(\lambda), &\quad \lambda\in \partial U(1,\delta),\\
		M_{-}(\lambda)P^{(-1)}(\lambda)\left\{P^{(\infty)}(\lambda)\right\}^{-1}M_{+}^{-1}(\lambda), &\quad \lambda\in \partial U(-1,\delta),\\
		M(\lambda)P^{(\infty)}(\lambda)\begin{pmatrix} 1 & 2\pi i e^{-2i\theta(\lambda)}\\ 0&1 \end{pmatrix}\left\{P^{(\infty)}(\lambda)\right\}^{-1}M^{-1}(\lambda), &\quad \lambda\in \Sigma_{1}\cup \Sigma_{2}\cup \Sigma_{5},\\
		M(\lambda)P^{(\infty)}(\lambda)\begin{pmatrix} 1 &0 \\ -\frac{2i}{\pi}e^{2i\theta(\lambda)} &1 \end{pmatrix}\left\{P^{(\infty)}(\lambda)\right\}^{-1}M^{-1}(\lambda), &\quad \lambda\in \Sigma_{3}.
	\end{array}\right.		
\end{equation}

\item(3)
$R(\lambda )=I+O\left(\frac{1}{\lambda}\right)$, as $\lambda\to\infty$.
\end{description}

From the matching conditions \eqref{1P1n-1}-\eqref{1P0n-1}, we have as $x\to+\infty$,
\begin{equation}
	J_{R}(\lambda)=\left\{\begin{array}{ll}
	I+O(x^{-1}),&\quad \lambda\in\partial U(\pm1,\delta),\\
	I+O(x^{-\frac{1}{2}}),&\quad \lambda\in\partial U(\lambda_{0},\delta),\\
	I+O(e^{-c_{1}x}),&\quad \lambda\in\Sigma_{1}\cup\Sigma_{2}\cup\Sigma_{3}\cup\Sigma_{5},
	\end{array}\right.
\end{equation}
where $c_{1}$ is some positive constant. Then we have as $ x \to+ \infty$,
\begin{equation}
	R(\lambda)=I+O(x^{-\frac{1}{2}}),
\end{equation}
where the error term is uniform for $\lambda$ bounded away from the jump contour for $R$.

\subsection{Large $x$ asymptotics in the space-like region}\label{1.7}
By tracing back the series of invertible transformations
\begin{equation}
	Y \mapsto T \mapsto R,
\end{equation}
we obtain that as $x\to+\infty$,
\begin{equation}
	Y(\lambda)=R(\lambda)A(\lambda)P^{(\infty)}(\lambda),\quad\lambda\in\mathbb{C}\setminus\left(\cup_{i=1}^{4}\Omega_{i}\cup U(1,\delta)\cup U(-1,\delta)\cup U(\lambda_{0},\delta)\right),
\end{equation}
where $P^{(\infty)}$ and $A$ are defined in \eqref{N} and \eqref{A}, and the regions $\Omega_{i}$, $i=1,\dots,4$, are shown in Fig. \ref{f1}. From \eqref{N}, we have
\begin{equation}\label{Pinfty}
	P^{(\infty)}(\lambda)=I+\frac{P^{(\infty)}_{1}}{\lambda}+\frac{P^{(\infty)}_{2}}{\lambda^{2}}+O\left(\frac{1}{\lambda^{3}}\right), \quad \lambda\to \infty,
\end{equation}
where
\begin{equation}\label{1P1P2}
    P^{(\infty)}_{1}=-\sigma_3, \quad P^{(\infty)}_{2}=
		\frac{1}{2}	I.
\end{equation}
From \eqref{A}, we have
\begin{equation}\label{Ainfty}
	A(\lambda)=I+\frac{A_{1}}{\lambda}+\frac{A_{2}}{\lambda^{2}}+\left(\frac{1}{\lambda^{3}}\right), \quad \lambda\to \infty,
\end{equation}
where
\begin{equation}\label{1A1A2}
    A_{1}=\begin{pmatrix}
		\frac{2e^{4ix}}{1+e^{4ix}}	& -\frac{2\pi e^{2i(x-t)}}{1+e^{4ix}}\\
		-\frac{\frac{2}{\pi} e^{2i(x+t)}}{1+e^{4ix}}	& -\frac{2e^{4ix}}{1+e^{4ix}}
	\end{pmatrix},\quad A_{2}=\begin{pmatrix}
		\frac{2e^{4ix}}{1+e^{4ix}}	& \frac{2\pi e^{2i(x-t)}}{1+e^{4ix}}\\
		-\frac{\frac{2}{\pi} e^{2i(x+t)}}{1+e^{4ix}}	& \frac{2e^{4ix}}{1+e^{4ix}}
	\end{pmatrix}.
\end{equation}
We have the asymptotic expansion
\begin{equation}
R(\lambda)=I+\frac{R_{1}}{\lambda}+\frac{R_{2}}{\lambda^{2}}+O\left(\frac{1}{\lambda^{3}}\right),\quad \lambda\to\infty.
\end{equation}
As $x\to+\infty$, we have
\begin{equation}
	R(\lambda)=I+\frac{R^{(1)}(\lambda)}{x^{\frac{1}{2}}}+O\left(x^{-1}\right),
\end{equation}
where the error term is uniform for $\lambda$ bounded away from the jump contour for $R$. Here $R^{(1)}$ satisfies
\begin{equation}
	R^{(1)}_{+}(\lambda)-R^{(1)}_{-}(\lambda)=\Delta(\lambda),\quad \lambda\in\partial U(\lambda_{0},\delta),
\end{equation}
with
\begin{small}
\begin{equation}
    \Delta(\lambda)=-\sqrt{-\lambda_{0}\pi}\frac{e^{i\left(\frac{x^{2}}{2t}-\frac{\pi}{4}\right)}}{\lambda-\lambda_{0}}\begin{pmatrix}
	\frac{1}{\lambda+1}\frac{\frac{2}{\pi}e^{2i(x+t)}}{1+e^{4ix}}+\frac{1}{\lambda^{2}-1}\frac{\frac{4}{\pi}e^{2i(x+t)+4ix}}{(1+e^{4ix})^{2}}	& \frac{\lambda-1}{\lambda+1}+\frac{1}{\lambda+1}\frac{4e^{4ix}}{1+e^{4ix}} +\frac{1}{\lambda^{2}-1}\frac{4e^{8ix}}{(1+e^{4ix})^{2}} \\ -\frac{1}{\lambda^{2}-1}\frac{\frac{4}{\pi^{2}}e^{4i(x+t)}}{(1+e^{4ix})^{2}}
		&-\frac{1}{\lambda+1}\frac{\frac{2}{\pi}e^{2i(x+t)}}{1+e^{4ix}}-\frac{1}{\lambda^{2}-1}\frac{\frac{4}{\pi}e^{2i(x+t)+4ix}}{(1+e^{4ix})^{2}}
	\end{pmatrix}.
\end{equation}\end{small}
We obtain that
\begin{equation}
	R^{(1)}(\lambda)=	\left\{\begin{array}{ll}
	\frac{C}{\lambda-\lambda_{0}},&\quad\lambda\in \mathbb{C}\setminus U(\lambda_{0},\delta)	, \\	\frac{C}{\lambda-\lambda_{0}}-\Delta(\lambda),&\quad\lambda\in U(\lambda_{0},\delta),	
	\end{array}\right.
\end{equation}
where $C=\Res(\Delta(\lambda),\lambda_{0})$ is given by
\begin{small}
\begin{equation}
	C=-\sqrt{-\lambda_{0}\pi}e^{i\left(\frac{x^{2}}{2t}-\frac{\pi}{4}\right)}\begin{pmatrix}
	\frac{1}{\lambda_{0}+1}\frac{\frac{2}{\pi}e^{2i(x+t)}}{1+e^{4ix}}+\frac{1}{\lambda_{0}^{2}-1}\frac{\frac{4}{\pi}e^{2i(x+t)+4ix}}{(1+e^{4ix})^{2}}	& \frac{\lambda_{0}-1}{\lambda_{0}+1}+\frac{1}{\lambda_{0}+1}\frac{4e^{4ix}}{1+e^{4ix}} +\frac{1}{\lambda_{0}^{2}-1}\frac{4e^{8ix}}{(1+e^{4ix})^{2}} \\ -\frac{1}{\lambda_{0}^{2}-1}\frac{\frac{4}{\pi^{2}}e^{4i(x+t)}}{(1+e^{4ix})^{2}}
		&-\frac{1}{\lambda_{0}+1}\frac{\frac{2}{\pi}e^{2i(x+t)}}{1+e^{4ix}}-\frac{1}{\lambda_{0}^{2}-1}\frac{\frac{4}{\pi}e^{2i(x+t)+4ix}}{(1+e^{4ix})^{2}}
	\end{pmatrix}.
\end{equation}\end{small}
Expanding $R^{(1)}$ into the Taylor series at infinity, we obtain the asymptotics for $R_{1}$ and $R_{2}$:
\begin{equation}\label{1R1R2}
	R_{1}=\frac{C}{x^{\frac{1}{2}}}+O(x^{-1}), \quad 	R_{2}=\frac{\lambda_{0}C}{x^{\frac{1}{2}}}+O(x^{-1}), \quad x\to+\infty.
\end{equation}
Then, $Y$ can be expressed in the following form
\begin{equation}\label{1Y}
	Y(\lambda)=I+\frac{Y_{1}}{\lambda}+\frac{Y_{2}}{\lambda^{2}}+O\left(\frac{1}{\lambda^{3}}\right), \quad \lambda\to\infty,
\end{equation}
where
\begin{equation}\label{1Y1Y2}
	Y_{1}=R_{1}+A_{1}+P^{(\infty)}_{1}, \quad Y_{2}=R_{1}A_{1}+R_{1}P^{(\infty)}_{1}+A_{1}P^{(\infty)}_{1}+R_{2}+A_{2}+P^{(\infty)}_{2}.
\end{equation}
Here $P^{(\infty)}_{1}$, $P^{(\infty)}_{2}$, $A_{1}$, $A_{2}$, $R_{1}$ and $R_{2}$ are defined in \eqref{1P1P2}, \eqref{1A1A2} and \eqref{1R1R2}.

From \eqref{1Y}, \eqref{1Y1Y2} and Proposition \ref{expression}, we obtain the asymptotics of $\partial_{t} D$, $\partial_{x} D$, $b_{++}$ and $B_{--}$ as $x\to+\infty$ in the space-like region, as given in  \eqref{th1dt}-\eqref{th1B}, which complete the proof of Theorem \ref{space}.

\section{Asymptotic analysis in the time-like region }\label{sec.t}
In this section, we derive the large $t$ asymptotics of the derivatives of $D$ defined in \eqref{D} and the corresponding solutions of the separated NLS equations in the time-like region, where $\frac{x}{2t}<1-\delta$ for any small but fixed $\delta>0$, by performing Deift-Zhou nonlinear steepest descent analysis of the RH problem for $Y$.

\subsection{Deformation of the jump contour}\label{2.1}
Define 
\begin{figure}
    \centering
    \includegraphics[width=0.7\linewidth]{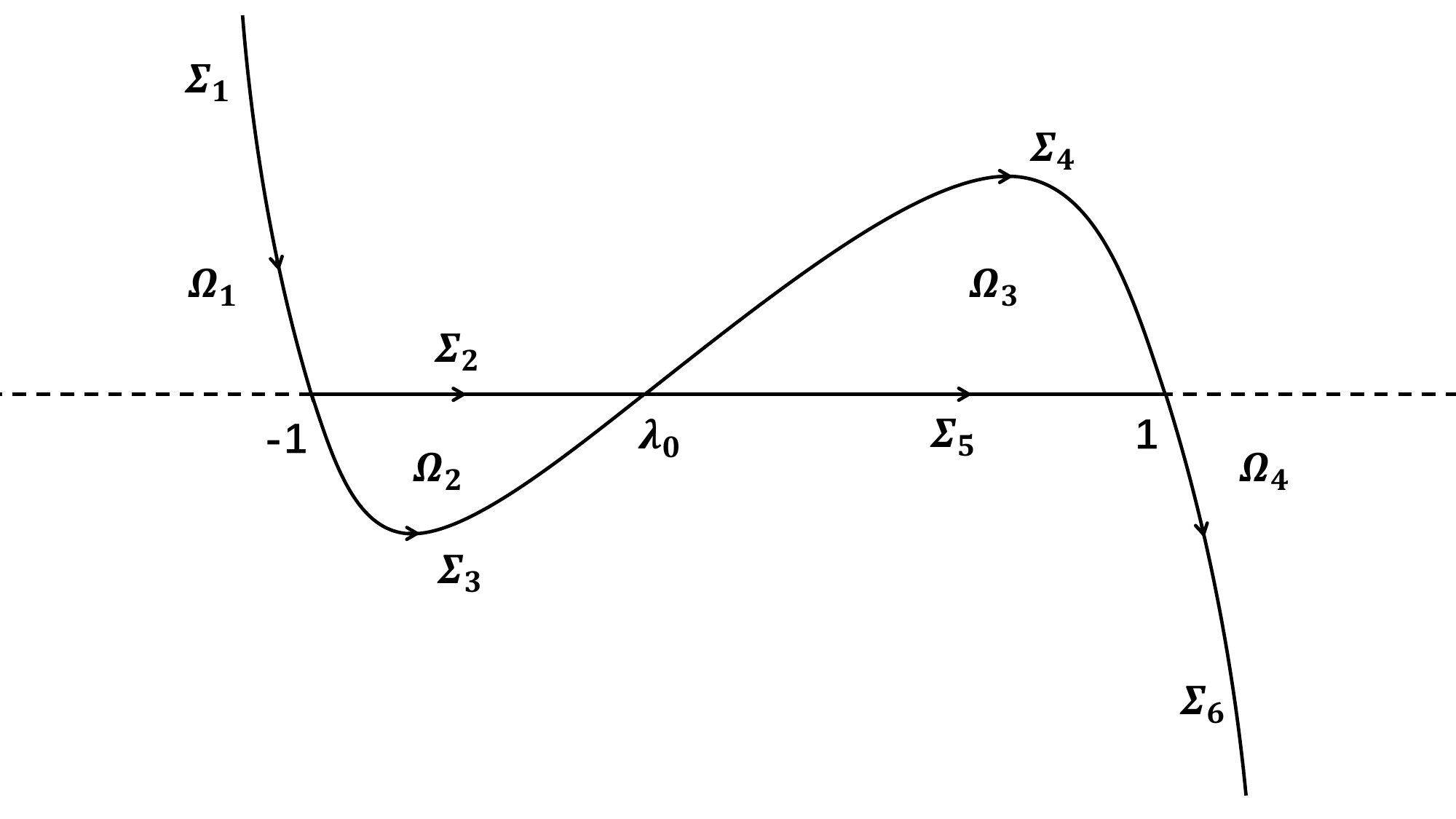}
    \caption{Deformation of the jump contour}
    \label{f2}
\end{figure}
\begin{equation}
	T(\lambda)=
	\left\{\begin{array}{ll}
		Y(\lambda)\begin{pmatrix}
			1& -2\pi ie^{-2i\theta(\lambda)}\\
			0&1
		\end{pmatrix},&\quad \lambda\in \Omega_{1},\\
	    Y(\lambda)\begin{pmatrix}
	    	1&0\\
	    	-\frac{2i}{\pi}e^{2i\theta(\lambda)}&1
	    \end{pmatrix}, &\quad \lambda\in\Omega_{2},\\
		Y(\lambda)\begin{pmatrix}
			1& 0\\
			\frac{2i}{\pi}e^{2i\theta(\lambda)}&1
		\end{pmatrix}, &\quad \lambda\in\Omega_{3},\\	
		Y(\lambda)\begin{pmatrix}
		1& 2\pi ie^{-2i\theta(\lambda)}\\
		0&1
		\end{pmatrix},&\quad \lambda\in \Omega_{4},\\
         Y(\lambda),&\quad \lambda\in\mathbb{C}\setminus\bigcup_{i=1}^{4}\Omega_{i},
	\end{array}\right.
\end{equation}
where the regions $\Omega_{i}$, $i=1,\dots,4$, are illustrated in Fig. \ref{f2}. Then $T$ solves the following RH problem.

\subsubsection*{RH problem for $T$}
\begin{description}
	\item(1)
    $T(\lambda)$ is analytic for $\lambda\in\mathbb{C}\setminus \Sigma$, where $\Sigma$ is shown in Fig \ref{f2}.

\item(2)
$T_{+}(\lambda)=T_{-}(\lambda)J_{T}(\lambda)$ , $\lambda\in\Sigma$, where
\begin{equation}\label{2JT}
	J_{T}(\lambda)=\left\{\begin{array}{ll}
		\begin{pmatrix}
			1	& 2\pi ie^{-2i\theta(\lambda)}\\
			0&1
		\end{pmatrix}, &\quad \lambda\in\Sigma_{1}\cup\Sigma_{6},	\\
		-I, &\quad\lambda\in\Sigma_{2}\cup\Sigma_{5},\\
		\begin{pmatrix}
		1& 0\\
		-\frac{2i}{\pi}e^{2i\theta(\lambda)}	&1
		\end{pmatrix}, &\quad\lambda\in\Sigma_{3}\cup\Sigma_{4}
		.
	\end{array}\right.
\end{equation}

\item(3)
  As $\lambda\to\infty$, $T(\lambda )= I+O\left(\frac{1}{\lambda}\right)$.

 \item(4)
As $\lambda\to\pm1$, $T(\lambda)=O(\ln(\lambda\mp1))$.

\end{description}

From \eqref{2JT}, we have $J_{T}(\lambda)\to I$, as $t\to+\infty$ for $\lambda\in\Sigma\setminus(\Sigma_{2}\cup\Sigma_{5})$. As $t\to+\infty$, it is expected that $T$ can be approximated by a solution to a RH problem with the jump matrix along the line segment $(-1,1)$. Therefore, we construct the same global parametrix as given in \eqref{N}.

\subsection{Local parametrix near $\lambda=-1$}\label{2.2}
In this subsection, we seek a parametrix $P^{(-1)}$ that satisfies the same jump conditions as $T$ on $\Sigma$ in the neighborhood $U(- 1,\delta) $, for some $\delta>0$.

\subsubsection*{RH problem for $P^{(-1)}$}
\begin{description}
	\item(1)
    $P^{(-1)}(\lambda)$ is analytic for $\lambda\in U(-1,\delta)\setminus \Sigma$.

\item(2)
$P^{(-1)}(\lambda)$ has the same jumps as $T(\lambda)$ on $U(-1,\delta)\cap\Sigma$.

\item(3)
 On the boundary $\partial U(-1,\delta)$,$P^{(-1)}(\lambda)$ satisfies 
\begin{equation}\label{2P-1N-1}
	P^{(-1)}(\lambda)\{P^{(\infty)}(\lambda)\}^{-1}M_{+}^{-1}(\lambda)=I+O(t^{-1}), \quad t \to +\infty,
\end{equation}
 where $M$ is defined in \eqref{M}.
\end{description}

We define the following conformal mapping
\begin{equation}\label{2CM}
	\xi(\lambda)=-2[\theta(\lambda)-\theta(-1)]=-2(x-2t)(\lambda+1)-2t(\lambda+1)^{2}.
\end{equation}
As $\lambda\to -1$, we have
\begin{equation}\label{2cm}
	\xi(\lambda)\sim2(2t-x)(\lambda+1).
\end{equation}

Let $\Phi^{(CHF)}$ be the confluent hypergeometric parametrix with the parameter $\beta=-\frac{1}{2}$, as given in Appendix \ref{CHF}. The solution to the above RH problem can be constructed as follows:
\begin{equation}\label{2P-1}
	P^{(-1)}(\lambda)=E^{(-1)}(\lambda)P^{(-1)}_{0}(\xi(\lambda))e^{i(t\lambda^{2}+x\lambda)\sigma_{3}},\quad \lambda\in U(-1,\delta),
\end{equation}
where
\begin{equation}\label{2P0-1}
	P^{(-1)}_{0}(\xi)=
	\left\{\begin{array}{ll}
		\Phi^{(CHF)}(\xi)\pi^{\frac{1}{2}\sigma_{3}}\sigma_{1}, &\quad \arg\xi\in(0,\frac{\pi}{3}) \cup (\frac{2\pi}{3},\pi),  \\
		\Phi^{(CHF)}(\xi)\begin{pmatrix} 1 & 0\\ i&1 \end{pmatrix}\pi^{\frac{1}{2}\sigma_{3}}\sigma_{1}, &\quad \arg\xi\in(\frac{\pi}{3},\frac{\pi}{2}),  \\
		\Phi^{(CHF)}(\xi)\begin{pmatrix} 1 & 0\\ -i&1 \end{pmatrix}\pi^{\frac{1}{2}\sigma_{3}}\sigma_{1}, &\quad \arg\xi\in(\frac{\pi}{2},\frac{2\pi}{3}),  \\
		\Phi^{(CHF)}(\xi)\begin{pmatrix}  0& -i \\ -i&0 \end{pmatrix}\pi^{\frac{1}{2}\sigma_{3}}\sigma_{1}, &\quad \arg\xi\in(\pi,\frac{4\pi}{3}) \cup (-\frac{\pi}{3}, 0),  \\
		\Phi^{(CHF)}(\xi)\begin{pmatrix}  0& -i \\ -i&1 \end{pmatrix}\pi^{\frac{1}{2}\sigma_{3}}\sigma_{1}, &\quad \arg\xi\in(\frac{4\pi}{3},\frac{3\pi}{2}), \\
		\Phi^{(CHF)}(\xi)\begin{pmatrix}  0& -i \\ -i&-1 \end{pmatrix}\pi^{\frac{1}{2}\sigma_{3}}\sigma_{1}, &\quad \arg\xi\in(-\frac{\pi}{2},-\frac{\pi}{3}),
	\end{array}\right.
\end{equation}
and
\begin{equation}\label{2E-1}
	E^{(-1)}(\lambda)=M(\lambda)P^{(\infty)}(\lambda)\sigma_{1}\xi(\lambda)^{-\frac{1}{2}\sigma_{3}}e^{i(t-x)\sigma_{3}}\pi^{-\frac{1}{2}\sigma_{3}}.
\end{equation}
It follows from \eqref{N}, \eqref{M} and \eqref{2cm} that $E^{(-1)}(\lambda)$ is analytic for $\lambda\in U(-1,\delta)$. From \eqref{N}, \eqref{M}, \eqref{2P-1}-\eqref{2E-1} and \eqref{CHFat00}, the matching condition \eqref{2P-1N-1} is fulfilled.

As $\lambda\to 1$, $T$ has the same jumps as the one in the space-like region in Section \ref{1.1}. So we can construct the local parametrix $P^{(1)}$
\begin{equation}\label{2P1}
	P^{(1)}(\lambda)=M(\lambda)E^{(1)}(\lambda)P^{(1)}_{0}(\xi(\lambda))e^{i(t\lambda^{2}+x\lambda)\sigma_{3}}, \quad \lambda\in U(1,\delta),
\end{equation}
where $P^{(1)}_{0}$, $E^{(1)}$ and $M$ are defined in \eqref{1P01}, \eqref{1E1} and \eqref{M}.  Then as $t\to+\infty$, we have the matching condition
\begin{equation}\label{2P1N-1}
    P^{(1)}(\lambda)\{P^{(\infty)}(\lambda)\}^{-1}M_{+}^{-1}(\lambda)=I+O(t^{-1}).
\end{equation}

\subsection{Local parametrix near the stationary point}\label{2.3}
In this subsection, we seek a parametrix $P^{(0)}$ that satisfies the same jump conditions as $T$ on $\Sigma$ in the neighborhood $U(\lambda_{0},\delta)$, for some $\delta>0$, where $\lambda_{0}=-\frac{x}{2t}$.

\subsubsection*{RH problem for $P^{(0)}$}
\begin{description}
	\item(1)
    $P^{(0)}(\lambda)$ is analytic for $\lambda\in U(\lambda_{0},\delta)\setminus \Sigma$.

\item(2)
 $P^{(0)}(\lambda)$ has the same jumps as $T(\lambda)$ on $U(\lambda_{0},\delta)\cap\Sigma$.

\item(3)
 On the boundary $\partial U(\lambda_{0},\delta)$, $P^{(0)}(\lambda)$ satisfies 
\begin{equation}\label{2P0N-1}
	P^{(0)}(\lambda)\{P^{(\infty)}(\lambda)\}^{-1}M^{-1}(\lambda)=I+O(t^{-\frac{1}{2}}), \quad t \to +\infty,
\end{equation}
where $M$ is defined in \eqref{M}.
\end{description}

Similarly, the solution to the above RH problem can be constructed by using the Cauchy integral
\begin{equation}\label{2P0}
	P^{(0)}(\lambda)=M(\lambda)P^{(\infty)}(\lambda)\left(I-\frac{1}{2\pi i}\int_{\Gamma_{0}} \frac{2ie^{2i(ts^{2}+xs)}}{\pi(s-\lambda)}ds\begin{pmatrix}
		0	& 0\\
		1	& 0
	\end{pmatrix}\right),
\end{equation}
where $P^{(\infty)}$ is defined in \eqref{N}. The integral contour is defined as $\Gamma_{0}=U(\lambda_{0},\delta)\cap(\Sigma_{3}\cup\Sigma_{4})$, where $\Sigma_{3}$ and $\Sigma_{4}$ are shown in Fig. \ref{f2}. As $t\to+\infty$, we have the asymptotics of the integral by using the steepest descent method
\begin{equation}\label{2p0n-1}
	\begin{aligned}
	\frac{1}{2\pi i}\int_{\Gamma_{0}} \frac{2ie^{2i(ts^{2}+xs)}}{\pi(s-\lambda)}ds&=\frac{e^{-i\frac{x^{2}}{2t}}}{\pi^{2}}\int_{\Gamma_{0}}\frac{e^{2it\left(s+\frac{x}{2t}\right)^{2}}}{\left(s+\frac{x}{2t}\right)-\left(\lambda+\frac{x}{2t}\right)}ds\\&=-\frac{e^{i\left(-\frac{x^{2}}{2t}+\frac{\pi}{4}\right)}}{\sqrt{2t}\pi^{\frac{3}{2}}(\lambda-\lambda_{0})}+O(t^{-\frac{3}{2}}).
\end{aligned}\end{equation}
From \eqref{N}, \eqref{M}, \eqref{2P0} and \eqref{2p0n-1}, the matching condition \eqref{2P0N-1} is fulfilled.

\subsection{Final transformation}\label{2.5}
The final transformation is defined as

\begin{equation}
	R(\lambda)=
	\left\{\begin{array}{ll}
		T(\lambda)\left\{M(\lambda)P^{(\infty)}(\lambda)\right\}^{-1}, &\quad \lambda\in\mathbb{C}\setminus (U(\lambda_{0},\delta)\cup U(1,\delta)\cup U(-1,\delta) ),  \\
		T(\lambda)\left\{P^{(0)}(\lambda)\right\}^{-1}, &\quad\lambda\in U(\lambda_{0},\delta)\setminus\Sigma,\\
		T(\lambda)\left\{P^{(1)}(\lambda)\right\}^{-1}, &\quad\lambda\in U(1,\delta)\setminus\Sigma,\\
		T(\lambda)\left\{P^{(-1)}(\lambda)\right\}^{-1}, &\quad\lambda\in U(-1,\delta)\setminus\Sigma,
	\end{array}\right.
\end{equation}
where $P^{(\infty)}$ and $M$ are defined in \eqref{N} and \eqref{M}. Then $R$ fulfills the following RH problem.

\subsubsection*{RH problem for $R$}
\begin{description}
	\item(1)
     $R(\lambda)$ is analytic for $\lambda\in\mathbb{C}\setminus \Sigma$, where the contour is shown in Fig. \ref{f2R}.
\begin{figure}
    \centering
    \includegraphics[width=0.7\linewidth]{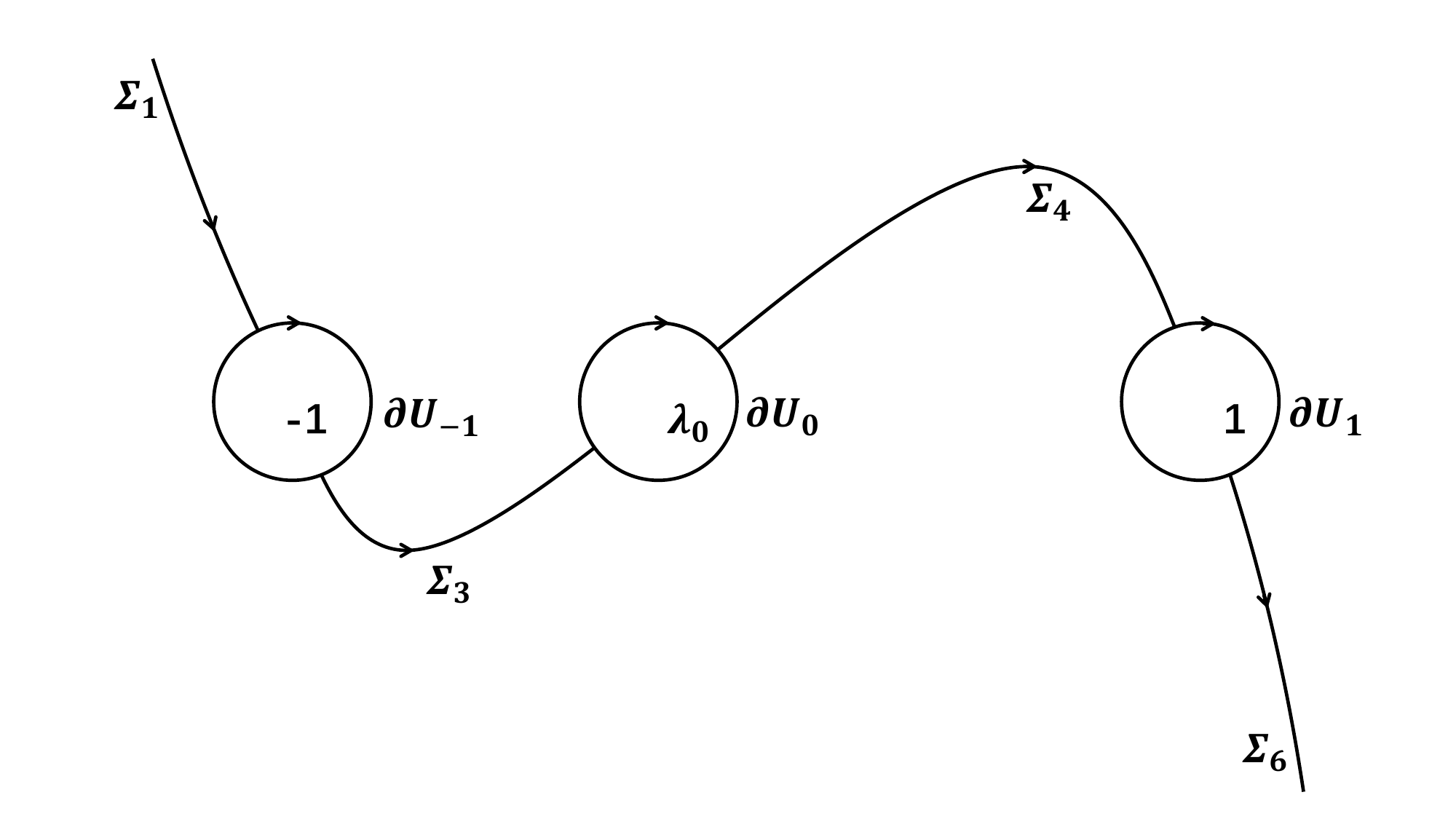}
    \caption{The jump contour of the RH problem for $R$}
    \label{f2R}
\end{figure}

\item(2)
$R_{+}(\lambda)= R_{-}(\lambda)J_{R}(\lambda), \lambda\in  \Sigma $, where
\begin{equation}
	J_{R}(\lambda)=\left\{\begin{array}{ll}
		P^{(0)}(\lambda)\left\{P^{(\infty)}(\lambda)\right\}^{-1}M^{-1}(\lambda), &\quad \lambda\in \partial U(\lambda_{0},\delta),\\
		P^{(1)}(\lambda)\left\{P^{(\infty)}(\lambda)\right\}^{-1}M_{+}^{-1}(\lambda), &\quad \lambda\in \partial U(1,\delta),\\
		P^{(-1)}(\lambda)\left\{P^{(\infty)}(\lambda)\right\}^{-1}M_{+}^{-1}(\lambda), &\quad \lambda\in \partial U(-1,\delta),\\
		M(\lambda)P^{(\infty)}(\lambda) \begin{pmatrix} 1 & 2\pi i e^{-2i\theta(\lambda)}\\0 &1 \end{pmatrix}\left\{P^{(\infty)}(\lambda)\right\}^{-1}M^{-1}(\lambda), &\quad \lambda\in \Sigma_{1}\cup \Sigma_{6},\\
		M(\lambda)P^{(\infty)}(\lambda)\begin{pmatrix} 1 &0 \\ -\frac{2i}{\pi}e^{2i\theta(\lambda)} &1 \end{pmatrix}\left\{P^{(\infty)}(\lambda)\right\}^{-1}M^{-1}(\lambda), &\quad \lambda\in \Sigma_{3}\cup \Sigma_{4}.
	\end{array}\right.		
\end{equation}

\item(3)
$R(\lambda )=I+O\left(\frac{1}{\lambda}\right)$, as $\lambda\to\infty$.
\end{description}

From the matching conditions \eqref{2P-1N-1}, \eqref{2P1N-1} and \eqref{2P0N-1}, we have as $t\to+\infty$,
\begin{equation}
	J_{R}(\lambda)=\left\{\begin{array}{ll}
	I+O(t^{-1}),&\quad \lambda\in\partial U(\pm1,\delta),\\
	I+O(t^{-\frac{1}{2}}),&\quad \lambda\in\partial U(\lambda_{0},\delta),\\
	I+O(e^{-c_{2}t}),&\quad \lambda\in\Sigma_{1}\cup\Sigma_{3}\cup\Sigma_{4}\cup\Sigma_{6},
	\end{array}\right.
\end{equation}
where $c_{2}$ is some positive constant. Then we have as $ t\to+ \infty$,
\begin{equation}
	R(\lambda)=I+O(t^{-\frac{1}{2}}),
\end{equation}
where the error term is uniform for $\lambda$ bounded away from the jump contour for $R$.

\subsection{Large $t$ asymptotics in the time-like region }\label{2.6}
By tracing back the series of invertible transformations
\begin{equation}
	Y \mapsto T \mapsto R,
\end{equation}
we obtain that as $t\to+\infty$,
\begin{equation}
	Y(\lambda)=R(\lambda)A(\lambda)P^{(\infty)}(\lambda),\quad\lambda\in\mathbb{C}\setminus\left(\cup_{i=1}^{4}\Omega_{i}\cup U(1,\delta)\cup U(-1,\delta)\cup U(\lambda_{0},\delta) \right),
\end{equation}
where $P^{(\infty)}$ and $A$ are defined in \eqref{N} and \eqref{A}, and the regions $\Omega_{i}$, $i=1,\dots,4$, are shown in Fig. \ref{f2}. As $\lambda\to\infty$, the asymptotic expansions of $P^{(\infty)}$ and $A$  are given in \eqref{Pinfty} and \eqref{Ainfty}. 

We expand $R$ as $\lambda\to\infty$,
\begin{equation}
R(\lambda)=I+\frac{R_{1}}{\lambda}+\frac{R_{2}}{\lambda^{2}}+O\left(\frac{1}{\lambda^{3}}\right).
\end{equation}
 As $t\to+\infty$, we have
\begin{equation}
	R(\lambda)=I+\frac{R^{(1)}(\lambda)}{t^{\frac{1}{2}}}+O(t^{-1}),
\end{equation}
where the error term is uniform for $\lambda$ bounded away from the jump contour for $R$. Here $R^{(1)}$ satisfies 
\begin{equation}
	R^{(1)}_{+}(\lambda)-R^{(1)}_{-}(\lambda)=\Delta(\lambda),\quad \lambda\in\partial U(\lambda_{0},\delta),
\end{equation}
with
\begin{small}
    \begin{equation}
\Delta(\lambda)=\frac{e^{-i\left(\frac{x^{2}}{2t}-\frac{\pi}{4}\right)}}{\sqrt{2}\pi^{\frac{3}{2}}(\lambda-\lambda_{0})}\begin{pmatrix}
		-\frac{1}{\lambda-1}\frac{2\pi e^{2i(x-t)}}{1+e^{4ix}}+\frac{1}{\lambda^{2}-1}\frac{4\pi e^{2i(x-t)+4ix}}{(1+e^{4ix})^{2}}	& -\frac{1}{\lambda^{2}-1}\frac{4\pi^{2}e^{4i(x-t)}}{(1+e^{4ix})^{2}} \\ \frac{\lambda+1}{\lambda-1}-\frac{1}{\lambda-1}\frac{4e^{4ix}}{1+e^{4ix}} +\frac{1}{\lambda^{2}-1}\frac{4e^{8ix}}{(1+e^{4ix})^{2}}
		&\frac{1}{\lambda-1}\frac{2\pi e^{2i(x-t)}}{1+e^{4ix}}-\frac{1}{\lambda^{2}-1}\frac{4\pi e^{2i(x-t)+4ix}}{(1+e^{4ix})^{2}}\end{pmatrix}.
\end{equation}\end{small}
We obtain that
\begin{equation}
	R^{(1)}(\lambda)=	\left\{\begin{array}{ll}
	\frac{C}{\lambda-\lambda_{0}},&\quad\lambda\in \mathbb{C}\setminus U(\lambda_{0},\delta)	, \\	\frac{C}{\lambda-\lambda_{0}}-\Delta(\lambda),&\quad\lambda\in U(\lambda_{0},\delta),	
	\end{array}\right.
\end{equation}
where $C=\Res(\Delta(\lambda),\lambda_{0})$ is given by
\begin{small}
\begin{equation}
	C=\frac{e^{-i\left(\frac{x^{2}}{2t}-\frac{\pi}{4}\right)}}{\sqrt{2}\pi^{\frac{3}{2}}}\begin{pmatrix}
		-\frac{1}{\lambda_{0}-1}\frac{2\pi e^{2i(x-t)}}{1+e^{4ix}}+\frac{1}{\lambda_{0}^{2}-1}\frac{4\pi e^{2i(x-t)+4ix}}{(1+e^{4ix})^{2}}	& -\frac{1}{\lambda_{0}^{2}-1}\frac{4\pi^{2}e^{4i(x-t)}}{(1+e^{4ix})^{2}} \\ \frac{\lambda_{0}+1}{\lambda_{0}-1}-\frac{1}{\lambda_{0}-1}\frac{4e^{4ix}}{1+e^{4ix}} +\frac{1}{\lambda_{0}^{2}-1}\frac{4e^{8ix}}{(1+e^{4ix})^{2}}
		&\frac{1}{\lambda_{0}-1}\frac{2\pi e^{2i(x-t)}}{1+e^{4ix}}-\frac{1}{\lambda_{0}^{2}-1}\frac{4\pi e^{2i(x-t)+4ix}}{(1+e^{4ix})^{2}}
	\end{pmatrix}.
\end{equation}\end{small}
Expanding $R^{(1)}$ into the Taylor series at infinity, we obtain the asymptotics for $R_{1}$ and $R_{2}$:
\begin{equation}\label{2R1R2}
	R_{1}=\frac{C}{t^{\frac{1}{2}}}+O(t^{-1}), \quad 	R_{2}=\frac{\lambda_{0}C}{t^{\frac{1}{2}}}+O(t^{-1}), \quad t\to+\infty.
\end{equation}
Then,  it follows that $Y$ can be expressed in the following form
\begin{equation}\label{2Y}
	Y(\lambda)=I+\frac{Y_{1}}{\lambda}+\frac{Y_{2}}{\lambda^{2}}+O\left(\frac{1}{\lambda^{3}}\right), \quad \lambda\to\infty,
\end{equation}
where
\begin{equation}\label{2Y1Y2}
	Y_{1}=R_{1}+A_{1}+P^{(\infty)}_{1}, \quad Y_{2}=R_{1}A_{1}+R_{1}P^{(\infty)}_{1}+A_{1}P^{(\infty)}_{1}+R_{2}+A_{2}+P^{(\infty)}_{2}.
\end{equation}
Here $P^{(\infty)}_{1}$, $P^{(\infty)}_{2}$, $A_{1}$, $A_{2}$, $R_{1}$ and $R_{2}$ are defined in \eqref{1P1P2}, \eqref{1A1A2} and \eqref{2R1R2}.

From \eqref{2Y}, \eqref{2Y1Y2} and Proposition \ref{expression}, we obtain the asymptotics of $\partial_{t} D$, $\partial_{x} D$, $b_{++}$ and $B_{--}$ as $t\to+\infty$ in the time-like region, as given in  \eqref{th2dt}-\eqref{th2B}, which complete the proof of Theorem \ref{time}.

\section{Asymptotic analysis in the transition region }\label{sec.3}
In this section, we derive the large $t$ asymptotics of the derivatives of $D$ defined in \eqref{D} and the corresponding solutions of the separated NLS equations in the transition region, where $t^{\frac{1}{2}}\left|\frac{x}{2t}-1\right|\le C$ for any constant $C>0$, by performing Deift-Zhou nonlinear steepest descent analysis of the RH problem for $Y$.

\subsection{Deformation of the jump contour}\label{3.1}
Define 

\begin{figure}
    \centering
    \includegraphics[width=0.7\linewidth]{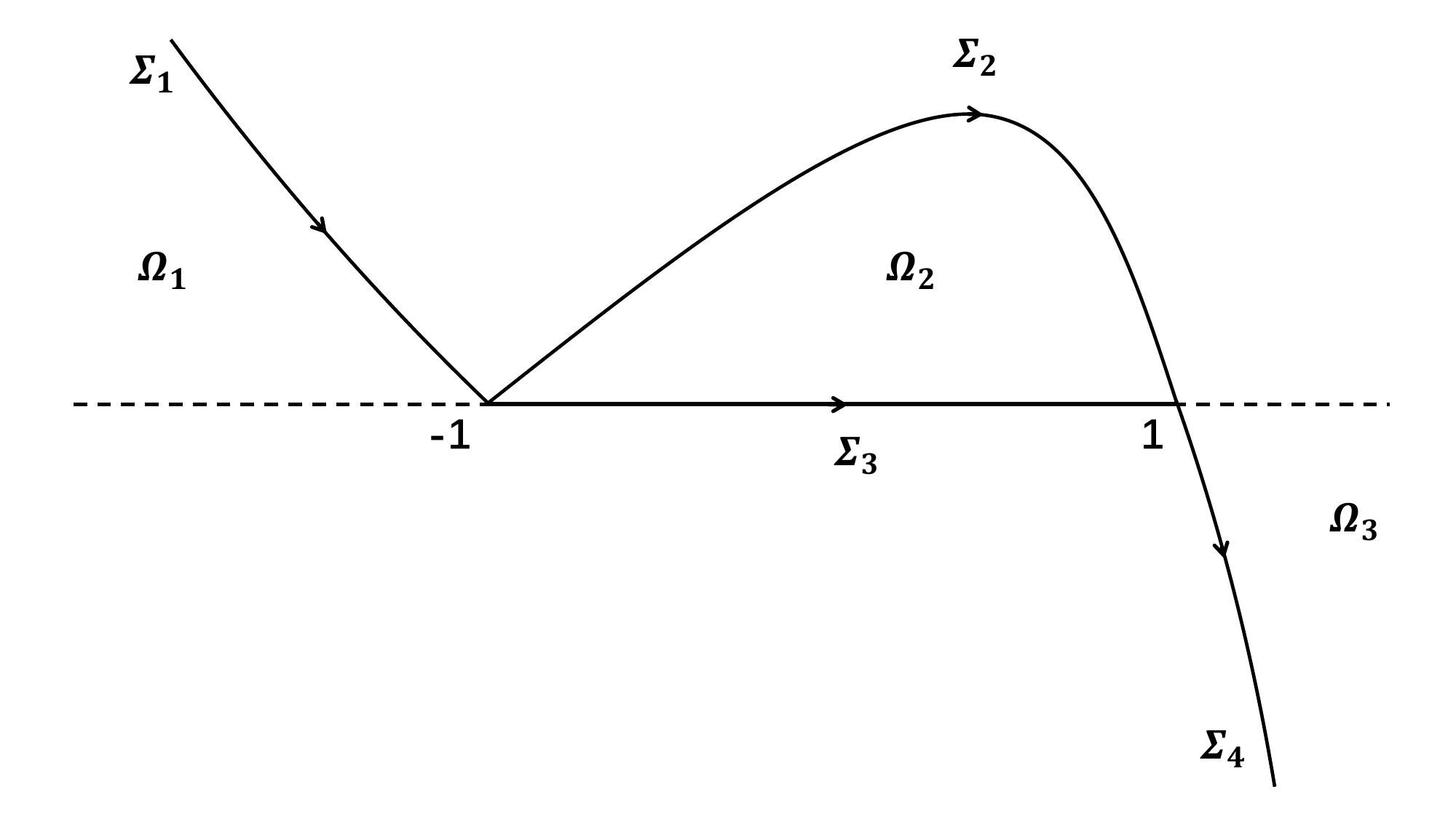}
    \caption{Deformation of the jump contour}
    \label{f3}
\end{figure}
\begin{equation}
	T(\lambda)=
	\left\{\begin{array}{ll}
		Y(\lambda)\begin{pmatrix}
			1& -2\pi ie^{-2i\theta(\lambda)}\\
			0&1
		\end{pmatrix},&\quad \lambda\in \Omega_{1},\\
		Y(\lambda)\begin{pmatrix}
			1& 0\\
			\frac{2i}{\pi}e^{2i\theta(\lambda)}&1
		\end{pmatrix}, &\quad \lambda\in\Omega_{2},\\
		Y(\lambda)\begin{pmatrix}
		1& 2\pi ie^{-2i\theta(\lambda)}\\
		0&1
		\end{pmatrix},&\quad \lambda\in \Omega_{3}, \\
         Y(\lambda),&\quad \lambda\in\mathbb{C}\setminus\bigcup_{i=1}^{3}\Omega_{i},
	\end{array}\right.
\end{equation}
where the regions $\Omega_{i}$, $i=1,2,3$, are illustrated in Fig. \ref{f3}. Then $T$ solves the following RH problem.

\subsubsection*{RH problem for $T$}
\begin{description}
	\item(1)
    $T(\lambda)$ is analytic for $\lambda\in\mathbb{C}\setminus \Sigma$,  where $\Sigma$ is shown in Fig \ref{f3}.

\item(2)
    $T_{+}(\lambda)=T_{-}(\lambda)J_{T}(\lambda)$ , $\lambda\in\Sigma$, where
\begin{equation}\label{3JT}
	J_{T}(\lambda)=\left\{\begin{array}{ll}
		\begin{pmatrix}
			1	& 2\pi ie^{-2i\theta(\lambda)}\\
			0&1
		\end{pmatrix}, &\quad \lambda\in\Sigma_{1}\cup\Sigma_{4},\\
		\begin{pmatrix}
			1& 0\\
			-\frac{2i}{\pi}e^{2i\theta(\lambda)}	&1
		\end{pmatrix}, &\quad\lambda\in\Sigma_{2},	\\
		-I, &\quad\lambda\in\Sigma_{3}
		.
	\end{array}\right.
\end{equation}

\item(3)
As $\lambda\to\infty$, $T(\lambda )=I+O\left(\frac{1}{\lambda}\right)$.

\item(4)
As $\lambda\to\pm1$, $T(\lambda)=O(\ln(\lambda\mp1))$.
\end{description}

From \eqref{3JT}, we have $J_{T}(\lambda)\to I$, as $t\to+\infty$ for $\lambda\in\Sigma\setminus\Sigma_{3}$. As $t\to+\infty$, it is expected that $T$ can be approximated by a solution to the RH problem with the jump matrix along the line segment $(-1,1)$. Therefore, we construct the same global parametrix as given in \eqref{N}.

\subsection{Local parametrix near $\lambda=-1$}\label{3.2}
In this subsection, we seek a parametrix $P^{(-1)}$ that satisfies the same jump conditions as $T$ on $\Sigma$ in the neighborhood $U(- 1,\delta)$, for some $\delta>0$.

\subsubsection*{RH problem for $P^{(-1)}$}
\begin{description}
	\item(1)
    $P^{(-1)}(\lambda)$ is analytic for $\lambda\in U(-1,\delta)\setminus \Sigma$.

\item(2)
$P^{(-1)}(\lambda)$ has the same jumps as $T(\lambda)$ on $U(-1,\delta)\cap\Sigma$.

\item(3)
On the boundary $\partial  U(-1,\delta)$, $P^{(-1)}(\lambda)$ satisfies 
\begin{equation}\label{3P-1N-1}
	P^{(-1)}(\lambda)\{P^{(\infty)}(\lambda)\}^{-1}=\begin{pmatrix}
		1&\frac{y}{2}\frac{\lambda-1}{\lambda+1}\pi e^{2i(x-t)}\\0 &1
	\end{pmatrix}+O(t^{-\frac{1}{2}}), \quad  t \to +\infty.
\end{equation}
\end{description}

We define the following conformal mapping
\begin{equation}\label{3CM}
	\xi(\lambda)=e^{-\frac{\pi i}{4}}\sqrt{2t}(\lambda+1).
\end{equation}
Let
\begin{equation}
	s=e^{-\frac{\pi i}{4}}\sqrt{2t}\left(\frac{x}{2t}-1\right).
\end{equation}
As $\lambda\to-1$, we have
\begin{equation}\label{3cm}
	\frac{\xi^{2}(\lambda)}{2}+s\xi(\lambda)=-i\theta(\lambda)-i(x-t).
\end{equation}

Let $\Psi$ be the solution to the RH problem associated with PIV equation with the parameters $\Theta_{\infty}=\frac{1}{2}$, $\Theta=0$ and the four Stokes multipliers $s_{1}=s_{2}=2i,s_{3}=s_{4}=0$; see Section \ref{PainleveIV}. The solution to the above RH problem can be constructed as follows:
\begin{equation}\label{3P-1}
	P^{(-1)}(\lambda)=E^{(-1)}(\lambda)P^{(-1)}_{0}(\xi(\lambda))e^{i(t\lambda^{2}+x\lambda)\sigma_{3}}, \quad \lambda\in U(-1,\delta),
\end{equation}
where
\begin{equation}\label{3P0-1}
    P_{0}^{(-1)}(\xi)=\left\{\begin{array}{ll}
    	\Psi(\xi,s)(e^{\pi i}\pi)^{-\frac{1}{2}\sigma_{3}}, &\quad \arg\xi\in(-\frac{\pi}{4},\frac{3\pi}{2}),\\
    	-\Psi(\xi,s)(e^{\pi i}\pi)^{-\frac{1}{2}\sigma_{3}}, &\quad \arg\xi\in(-\frac{\pi}{2},-\frac{\pi}{4}),
    		\end{array}\right.
\end{equation}
and 
\begin{equation}\label{3E-1}
	E^{(-1)}(\lambda)=P^{(\infty)}(\lambda)e^{i(x-t)\sigma_{3}}\xi^{\frac{1}{2}\sigma_{3}}(e^{\pi i}\pi)^{\frac{1}{2}\sigma_{3}}.
\end{equation}
It follows from \eqref{N} and \eqref{3cm} that $E^{(-1)}(\lambda)$ is analytic for $\lambda\in U(-1,\delta)$. From \eqref{Psiinfty}-\eqref{1u(s)}, \eqref{N} and \eqref{3P-1}-\eqref{3E-1}, we have \eqref{3P-1N-1}. We mention that in \cite{MLS2020} a model problem which is similar to the RH problem associated with the PIV equation has been used in the studies of the asymptotics of the focusing NLS equation.

As $\lambda\to 1$, $T$ has the same jumps as the one in the space-like region in Section \ref{1.1}. Therefore, for $\lambda\in U(1,\delta)$ with some $\delta>0$, we can construct the same local parametrix $P^{(1)}$ as given in \eqref{1P1}. From \eqref{N}, \eqref{1P1}-\eqref{1E1} and \eqref{CHFat00}, we have
\begin{equation}\label{3P1N-1}
    P^{(1)}(\lambda)\{P^{(\infty)}(\lambda)\}^{-1}=\begin{pmatrix}  1&  0\\-\frac{\lambda+1}{(\lambda-1)\pi} e^{2i(x+t)} &1 \end{pmatrix}+O(t^{-1}), \quad t \to +\infty.
\end{equation}

\subsection{RH problem for $M$}\label{3.3}
As $t\to+\infty$, it follows from \eqref{3P-1N-1} and \eqref{3P1N-1}  that  $P^{(-1)}\left\{P^{(\infty)}\right\}^{-1}$ and $P^{(1)}\left\{P^{(\infty)}\right\}^{-1}$ do not tend to the identity matrix on $\partial U(-1,\delta)$ and $\partial U(1,\delta)$, respectively. To resolve this issue, we construct a matrix-valued function $M(\lambda)$, which solves the remaining jumps along $\partial U(1,\delta)$ and $\partial U(-1,\delta)$.

\subsubsection*{RH problem for $M$}
\begin{description}
    \item(1)
    $M(\lambda)$ is analytic for $\lambda\in\mathbb{C}\setminus\left(\partial U(-1,\delta)\cup \partial U(1,\delta)\right)$.
 
    \item(2) 
    On the boundaries $\partial U(-1,\delta)$ and $\partial U(1,\delta)$, we have
    \begin{equation}
        M_{+}(\lambda)=M_{-}(\lambda)\begin{pmatrix}
		1&\frac{y}{2}\frac{\lambda-1}{\lambda+1}\pi e^{2i(x-t)}\\0 &1
	\end{pmatrix}, \quad \lambda\in\partial U(-1,\delta),
        \end{equation}
        \begin{equation}
        M_{+}(\lambda)=M_{-}(\lambda)\begin{pmatrix}  1&  0\\-\frac{\lambda+1}{(\lambda-1)\pi} e^{2i(x+t)} &1 \end{pmatrix}, \quad\lambda\in\partial U(1,\delta).
        \end{equation}
 
    \item(3)
    As $\lambda\to\infty$, $M(\lambda)=I+O\left(\frac{1}{\lambda}\right)$.
\end{description}

Let
\begin{equation}\label{3A}
	A(\lambda)=I+\frac{B}{\lambda+1}+\frac{C}{\lambda-1},
\end{equation}
then we seek a solution to the above RH problem of the form:
\begin{equation}\label{3M}
    M(\lambda)=\left\{\begin{array}{ll}
A(\lambda)\begin{pmatrix}  1& -\frac{y}{2}\frac{\lambda-1}{\lambda+1}\pi e^{2i(x-t)} \\ 0&1 \end{pmatrix}, &\quad \lambda\in U(-1,\delta),\\
A(\lambda)\begin{pmatrix}  1&  0\\\frac{\lambda+1}{\pi(\lambda-1)} e^{2i(x+t)} &1 \end{pmatrix}, &\quad  \lambda\in U(1,\delta),\\
A(\lambda), &\quad \lambda\in \mathbb{C}\setminus(U(-1,\delta)\cup U(1,\delta)).
    \end{array}\right.
\end{equation}
By the condition that $M$ is analytic near $\lambda=\pm1$, we derive the coefficients in \eqref{3A}
\begin{equation}
	B=\begin{pmatrix}
		0	& -\frac{\pi y e^{2i(x-t)}}{1+\frac{y}{2}e^{4ix}}\\
		0	&-\frac{ye^{4ix}}{1+\frac{y}{2}e^{4ix}}
	\end{pmatrix},\quad C=\begin{pmatrix}
		\frac{ye^{4ix}}{1+\frac{y}{2}e^{4ix}}	& 0 \\
		-\frac{\frac{2}{\pi} e^{2i(x+t)}}{1+\frac{y}{2}e^{4ix}}	& 0
	\end{pmatrix}.
\end{equation}
From \eqref{N}, \eqref{3P-1N-1}, \eqref{3P1N-1} and \eqref{3M}, we have as $t\to+\infty$,
\begin{equation}\label{3P1n-1}
    M_{-}(\lambda)P^{(1)}(\lambda)\{P^{(\infty)}(\lambda)\}^{-1}M^{-1}_{+}(\lambda)=I+O\left(t^{-1}\right),\quad \lambda\in\partial U(1,\delta),
\end{equation}
\begin{equation}\label{3P-1n-1}
   M_{-}(\lambda) P^{(-1)}(\lambda)\{P^{(\infty)}(\lambda)\}^{-1}M^{-1}_{+}(\lambda)=I+O\left(t^{-\frac{1}{2}}\right),\quad \lambda\in\partial U(-1,\delta).
\end{equation}

\subsection{Final transformation}
The final transformation is defined as
\begin{equation}
	R(\lambda)=
	\left\{\begin{array}{ll}
		T(\lambda)\left\{M(\lambda)P^{(\infty)}(\lambda)\right\}^{-1}, &\quad \lambda\in\mathbb{C}\setminus ( U(1,\delta)\cup U(-1,\delta) ),  \\
		T(\lambda)\left\{M(\lambda)P^{(1)}(\lambda)\right\}^{-1}, &\quad\lambda\in U(1,\delta)\setminus\Sigma,\\
		T(\lambda)\left\{M(\lambda)P^{(-1)}(\lambda)\right\}^{-1}, &\quad\lambda\in U(-1,\delta)\setminus\Sigma,
	\end{array}\right.
\end{equation}
where $P^{(\infty)}$ is defined in \eqref{N}.

Then $R$ fulfills the following RH problem.

\subsubsection*{RH problem for $R$}
\begin{description}
	\item(1)
     $R(\lambda)$ is analytic for $\lambda\in\mathbb{C}\setminus \Sigma$, where the contour is shown in Fig. \ref{f3R}.
\begin{figure}
    \centering
    \includegraphics[width=0.7\linewidth]{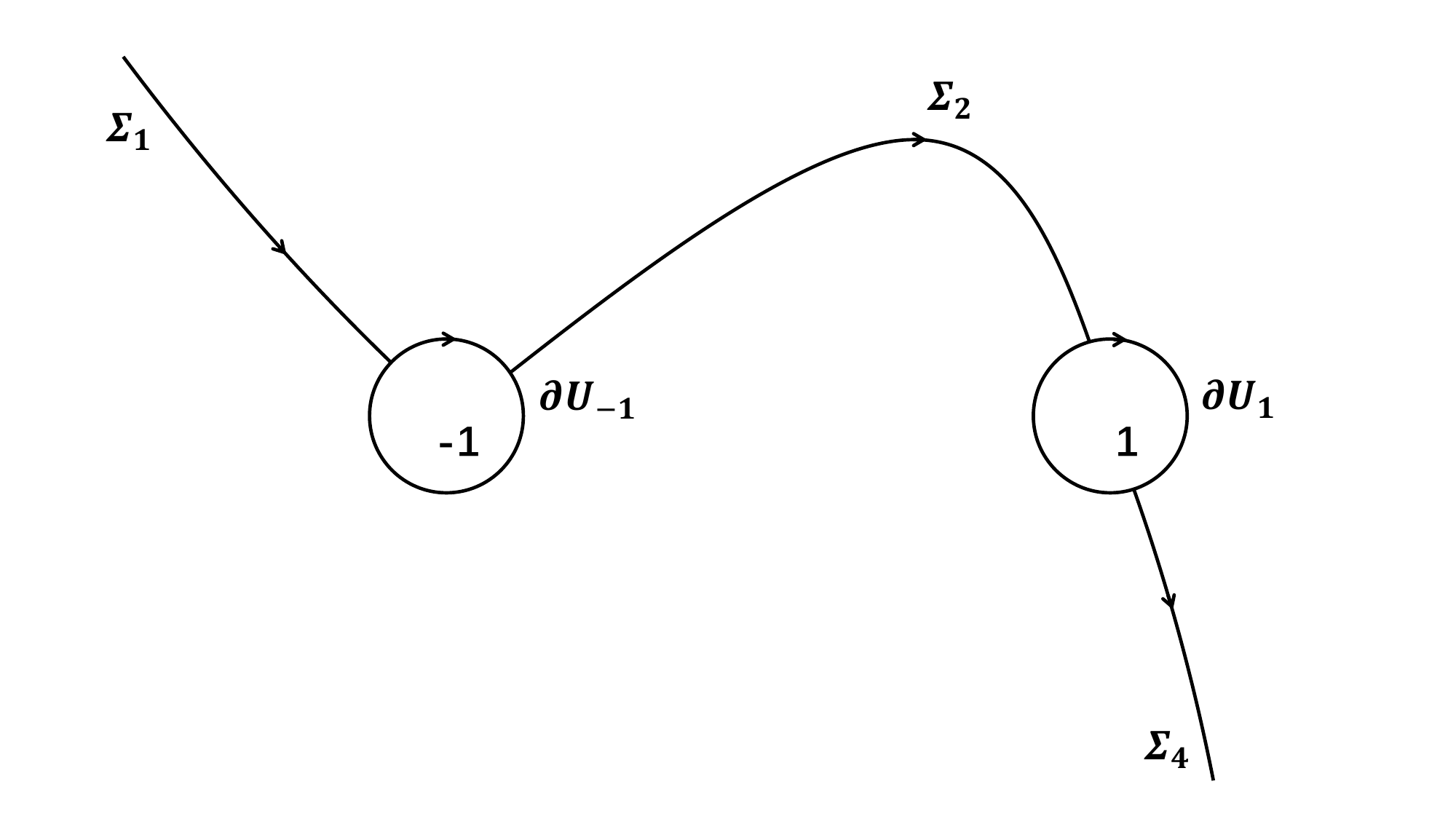}
    \caption{The jump contour of the RH problem for $R$}
    \label{f3R}
\end{figure}

\item(2)
$R_{+}(\lambda)= R_{-}(\lambda)J_{R}(\lambda), \lambda\in  \Sigma $, where
\begin{equation}
J_{R}(\lambda)=\left\{\begin{array}{ll}
		M_{-}(\lambda)P^{(1)}(\lambda)\left\{P^{(\infty)}(\lambda)\right\}^{-1}M_{+}^{-1}(\lambda), &\quad \lambda\in \partial U(1,\delta),\\
		M_{-}(\lambda)P^{(-1)}(\lambda)\left\{P^{(\infty)}(\lambda)\right\}^{-1}M_{+}^{-1}(\lambda), &\quad \lambda\in \partial U(-1,\delta),\\
		M(\lambda)P^{(\infty)}(\lambda) \begin{pmatrix} 1 & 2\pi i e^{-2i\theta}\\0 &1 \end{pmatrix}\left\{P^{(\infty)}(\lambda)\right\}^{-1}M^{-1}(\lambda), &\quad \lambda\in \Sigma_{1}\cup \Sigma_{4},\\
		M(\lambda)P^{(\infty)}(\lambda) \begin{pmatrix} 1 & 0\\ -\frac{2i}{\pi}e^{2i\theta}&1 \end{pmatrix}\left\{P^{(\infty)}(\lambda)\right\}^{-1}M^{-1}(\lambda), &\quad \lambda\in \Sigma_{2}.
	\end{array}\right.		
\end{equation}

\item(3)
$R(\lambda )= I+O\left(\frac{1}{\lambda}\right)$, as $\lambda\to\infty$.
\end{description}

From the matching conditions \eqref{3P1n-1} and \eqref{3P-1n-1}, we have as $t\to+\infty$, 
\begin{equation}
	J_{R}(\lambda)=\left\{\begin{array}{ll}
	I+O(t^{-1}),&\quad \lambda\in\partial U(1,\delta),\\
	I+O(t^{-\frac{1}{2}}),&\quad \lambda\in\partial U(-1,\delta),\\
	I+O(e^{-c_{3}t}),&\quad \lambda\in\Sigma_{1}\cup\Sigma_{2}\cup\Sigma_{4},
	\end{array}\right.
\end{equation}
where $c_{3}$ is some positive constant. Then we have as $ t \to+ \infty$,
\begin{equation}
	R(\lambda)=I+O(t^{-\frac{1}{2}}),
\end{equation}
where the error term is uniform for $\lambda$ bounded away from the jump contour for $R$.

\subsection{Painlev\'e IV asymptotics in the transition region}
By tracing back the series of invertible transformations
\begin{equation}
	Y \mapsto T \mapsto R,
\end{equation}
we obtain that as $t\to+\infty$,
\begin{equation}
	Y(\lambda)=R(\lambda)A(\lambda)P^{(\infty)}(\lambda),\quad\lambda\in\mathbb{C}\setminus\left(\cup_{i=1}^{3}\Omega_{i}\cup\partial U(1,\delta)\cup\partial(-1,\delta)\right),
\end{equation}
where $P^{(\infty)}$ and $A$ are defined in \eqref{N} and \eqref{3A}, and the regions $\Omega_{i}$, $i=1,2,3$, are shown in Fig. \ref{f3}. As $\lambda\to\infty$, the asymptotic expansion of $P^{(\infty)}$ is given in \eqref{1P1P2}. From \eqref{3A}, we have

\begin{equation}\label{3Ainfty}
	A(\lambda)=I+\frac{A_{1}}{\lambda}+\frac{A_{2}}{\lambda^{2}}+\left(\frac{1}{\lambda^{3}}\right), \quad \lambda\to \infty,
\end{equation}
where
\begin{equation}\label{3A1A2}
    A_{1}=\begin{pmatrix}
		\frac{ye^{4ix}}{1+\frac{y}{2}e^{4ix}}	& -\frac{\pi y e^{2i(x-t)}}{1+\frac{y}{2}e^{4ix}}\\
		-\frac{\frac{2}{\pi} e^{2i(x+t)}}{1+\frac{y}{2}e^{4ix}}	& -\frac{ye^{4ix}}{1+\frac{y}{2}e^{4ix}}
	\end{pmatrix},\quad A_{2}=\begin{pmatrix}
		\frac{ye^{4ix}}{1+\frac{y}{2}e^{4ix}}	& \frac{\pi y e^{2i(x-t)}}{1+\frac{y}{2}e^{4ix}}\\
		-\frac{\frac{2}{\pi} e^{2i(x+t)}}{1+\frac{y}{2}e^{4ix}}	& \frac{ye^{4ix}}{1+\frac{y}{2}e^{4ix}}
	\end{pmatrix}.
\end{equation}

We have the asymptotic expansion
\begin{equation}
R(\lambda)=I+\frac{R_{1}}{\lambda}+\frac{R_{2}}{\lambda^{2}}+O\left(\frac{1}{\lambda^{3}}\right),\quad \lambda\to\infty.
\end{equation}
As $t\to+\infty$, we have
\begin{equation}
	R(\lambda)=I+\frac{R^{(1)}(\lambda)}{t^{\frac{1}{2}}}+O\left(t^{-1}\right),
\end{equation}
where the error term is uniform for $\lambda$ bounded away from the jump contour for $R$. Here $R^{(1)}$ satisfies
\begin{equation}
	R^{(1)}_{+}(\lambda)-R^{(1)}_{-}(\lambda)=\Delta(\lambda),\quad \lambda\in\partial U(-1,\delta),
\end{equation}
with
\begin{small}
    \begin{equation}
    \begin{aligned}
\Delta(\lambda)
&    =
	\frac{e^{\frac{\pi}{4}i}}{\sqrt{2}(\lambda+1)}\left[-HA(\lambda)\sigma_{3}A^{-1}(\lambda)-\frac{\pi y}{2}\left(\frac{u}{2}+s\right)\frac{\lambda-1}{\lambda+1}e^{2i(x-t)}A(\lambda)
    \begin{pmatrix}
        0&1\\0&0
    \end{pmatrix}A^{-1}(\lambda)\right]\\
   & =- \frac{e^{\frac{\pi}{4}i}}{\sqrt{2}(\lambda+1)}\left[{H\begin{pmatrix}
	1+\frac{4ye^{4ix}}{(\lambda^{2}-1)\left(1+\frac{y}{2}e^{4ix}\right)^{2}} &\frac{2y\pi e^{2i(x-t)}}{(\lambda+1)\left(1+\frac{y}{2}e^{4ix}\right)}+\frac{2\pi y^{2}e^{2i(x-t)+4ix}}{(\lambda^{2}-1)\left(1+\frac{y}{2}e^{4ix}\right)^{2}} \\
	-\frac{4e^{2i(x+t)}}{\pi(\lambda-1)\left(1+\frac{y}{2}e^{4ix}\right)}+\frac{4ye^{2i(x+t)+4ix}}{\pi(\lambda^{2}-1)\left(1+\frac{y}{2}e^{4ix}\right)^{2}} & -1-\frac{4ye^{4ix}}{(\lambda^{2}-1)\left(1+\frac{y}{2}e^{4ix}\right)^{2}}
	\end{pmatrix}} +	\right.\\
&\quad \left.{\frac{\pi y}{2}\left(\frac{u}{2}+s\right) e^{2i(x-t)}
	\begin{pmatrix}
		\frac{\frac{2}{\pi}e^{2i(x+t)}}{(\lambda+1)\left(1+\frac{y}{2}e^{4ix}\right)}+\frac{\frac{2}{\pi} ye^{2i(x+t)+4ix}}{(\lambda^{2}-1)\left(1+\frac{y}{2}e^{4ix}\right)^{2}}& \frac{\lambda-1}{\lambda+1}+\frac{2ye^{4ix}}{(\lambda+1)\left(1+\frac{y}{2}e^{4ix}\right)}+\frac{y^{2}e^{8ix}}{(\lambda^{2}-1)\left(1+\frac{y}{2}e^{4ix}\right)^{2}} \\ -\frac{\frac{4}{\pi^{2}}e^{4i(x+t)}}{(\lambda^{2}-1)\left(1+\frac{y}{2}e^{4ix}\right)^{2}}  &-\frac{\frac{2}{\pi}e^{2i(x+t)}}{(\lambda+1)\left(1+\frac{y}{2}e^{4ix}\right)}-\frac{\frac{2}{\pi} ye^{2i(x+t)+4ix}}{(\lambda^{2}-1)\left(1+\frac{y}{2}e^{4ix}\right)^{2}}
	\end{pmatrix}   }\right].
    \end{aligned}
\end{equation}\end{small}
We obtain that
\begin{equation}
	R^{(1)}(\lambda)=	\left\{\begin{array}{ll}
		\frac{C}{\lambda+1}+\frac{D}{(\lambda+1)^{2}},&\quad\lambda\in \mathbb{C}\setminus U(-1,\delta)	, \\	\frac{C}{\lambda+1}+\frac{D}{(\lambda+1)^{2}}-\Delta(\lambda),&\quad\lambda\in U(-1,\delta),	
	\end{array}\right.
\end{equation}
where
    \begin{equation}\begin{aligned}
	C=&-\frac{e^{\frac{\pi i}{4}}H}{\sqrt{2}}\begin{pmatrix}
		1-\frac{ye^{4ix}}{(1+\frac{y}{2}e^{4ix})^{2}}	& -\frac{\pi y^{2}e^{2i(x-t)+4ix}}{2(1+\frac{y}{2}e^{4ix})^{2}} \\ \frac{2e^{2i(x+t)}}{\pi(1+\frac{y}{2}e^{4ix})^{2}} 
		&-1+\frac{ye^{4ix}}{(1+\frac{y}{2}e^{4ix})^{2}}
	\end{pmatrix}
   \\& -\frac{\pi y e^{2i(x-t)+\frac{\pi i}{4}}}{2\sqrt{2}}\left(\frac{u}{2}+s\right)\begin{pmatrix}
	-\frac{y e^{2i(x+t)+4ix}}{2\pi(1+\frac{y}{2}e^{4ix})^{2}} & 1-\frac{y^{2}e^{8ix}}{4(1+\frac{y}{2}e^{4ix})^{2}} \\ \frac{e^{4i(x+t)}}{\pi^{2}(1+\frac{y}{2}e^{4ix})^{2}} &\frac{y e^{2i(x+t)+4ix}}{2\pi(1+\frac{y}{2}e^{4ix})^{2}}
	\end{pmatrix},
\end{aligned}\end{equation}

    \begin{equation}\begin{aligned}
   D=&-\frac{e^{\frac{\pi i}{4}}H}{\sqrt{2}}\begin{pmatrix}
        -\frac{2ye^{4ix}}{(1+\frac{y}{2}e^{4ix})^{2}} &\frac{2\pi y e^{2i(x-t)}}{(1+\frac{y}{2}e^{4ix})^{2}} \\-\frac{2y e^{2i(x+t)+4ix}}{\pi(1+\frac{y}{2}e^{4ix})^{2}} & \frac{2ye^{4ix}}{(1+\frac{y}{2}e^{4ix})^{2}}
    \end{pmatrix}
   \\&-\frac{\pi ye^{2i(x-t)+\frac{\pi i}{4}}}{2\sqrt{2}}\left( \frac{u}{2}+s\right)\begin{pmatrix}
        \frac{2e^{2i(x+t)}}{\pi(1+\frac{y}{2}e^{4ix})^{2}} & -2+\frac{2ye^{4ix}}{1+\frac{y}{2}e^{4ix}}-\frac{y^{2}e^{8ix}}{2(1+\frac{y}{2}e^{4ix})^{2}} \\\frac{2e^{4i(x+t)}}{\pi^{2}(1+\frac{y}{2}e^{4ix})^{2}} & -\frac{2e^{2i(x+t)}}{\pi(1+\frac{y}{2}e^{4ix})^{2}}
    \end{pmatrix}.
 \end{aligned}\end{equation}
 Expanding $R^{(1)}$ into the Taylor series at infinity, we obtain the asymptotics for $R_{1}$ and $R_{2}$:
 \begin{equation}\label{3R1R2}
	R_{1}=\frac{C}{t^{\frac{1}{2}}}+O(t^{-1}), \quad 	R_{2}=\frac{-C+D}{t^{\frac{1}{2}}}+O(t^{-1}), \quad t\to+\infty.
\end{equation}
Then, $Y$ can be expressed in the following form
\begin{equation}\label{3Y}
	Y(\lambda)=I+\frac{Y_{1}}{\lambda}+\frac{Y_{2}}{\lambda^{2}}+O\left(\frac{1}{\lambda^{3}}\right), \quad \lambda\to\infty,
\end{equation}
where
\begin{equation}\label{3Y1Y2}
	Y_{1}=R_{1}+A_{1}+P^{(\infty)}_{1}, \quad Y_{2}=R_{1}A_{1}+R_{1}P^{(\infty)}_{1}+A_{1}P^{(\infty)}_{1}+R_{2}+A_{2}+P^{(\infty)}_{2}.
\end{equation}
Here $P^{(\infty)}_{1}$, $P^{(\infty)}_{2}$, $A_{1}$, $A_{2}$, $R_{1}$ and $R_{2}$ are defined in \eqref{1P1P2}, \eqref{3A1A2} and \eqref{3R1R2}.

From \eqref{3Y}, \eqref{3Y1Y2} and Proposition \ref{expression}, we obtain the asymptotics of $\partial_{t} D$, $\partial_{x} D$, $b_{++}$ and $B_{--}$ as $t\to+\infty$ in the transition region, as given in  \eqref{th3dt}-\eqref{th3B}, which complete the proof of Theorem \ref{transition}.

\section{Asymptotic analysis of the PIV equation }\label{sec.pIV}
In this section, we derive the asymptotics of a special solution of the PIV equation \eqref{p4} with  the parameters $\Theta_{\infty}=\frac{1}{2}$, $\Theta=0$, which is corresponding to the special Stokes multipliers $s_{1}=s_{2}=2i,s_{3}=s_{4}=0$, by using the RH problem given in Section \ref{PainleveIV}.

Define 
\begin{equation}
\tau=e^{\frac{\pi i}{4}}s\in\mathbb{R}.
\end{equation}
We introduce the transformation
\begin{equation}
	\Phi(z)=\left(e^{-\frac{\pi i}{4}}|\tau|\right)^{\frac{1}{2}\sigma_{3}}\Psi(e^{-\frac{\pi i}{4}}|\tau|z,e^{-\frac{\pi i}{4}}\tau)e^{-\frac{i\tau^{2}}{2}\varphi(z)\sigma_{3}},\arg z\in(-\frac{\pi}{4},\frac{7\pi}{4}),
\end{equation}
where 
\begin{equation}
\varphi(z)=z^{2}+2\sgn(\tau) z.
\end{equation} 
Then $\Phi$ solves the following RH problem.

\subsubsection*{RH problem for $\Phi$}
\begin{description}
	\item(1)
    $\Phi(z)$ is analytic for $z\in\mathbb{C}\setminus \Sigma_{\Phi}$, shown in Fig. \ref{f4}.
\begin{figure}
    \centering
    \includegraphics[width=0.7\linewidth]{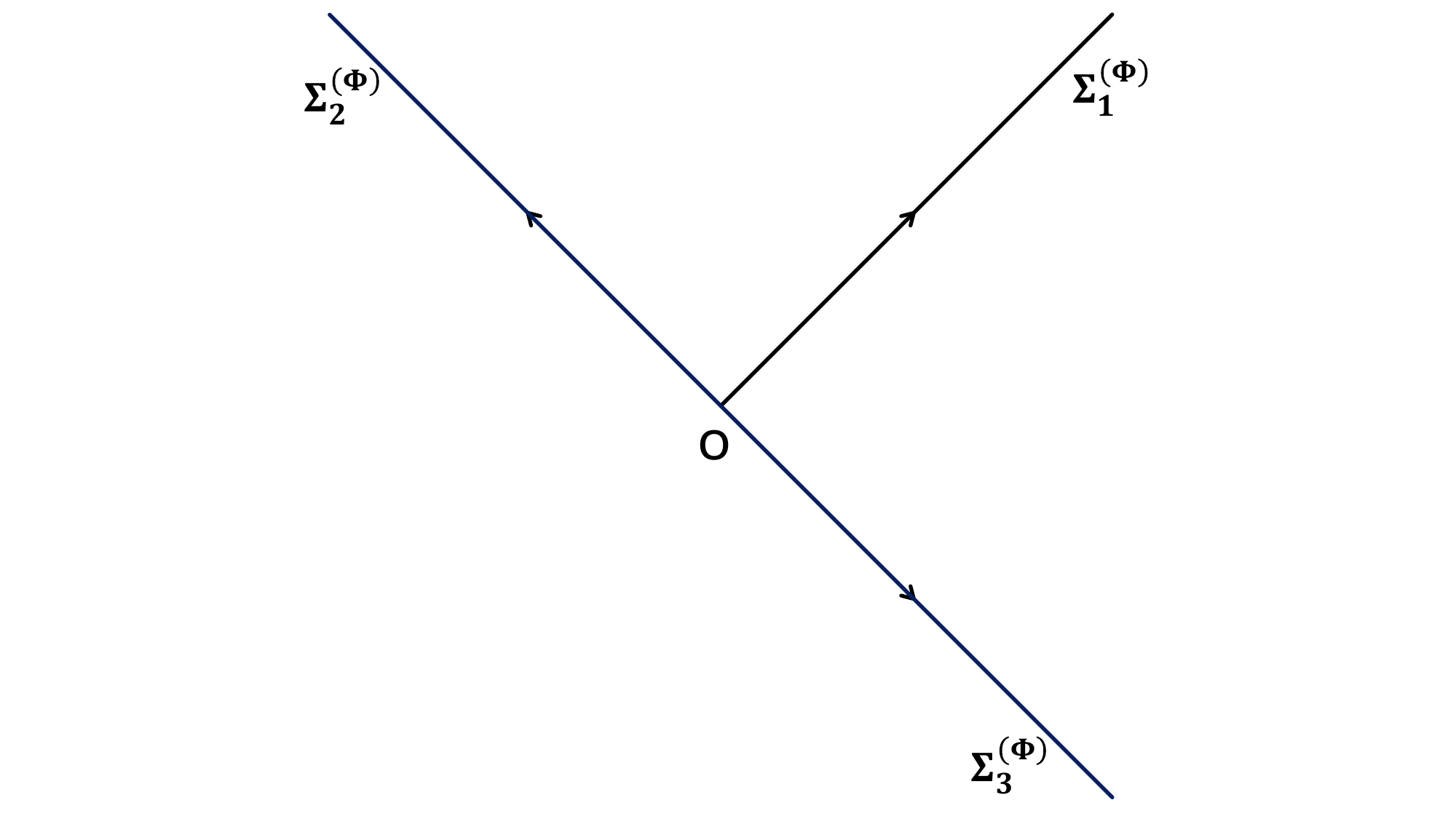}
    \caption{The jump contour of the RH problem for $\Phi$}
    \label{f4}
\end{figure}

\item(2)
$\Phi_{+}(z)=\Phi_{-}(z)J_{\Phi}(z)$, $z\in\Sigma^{(\Phi)}$, where
\begin{equation}
	J_{\Phi}(z)=\left\{\begin{array}{ll}
		\begin{pmatrix}
			1	& 0\\
		2ie^{i\tau^{2}\varphi(z)}	&1
		\end{pmatrix}, &\quad z\in\Sigma_{1}^{(\Phi)},\\
		\begin{pmatrix}
			1& 2ie^{-i\tau^{2}\varphi(z)} \\
	0	&1
		\end{pmatrix}, &\quad z \in\Sigma_{2}^{(\Phi)},	\\
		-I, &\quad z\in\Sigma_{3}^{(\Phi)}
		.
	\end{array}\right.
\end{equation}

\item(3)
As $z\to\infty$,
\begin{equation}
	\Phi(z)=\left[I+O\left(\frac{1}{z}\right)\right]z^{-\frac{1}{2}\sigma_{3}},
\end{equation}
where the branch for $z^{\frac{1}{2}}$ is taken such that $\arg z\in(-\frac{\pi}{4},\frac{7\pi}{4})$.

\item(4)
As $z\to0$, $\Phi(z)=O(\ln|z|)$.
\end{description}

Since the position of the stationary point of the phase function $\varphi$ depends on the sign of $\tau$, we will consider these two cases separately in Sections \ref{4.1} and \ref{4.2}.

\subsection{Asymptotic analysis of the PIV equation as $\tau\to+\infty$}\label{4.1}
For $\tau>0$, the function $\varphi(z)=z^{2}+2z$  possesses the stationary point $z=-1$.

\subsubsection{Deformation of the jump contour}\label{4.1.1}
Define 
\begin{figure}
    \centering
    \includegraphics[width=0.7\linewidth]{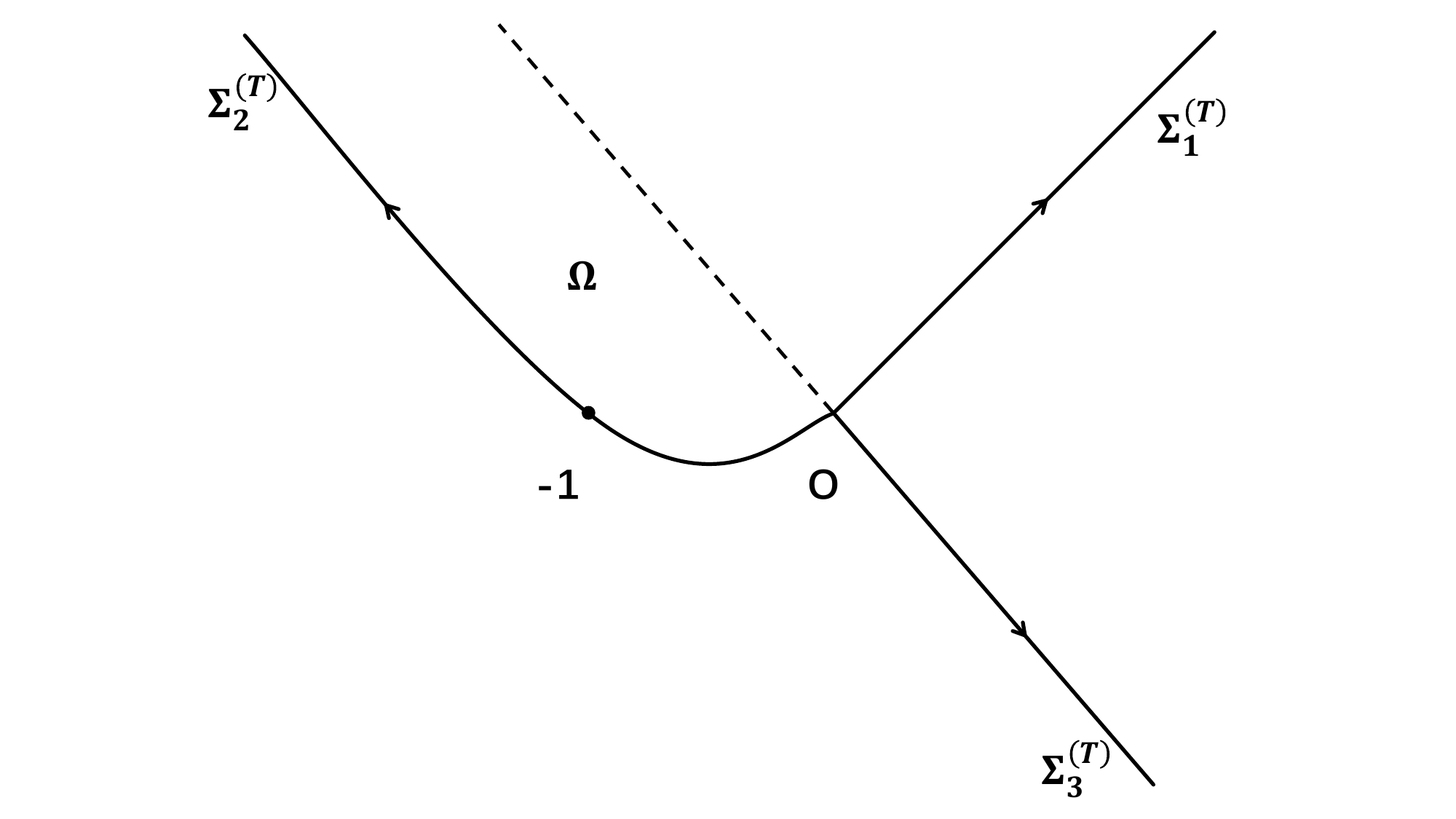}
    \caption{The jump contour of the RH problem for $T$}
    \label{f5}
\end{figure}
\begin{equation}
	T(z)=\left\{\begin{array}{ll}\Phi(z)\begin{pmatrix}
		1& -2ie^{-i\tau^{2}\varphi(z)} \\0 & 1
	\end{pmatrix} ,&\quad z\in \Omega, \\
	\Phi(z), &\quad z\in\mathbb{C}\setminus\Omega,
		\end{array}\right.
\end{equation}
where the region $\Omega$ is shown in Fig. \ref{f5}.

\subsubsection*{RH problem for $T$}
\begin{description}
	\item(1)
    $T(z)$ is analytic for $z\in\mathbb{C}\setminus \Sigma^{(T)}$, where $\Sigma^{(T)}$ is shown in Fig. \ref{f5}.

\item(2)
$T_{+}(z)=T_{-}(z)J_{T}(z)$ , $z\in\Sigma^{(T)}$, where
\begin{equation}\label{4JT}
	J_{T}(z)=\left\{\begin{array}{ll}
		\begin{pmatrix}
			1	&0 \\
			2ie^{i\tau^{2}\varphi(z)}	&1
		\end{pmatrix}, &\quad  z\in\Sigma_{1}^{(T)},\\
		\begin{pmatrix}
			1& 2ie^{-i\tau^{2}\varphi(z)} \\
			0&1
		\end{pmatrix}, &\quad z\in\Sigma_{2}^{(T)},	\\
		-I, &\quad z\in\Sigma_{3}^{(T)}
		.
	\end{array}\right.
\end{equation}

\item(3)
As $z\to\infty$, $T(z)z^{\frac{1}{2}\sigma_{3}}=I+O\left(\frac{1}{z}\right)$.

\item(4)
As $z\to0$, $T(z)=O(\ln|z|)$.
\end{description}

From \eqref{4JT}, we have $J_{T}(z)\to I$, as $\tau\to+\infty$, for $z\in\Sigma^{(T)}\setminus\Sigma_{3}^{(T)}$. As $\tau\to+\infty$, it is expected that $T$ can be approximated by $z^{-\frac{1}{2}\sigma_{3}}$. In order to fulfill the matching conditions later, we define the global parametrix as follows:
\begin{equation}\label{4N}
	P^{(\infty)}(z)=\begin{pmatrix}
	    1&-\frac{1}{z}\\0&1
	\end{pmatrix}z^{-\frac{1}{2}\sigma_{3}},
\end{equation}
where the branch for $z^{\frac{1}{2}}$ is taken such that $\arg z\in(-\frac{\pi}{4},\frac{7\pi}{4})$.

\subsubsection{Local parametrix near $z=0$}\label{4.1.2}
In this subsection, we seek a parametrix $P^{(0)}$ satisfying the same jump conditions as $T$ on $\Sigma^{(T)}$ in the neighborhood $U(0,\delta)$, for some $\delta>0$.

\subsubsection*{RH problem for $P^{(0)}$}
\begin{description}
	\item(1)
    $P^{(0)}(z)$ is analytic for $z\in U(0,\delta)\setminus \Sigma^{(T)}$.

\item(2)
$P^{(0)}(z)$ has the same jumps as $T(z)$ on $U(0,\delta)\cap\Sigma^{(T)}$.

\item(3)
On the boundary $\partial U(0,\delta)$, $P^{(0)}(z)$ satisfies
\begin{equation}\label{4P0}
 P^{(0)}(z)\{P^{(\infty)}(z)\}^{-1}=I+O\left(\tau^{-2}\right),\quad\tau\to+\infty.
\end{equation}
\end{description}

 We define the following conformal mapping 
\begin{equation}\label{4CM}
	\zeta(z)=2\tau^{2}\left(z+\frac{z^{2}}{2}\right).
\end{equation}
As $z\to0$, we have
\begin{equation}\label{4cm}
	\zeta(z)\sim	2\tau^{2}z.
\end{equation}

Let $\Phi^{(CHF)}$ be the confluent hypergeometric parametrix with the parameter $\beta=\frac{1}{2}$, as given in Appendix \ref{CHF}. The solution to the above RH problem can be constructed as follows:
\begin{equation}\label{4P0z}
	P^{(0)}(z)=E^{(0)}(z)P^{(0)}_{0}(\zeta(z))e^{\frac{i\tau^{2}}{2}\varphi(z)\sigma_{3}},
\end{equation}
where
\begin{equation}\label{4P00z}
	P^{(0)}_{0}(\zeta)=
	\left\{\begin{array}{ll}
		\Phi^{(CHF)}(\zeta), &\quad \arg\zeta\in(0,\frac{\pi}{4}) \cup (\frac{2\pi}{3},\pi),  \\
		\Phi^{(CHF)}(\zeta)\begin{pmatrix} 1 & 0\\ 2i&1 \end{pmatrix}, &\quad \arg\zeta\in(\frac{\pi}{4},\frac{\pi}{3}),  \\
		\Phi^{(CHF)}(\zeta)\begin{pmatrix} 1 &0 \\ i&1 \end{pmatrix}, &\quad \arg\zeta\in(\frac{\pi}{3},\frac{2\pi}{3}),  \\
		\Phi^{(CHF)}(\zeta)\begin{pmatrix} 0 & i \\ i& 0\end{pmatrix}, &\quad \arg\zeta\in(\pi,\frac{5\pi}{4}) \cup (-\frac{\pi}{3}, -\frac{\pi}{4}),  \\
		\Phi^{(CHF)}(\zeta)\begin{pmatrix}  0& i \\ i& -2\end{pmatrix}, &\quad \arg\zeta\in(\frac{5\pi}{4},\frac{4\pi}{3}),  \\
		\Phi^{(CHF)}(\zeta)\begin{pmatrix}  0& i \\ i&-1 \end{pmatrix}, &\quad \arg\zeta\in(\frac{4\pi}{3},\frac{3\pi}{2})\cup(-\frac{\pi}{2},-\frac{\pi}{3}),\\
		\Phi^{(CHF)}(\zeta)\begin{pmatrix}  0& -i \\ -i&0 \end{pmatrix}, &\quad\arg\zeta\in(-\frac{\pi}{4},0), 
	\end{array}\right.
\end{equation}
and
\begin{equation}\label{4E0z}
	E^{(0)}(z)=z^{-\frac{1}{2}\sigma_{3}}\zeta(z)^{\frac{1}{2}\sigma_{3}}.
\end{equation}
It follows from \eqref{4cm} that  $E^{(0)}(z)$ is analytic for $z\in U(0,\delta)$. From \eqref{4N},  \eqref{4P0z}-\eqref{4E0z} and \eqref{CHFat00}, the matching condition \eqref{4P0} is fulfilled.

\subsubsection{Local parametrix near $z=-1$}\label{4.1.3}

In this subsection, we seek a parametrix $P^{(-1)}$ that satisfies the same jump conditions as $T$ on $\Sigma^{(T)}$ in the neighborhood $U(-1,\delta)$, for some $\delta>0$.

\subsubsection*{RH problem for $P^{(-1)}$}
\begin{description}
	\item(1)
    $P^{(-1)}(z)$ is analytic for $z\in U(-1,\delta)\setminus \Sigma^{(T)}$.

\item(2)
$P^{(-1)}(z)$ has the same jumps as $T(z)$ on $U(-1,\delta)\cap\Sigma^{(T)}$.

\item(3)
On the boundary $\partial U(-1,\delta)$, $P^{(-1)}(z)$ satisfies
\begin{equation}\label{4P-1}
 P^{(-1)}(z)\{P^{(\infty)}(z)\}^{-1}=I+O\left(\tau^{-1}\right),\quad\tau\to+\infty.
\end{equation}
 \end{description}
The solution to the above RH problem can be constructed by using the Cauchy integral
\begin{equation}\label{4P-1z}
	P^{(-1)}(z)=P^{(\infty)}(z)\left(I+\frac{1}{\pi}\int_{\Gamma_{-1}}^{} \frac{e^{-i\tau^{2}(s^{2}+2s)}}{s-z}ds\begin{pmatrix}
		0	& 1\\
		0	& 0
	\end{pmatrix}\right),
\end{equation}
where $P^{(\infty)}$ is defined in \eqref{4N}. The integral contour is defined as $\Gamma_{-1}=U(-1,\delta)\cap\Sigma_{2}^{(T)}$, where $\Sigma_{2}^{(T)}$ is shown in Fig. \ref{f5}. As $\tau\to+\infty$, we have the asymptotics of the integral by using the steepest descent method
\begin{equation}\label{4p-10}
	\frac{1}{\pi}\int_{\Gamma_{-1}}^{} \frac{e^{-i\tau^{2}(s^{2}+2s)}}{s-z}ds=-\frac{e^{i\tau^{2}+\frac{3\pi}{4}i}}{\tau(z+1)\sqrt{\pi}}+O(\tau^{-2}).
\end{equation}
From \eqref{4N}, \eqref{4P-1z} and \eqref{4p-10}, the matching condition \eqref{4P-1} is fulfilled.

\subsubsection{Final transformation}\label{4.1.5}
The final transformation is defined as
\begin{equation}
	R(z)=
	\left\{\begin{array}{ll}
		T(z)\left\{P^{(\infty)}(z)\right\}^{-1}, &\quad z\in\mathbb{C}\setminus (U(0,\delta)\cup U(-1,\delta) ),\\
		T(z)\left\{P^{(0)}(z)\right\}^{-1}, &\quad z\in U(0,\delta)\setminus\Sigma^{(T)},\\
		T(z)\left\{P^{(-1)}(z)\right\}^{-1}, &\quad z\in U(-1,\delta)\setminus\Sigma^{(T)}.
	\end{array}\right.
\end{equation}
\begin{figure}
    \centering
    \includegraphics[width=0.7\linewidth]{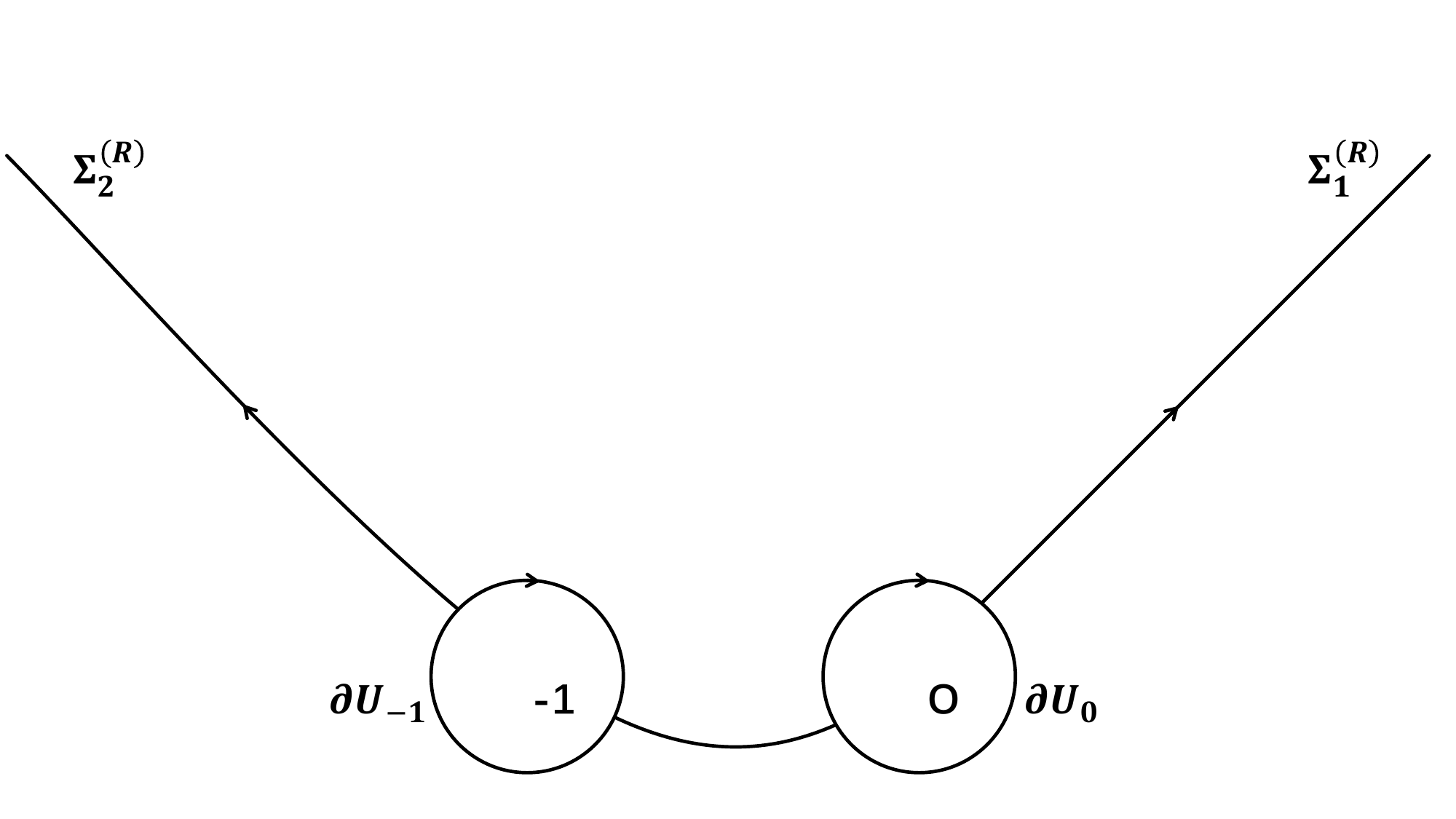}
    \caption{The jump contour of the RH problem for $R$}
    \label{f5R}
\end{figure}
Then $R$ fulfills the following RH problem.

\subsubsection*{RH problem for $R$}
\begin{description}
	\item(1)
    $R(z)$ is analytic for $z\in\mathbb{C}\setminus \Sigma^{(R)}$, where $\Sigma^{(R)}$ is shown in Fig. \ref{f5R}. 

\item(2)
$R_{+}(z)= R_{-}(z)J_{R}(z), z\in  \Sigma^{(R)} $, where
\begin{equation}
	J_{R}(z)=\left\{\begin{array}{ll}
		P^{(0)}(z)\left\{P^{(\infty)}(z)\right\}^{-1}, &\quad z\in \partial U(0,\delta),\\
		P^{(-1)}(z)\left\{P^{(\infty)}(z)\right\}^{-1}, &\quad z\in \partial U(-1,\delta),\\
		P^{(\infty)}(z) \begin{pmatrix} 1 & 0\\ 2ie^{i\tau^{2}\varphi(z)}&1 \end{pmatrix}\left\{P^{(\infty)}(z)\right\}^{-1}, &\quad z\in \Sigma_{1}^{(R)},\\
		P^{(\infty)}(z) \begin{pmatrix} 1 & 2ie^{-i\tau^{2}\varphi(z)}\\0  &1 \end{pmatrix}\left\{P^{(\infty)}(z)\right\}^{-1}, &\quad z\in \Sigma_{2}^{(R)}.
	\end{array}\right.		
\end{equation}

\item(3)
$R(z)=I+O\left(\frac{1}{z}\right)$, as $z\to\infty$.
\end{description}

From the matching conditions \eqref{4P0} and \eqref{4P-1}, we have as $\tau\to+\infty$,
\begin{equation}
J_{R}(z)=\left\{\begin{array}{ll}
	 I+O(\tau^{-2}),&\quad z\in \partial U(0,\delta),\\
     I+O(\tau^{-1}),&\quad z\in \partial U(-1,\delta),\\
    I+O(e^{-c_{4}\tau}), &\quad z\in \Sigma_{1}^{(R)}\cup\Sigma_{2}^{(R)},
	\end{array}\right.	\end{equation}
    where $c_{4}$ is some positive constant. Then we have as $\tau \to +\infty$,
\begin{equation}
	R(z)=I+O(\tau^{-1}),
\end{equation}
where the error term is uniform for $z$ bounded away from the jump contour for $R$.

\subsubsection{Proof of Proposition \ref{thPIV}: asymptotics of the PIV  as $\tau=e^{\frac{\pi i}{4}}s\to+\infty$}\label{4.1.6}
By tracing back the series of invertible transformations
\begin{equation}
	\Psi \mapsto \Phi\mapsto  T \mapsto R,
\end{equation}
we obtain that for $z\in \mathbb{C}\setminus(\Omega\cup U(0,\delta)\cup U(-1,\delta))$, where the region $\Omega$ is shown in Fig. \ref{f5}, as $\tau\to+\infty$,
\begin{equation}\label{4Psi}
	\Psi(sz,s)=s^{-\frac{1}{2}\sigma_{3}}\Phi(z)e^{\frac{s^{2}}{2}\varphi(z)\sigma_{3}}, \quad \Phi(z)=R(z)P^{(\infty)}(z).
\end{equation}
Here $s=e^{-\frac{\pi i}{4}}\tau$ and $P^{(\infty)}$ is defined in \eqref{4N}. From \eqref{4N}, we have as $z\to\infty$,
\begin{equation}\label{4Pinfty}
    P^{(\infty)}(z)=\left[I+\frac{ P^{(\infty)}_{1}}{z}+\frac{ P^{(\infty)}_{2}}{z^{2}}+O\left(\frac{1}{z^{3}}\right)\right]z^{-\frac{1}{2}\sigma_{3}},
\end{equation}
where 
\begin{equation}\label{4P1P2}
     P^{(\infty)}_{1}=\begin{pmatrix}
        0&-1\\0&0
    \end{pmatrix},\quad  P^{(\infty)}_{2}=0.
\end{equation}

As $z\to \infty$, we have the asymptotic expansion
\begin{equation}
    R(z)=I+\frac{R_{1}}{z}+\frac{R_{2}}{z^{2}}+O\left(\frac{1}{z^{3}}\right).
\end{equation}
As $\tau\to+\infty$, we have
\begin{equation}
	R(z)=I+\frac{R^{(1)}(z)}{\tau}+O(\tau^{-2}), 
\end{equation}
where the error term is uniform for $z$ bounded away from the jump contour for $R$. Here $R^{(1)}$ satisfies
\begin{equation}
	R^{(1)}_{+}(z)-R^{(1)}_{-}(z)=\Delta(z),\quad z\in\partial U(-1,\delta),
\end{equation}
with
\begin{equation}
    \Delta(z)=-\frac{e^{i\tau^{2}+\frac{3\pi i}{4}}}{\sqrt{\pi}z(z+1)}\begin{pmatrix}
        0&1\\0&0
    \end{pmatrix}, \quad z\in\partial U(-1,\delta).
\end{equation}
We obtain that
\begin{equation}
	R^{(1)}(z)=	\left\{\begin{array}{ll}
		\frac{C}{z+1},&\quad z\in \mathbb{C}\setminus U(-1,\delta)	, \\	\frac{C}{z+1}-\Delta(z),&\quad z\in U(-1,\delta),	
	\end{array}\right.
\end{equation}
where $C=\Res(\Delta(z),-1)$ is given by
\begin{equation}
	C=\frac{e^{i\tau^{2}+\frac{3\pi}{4}i}}{\sqrt{\pi}}\begin{pmatrix}
        0&1\\0&0
    \end{pmatrix}.
\end{equation}
Expanding $R^{(1)}$ into the Taylor series at infinity, we obtain the asymptotics for $R_{1}$ and $R_{2}$:
\begin{equation}\label{4R1R2}
	R_{1}=\frac{C}{\tau}+O(\tau^{-2}), \quad 	R_{2}=-\frac{C}{\tau}+O(\tau^{-2}), \quad \tau\to+\infty.
\end{equation}
Then, $\Phi$ can be expressed in the following form
\begin{equation}\label{4Y}
	\Phi(z)=\left[I+\frac{\Phi_{1}}{z}+\frac{\Phi_{2}}{z^{2}}+O(z^{-3})\right]z^{-\frac{1}{2}\sigma_{3}}, \quad z\to\infty,
\end{equation}
where
\begin{equation}\label{4Y1Y2}
	\Phi_{1}=R_{1}+P^{(\infty)}_{1}, \quad \Phi_{2}=R_{1}P^{(\infty)}_{1}+R_{2}+P^{(\infty)}_{2}.
\end{equation}
Here $P^{(\infty)}_{1}$, $P^{(\infty)}_{2}$, $R_{1}$ and $R_{2}$ are defined in \eqref{4P1P2} and \eqref{4R1R2}.

From \eqref{y(s)}, \eqref{H(s)}, \eqref{u(s)}, \eqref{4Psi}, \eqref{4Y} and \eqref{4Y1Y2}, we obtain the following asymptotics for $y$, $H$ and $u$ as $\tau=e^{\frac{\pi i}{4}}s\to+\infty$:
\begin{equation}\label{1y}
	y=-2(\Psi_{1})_{12}=-2(\Phi_{1})_{12}=2-\frac{2ie^{-s^{2}}}{\sqrt{\pi}s}+O(s^{-2}),
\end{equation}
\begin{equation}\label{1H}
	H=(\Psi_{1})_{22}=s(\Phi_{1})_{22}=O(s^{-1}),
\end{equation}
\begin{equation}\label{1u}
	u(s)=-2s-\frac{2ie^{-s^{2}}}{\sqrt{\pi}}+O(s^{-1}),
\end{equation}
as given in \eqref{th1u}, \eqref{th1y} and \eqref{th1H}.

\subsection{Asymptotic analysis of the PIV equation as $\tau\to-\infty$}\label{4.2}
For $\tau <0$, the phase function $\varphi(z)=z^{2}-2z$ possesses the stationary point $z=1$.

\subsubsection{Deformation of the jump contour}\label{4.2.1}
\begin{figure}
    \centering
    \includegraphics[width=0.6\linewidth]{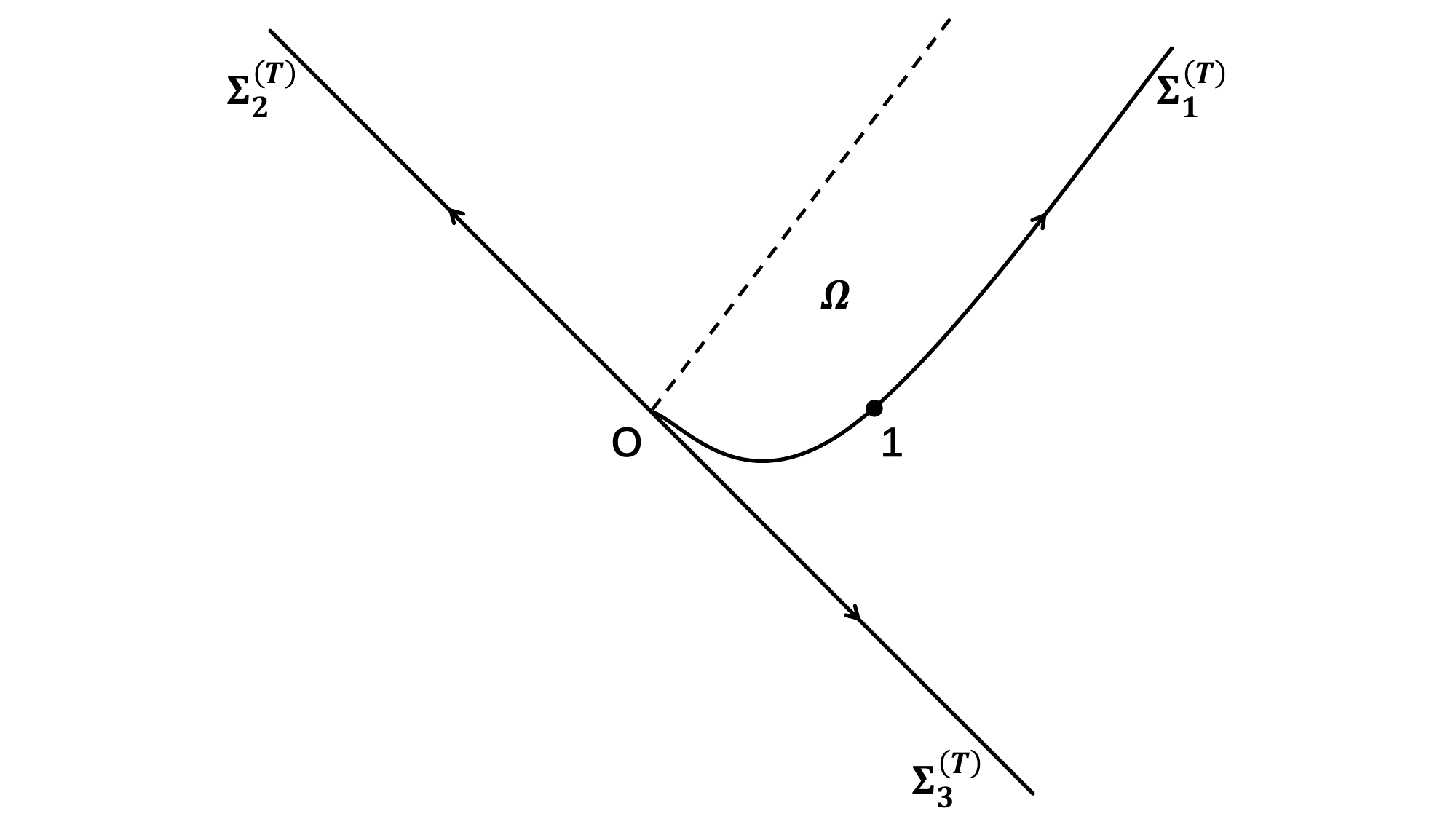}
    \caption{The jump contour of the RH problem of $T$}
    \label{f6}
\end{figure}

Define
\begin{equation}
	T(z)=\left\{\begin{array}{ll}\Phi(z)\begin{pmatrix}
			1&  0\\ 2ie^{i\tau^{2}\varphi(z)} & 1
		\end{pmatrix} ,&\quad z\in \Omega, \\
		\Phi(z), &\quad z\in\mathbb{C}\setminus\Omega,
	\end{array}\right.
\end{equation}
where the region $\Omega$ is shown in Fig. \ref{f6}.

\subsubsection*{RH problem for $T$}
\begin{description}
	\item(1)
    $T(z)$ is analytic for $z\in\mathbb{C}\setminus \Sigma^{(T)}$, where $\Sigma^{(T)}$ is shown in Fig. \ref{f6}.

\item(2)
$T_{+}(z)=T_{-}(z)J_{T}(z)$ , $z\in\Sigma^{(T)}$, where
\begin{equation}\label{5JT}
	J_{T}(z)=\left\{\begin{array}{ll}
		\begin{pmatrix}
			1	& 0\\
			2ie^{i\tau^{2}\varphi(z)}	&1
		\end{pmatrix}, &\quad  z\in\Sigma_{1}^{(T)},\\
		\begin{pmatrix}
			1& 2ie^{-i\tau^{2}\varphi(z)} \\
			0&1
		\end{pmatrix}, &\quad z\in\Sigma_{2}^{(T)},	\\
		-I, &\quad z\in\Sigma_{3}^{(T)}
		.
	\end{array}\right.
\end{equation}

\item(3)
As $z\to\infty$, $T(z)z^{\frac{1}{2}\sigma_{3}}=I+O\left(\frac{1}{z}\right)$.

\item(4)
As $z\to0$, $T(z)=O(\ln|z|)$.
\end{description}

From \eqref{5JT}, we have $J_{T}(z)\to I$, as $\tau\to-\infty$, for $z\in\Sigma^{(T)}\setminus\Sigma_{3}^{(T)}$. As $\tau\to-\infty$, it is expected that $T$ can be approximated by $z^{-\frac{1}{2}\sigma_{3}}$. Similarly, we can construct the same global parametrix $P^{(\infty)}$ given in \eqref{4N}.

\subsubsection{Local parametrix near $z=0$}\label{4.2.2}
In this subsection, we seek a parametrix $P^{(0)}$ satisfying the same jump conditions as $T$ on $\Sigma^{(T)}$ in the neighborhood $U(0,\delta)$, for some $\delta>0$.

\subsubsection*{RH problem for $P^{(0)}$}
\begin{description}
	\item(1)
    $P^{(0)}(z)$ is analytic for $z\in U(0,\delta)\setminus \Sigma^{(T)}$.

\item(2)
$P^{(0)}(z)$ has the same jumps as $T(z)$ on $U(0,\delta)\cap\Sigma^{(T)}$.

\item(3)
On the boundary $\partial U(0,\delta)$, $P^{(0)}(z)$ satisfies 
 \begin{equation}\label{5P0}
 P^{(0)}(z)\{P^{(\infty)}(z)\}^{-1}=I+O\left(\tau^{-2}\right),\quad\tau\to-\infty.
\end{equation}
\end{description}

We define the following conformal mapping
\begin{equation}\label{5CM}
	\zeta(z)=2\tau^{2}\left(z-\frac{z^{2}}{2}\right).
\end{equation}
As $z\to0$, we have
\begin{equation}\label{5cm}
	\zeta(z)\sim   2\tau^{2}z .
\end{equation}

Let $\Phi^{(CHF)}$ be the confluent hypergeometric parametrix with the parameter $\beta=-\frac{1}{2}$, as given in Appendix \ref{CHF}. The solution to the above RH problem can be constructed as follows:
\begin{equation}\label{5P0z}
	P^{(0)}(z)=E^{(0)}(z)P^{(0)}_{0}(\zeta(z))e^{\frac{i\tau^{2}}{2}\varphi(z)\sigma_{3}},
\end{equation}
where
\begin{equation}\label{5P00z}
	P^{(0)}_{0}(\zeta)=
	\left\{\begin{array}{ll}
		\Phi^{(CHF)}(\zeta)\sigma_{1}e^{-\frac{\pi i}{2}\sigma_{3}}, &\quad \arg\zeta\in(0,\frac{\pi}{3})\cup(\frac{3\pi}{4},\pi),  \\
		\Phi^{(CHF)}(\zeta)\begin{pmatrix} 1 &0 \\ i&1 \end{pmatrix}\sigma_{1}e^{-\frac{\pi i}{2}\sigma_{3}}, &\quad \arg\zeta\in(\frac{\pi}{3},\frac{2\pi}{3}),  \\
		\Phi^{(CHF)}(\zeta)\begin{pmatrix} 1 &0 \\ 2i&1 \end{pmatrix}\sigma_{1}e^{-\frac{\pi i}{2}\sigma_{3}}, &\quad \arg\zeta\in(\frac{2\pi}{3},\frac{3\pi}{4}),  \\
		\Phi^{(CHF)}(\zeta)\begin{pmatrix}  0&-i  \\-i & 0 \end{pmatrix}\sigma_{1}e^{-\frac{\pi i}{2}\sigma_{3}}, &\quad \arg\zeta\in(\pi,\frac{4\pi}{3}) ,\\
		\Phi^{(CHF)}(\zeta)\begin{pmatrix}  0& -i \\ -i&1 \end{pmatrix}\sigma_{1}e^{-\frac{\pi i}{2}\sigma_{3}}, &\quad \arg\zeta\in(\frac{4\pi}{3},\frac{3\pi}{2})\cup(-\frac{\pi}{2},-\frac{\pi}{3}),\\
		\Phi^{(CHF)}(\zeta)\begin{pmatrix} 0 & -i \\ -i& 2\end{pmatrix}\sigma_{1}e^{-\frac{\pi i}{2}\sigma_{3}}, &\quad \arg\zeta\in(-\frac{\pi}{3},-\frac{\pi}{4}),  	
  \\
		\Phi^{(CHF)}(\zeta)\begin{pmatrix} 0 & i \\ i& 0\end{pmatrix}\sigma_{1}e^{-\frac{\pi i}{2}\sigma_{3}}, &\quad \arg\zeta\in(-\frac{\pi}{4},0), 
	\end{array}\right.
\end{equation}
and
\begin{equation}\label{5E0z}
	E^{(0)}(z)=z^{-\frac{1}{2}\sigma_{3}}e^{\frac{\pi i}{2}\sigma_{3}}\sigma_{1}\zeta(z)^{-\frac{1}{2}\sigma_{3}}.
\end{equation}
 It follows from \eqref{4N} and \eqref{5cm} that $E^{(0)}(z)$ is analytic for $z\in U(0,\delta)$. From \eqref{4N}, \eqref{5P0z}-\eqref{5E0z} and \eqref{CHFat00}, the matching condition \eqref{5P0} is fulfilled.

\subsubsection{Local parametrix near $z=1$}\label{4.2.3}
In this subsection, we seek a parametrix $P^{(1)}$ satisfying the same jump conditions as $T$ on $\Sigma^{(T)}$ in the
neighborhood $U(1,\delta)$, for some $\delta>0$.

\subsubsection*{RH problem for $P^{(1)}$}
\begin{description}
	\item(1)
    $P^{(1)}(z)$ is analytic for $z\in U(1,\delta)\setminus \Sigma^{(T)}$.

\item(2)
$P^{(1)}(z)$ has the same jumps as $T(z)$ on $U(1,\delta)\cap\Sigma^{(T)}$.

\item(3)
On the boundary $\partial U(1,\delta)$, $P^{(1)}(z)$ satisfies
\begin{equation}\label{5P1}
 P^{(1)}(z)\{P^{(\infty)}(z)\}^{-1}=I+O\left(\tau^{-1}\right),\quad\tau\to-\infty.
\end{equation}
\end{description}

The solution to the above RH problem can be constructed by using the Cauchy integral
\begin{equation}\label{5P1z}
	P^{(1)}(z)=P^{(\infty)}(z)\left(I+\frac{1}{\pi}\int_{\Gamma_{1}}^{} \frac{e^{i\tau^{2}(s^{2}-2s)}}{s-z}ds\begin{pmatrix}
		0	& 0\\
		1	& 0
	\end{pmatrix}\right),
\end{equation}
where $P^{(\infty)}$ is defined in \eqref{4N}. The integral contour is defined as $\Gamma_{1}=U(1,\delta)\cap\Sigma_{1}^{(T)}$, where $\Sigma_{1}^{(T)}$ is shown in Fig. \ref{f6}. As $\tau\to-\infty$, we have the asymptotics of the integral by using the steepest descent method 
\begin{equation}\label{5p10}
\frac{1}{\pi}\int_{\Gamma_{1}}^{} \frac{e^{i\tau^{2}(s^{2}-2s)}}{s-z}ds=\frac{e^{-i\tau^{2}+\frac{\pi}{4}i}}{\tau(z-1)\sqrt{\pi}}+O(\tau^{-2}).
\end{equation}
From \eqref{4N}, \eqref{5P1z} and \eqref{5p10}, the matching condition \eqref{5P1} is fulfilled.

\subsubsection{Final transformation}\label{4.2.5}
The final transformation is defined as
\begin{equation}
	R(z)=
	\left\{\begin{array}{ll}
		T(z)\left\{P^{(\infty)}(z)\right\}^{-1}, &\quad z\in\mathbb{C}\setminus (U(0,\delta)\cup U(1,\delta) ),\\
		T(z)\left\{P^{(0)}(z)\right\}^{-1}, &\quad z\in U(0,\delta)\setminus\Sigma^{(T)},\\
		T(z)\left\{P^{(1)}(z)\right\}^{-1}, &\quad z\in U(1,\delta)\setminus\Sigma^{(T)}.
	\end{array}\right.
\end{equation}
Then $R$ fulfills the following RH problem.

\subsubsection*{RH problem for $R$}
\begin{description}
	\item(1)
    $R(z)$ is analytic for $z\in\mathbb{C}\setminus \Sigma^{(R)}$, where $\Sigma^{(R)}$ is shown in Fig. \ref{f6R}.
\begin{figure}
    \centering
    \includegraphics[width=0.7\linewidth]{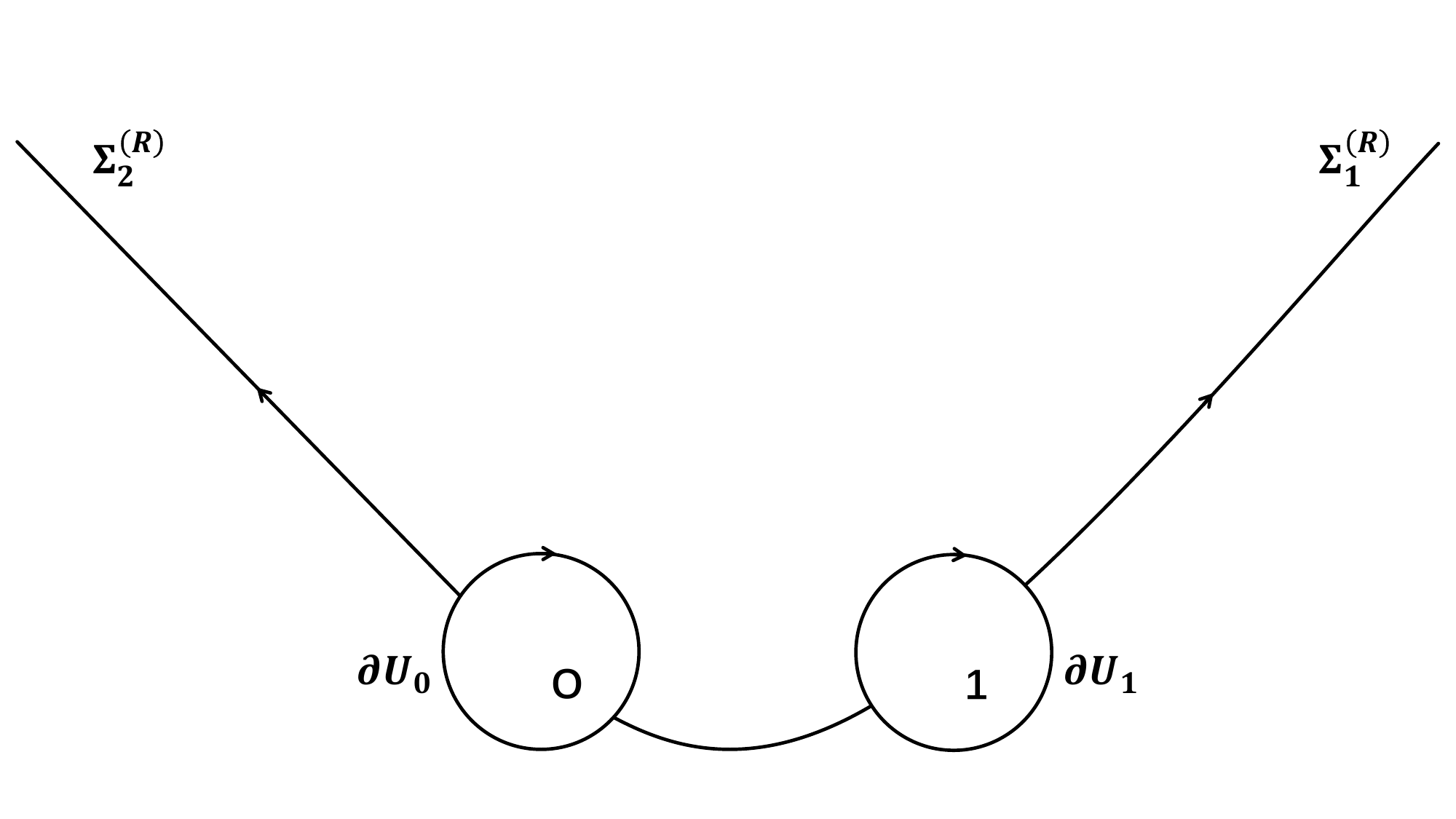}
    \caption{The jump contour of the RH problem for $R$}
    \label{f6R}
\end{figure}

\item(2)
$R_{+}(z)= R_{-}(z)J_{R}(z), z\in  \Sigma^{(R)} $, where
\begin{equation}
	J_{R}(z)=\left\{\begin{array}{ll}
		P^{(0)}(z)\left\{P^{(\infty)}(z)\right\}^{-1}, &\quad z\in \partial U(0,\delta),\\
		P^{(1)}(z)\left\{P^{(\infty)}(z)\right\}^{-1}, &\quad z\in \partial U(1,\delta),\\
		P^{(\infty)}(z) \begin{pmatrix} 1 & 0\\ 2ie^{i\tau^{2}\varphi(z)}&1 \end{pmatrix}\left\{P^{(\infty)}(z)\right\}^{-1}, &\quad z\in \Sigma_{1}^{(R)},\\
		P^{(\infty)}(z) \begin{pmatrix} 1 & 2ie^{-i\tau^{2}\varphi(z)}\\0  &1 \end{pmatrix}\left\{P^{(\infty)}(z)\right\}^{-1}, &\quad z\in \Sigma_{2}^{(R)}.
	\end{array}\right.		
\end{equation}

\item(3)
$R(z)=I+O\left(\frac{1}{z}\right)$, as $z\to\infty$.
\end{description}

From the matching conditions \eqref{5P0} and \eqref{5P1}, we have as $\tau\to-\infty$,
\begin{equation}
J_{R}(z)=\left\{\begin{array}{ll}
	 I+O(\tau^{-2}),\quad z\in \partial U(0,\delta),\\
     I+O(\tau^{-1}),\quad z\in \partial U(1,\delta),\\
    I+O(e^{c_{5}\tau}), \quad z\in \Sigma_{1}^{(R)}\cup\Sigma_{2}^{(R)},
	\end{array}\right.	\end{equation}
    where $c_{5}$ is some positive constant. Then we have as $\tau \to -\infty$,
\begin{equation}
	R(z)=I+O(\tau^{-1}),
\end{equation}
where the error term is uniform for $z$ bounded away from the jump contour for $R$.

\subsubsection{Proof of Proposition \ref{thPIV}: asymptotics of the PIV  as $\tau=e^{\frac{\pi i}{4}}s\to-\infty$}\label{4.2.6}
By tracing back the series of invertible transformations
\begin{equation}
	\Psi \mapsto \Phi\mapsto  T \mapsto R,
\end{equation}
we obtain that for $z\in\mathbb{C}\setminus(\Omega\cup U(0,\delta)\cup U(1,\delta))$, where the region $\Omega$ is shown in Fig. \ref{f6}, as $\tau\to-\infty$,
\begin{equation}\label{5Psi}
	\Psi(sz,s)=(e^{\pi i}s)^{-\frac{1}{2}\sigma_{3}}\Phi(z)e^{\frac{s^{2}}{2}\varphi(z)\sigma_{3}}, \quad \Phi(z)=R(z)P^{(\infty)}(z).
\end{equation}
Here $s=e^{-\frac{\pi i}{4}}\tau$, $P^{(\infty)}$ is defined in  \eqref{4N}, and the asymptotic expansion of $P^{(\infty)}$ is given in \eqref{4Pinfty}.

As $z\to \infty$, we have the asymptotic expansion
\begin{equation}
    R(z)=I+\frac{R_{1}}{z}+\frac{R_{2}}{z^{2}}+O\left(\frac{1}{z^{3}}\right).
\end{equation}
As $\tau\to-\infty$, we have
\begin{equation}
	R(z)=I+\frac{R^{(1)}(z)}{\tau}+O(\tau^{-2}), 
\end{equation}
where the error term is uniform for $z$ bounded away from the jump contour for $R$. Here $R^{(1)}$ satisfies
\begin{equation}
	R^{(1)}_{+}(z)-R^{(1)}_{-}(z)=\Delta(z),\quad z\in\partial U(1,\delta),
\end{equation}
with
\begin{equation}
    \Delta(z)=\frac{e^{-i\tau^{2}+\frac{\pi i}{4}}}{\sqrt{\pi}(z-1)}\begin{pmatrix}
        -1&-\frac{1}{z}\\z&1
    \end{pmatrix},\quad z\in\partial U(1,\delta).
\end{equation}
We obtain that
\begin{equation}
	R^{(1)}(z)=	\left\{\begin{array}{ll}
		\frac{C}{z-1},&\quad z\in \mathbb{C}\setminus U(1,\delta)	, \\	\frac{C}{z-1}-\Delta(z),&\quad z\in U(1,\delta),	
	\end{array}\right.
\end{equation}
where $C=\Res(\Delta(z),1)$ is given by
\begin{equation}
	C=-\frac{e^{-i\tau^{2}+\frac{\pi}{4}i}}{\sqrt{\pi}}\begin{pmatrix}
		1&1 \\-1&-1
	\end{pmatrix}.
\end{equation}
Expanding $R^{(1)}$ into the Taylor series at infinity, we obtain the asymptotics for $R_{1}$ and $R_{2}$:
\begin{equation}\label{5R1R2}
	R_{1}=\frac{C}{\tau}+O(\tau^{-2}), \quad 	R_{2}=\frac{C}{\tau}+O(\tau^{-2}), \quad \tau\to-\infty.
\end{equation}
Then, $\Phi$ can be expressed in the following form
\begin{equation}\label{5Y}
	\Phi(z)=\left[I+\frac{\Phi_{1}}{z}+\frac{\Phi_{2}}{z^{2}}+O(z^{-3})\right]z^{-\frac{1}{2}\sigma_{3}}, \quad z\to\infty,
\end{equation}
where
\begin{equation}\label{5Y1Y2}
	\Phi_{1}=R_{1}+P^{(\infty)}_{1}, \quad \Phi_{2}=R_{1}P^{(\infty)}_{1}+R_{2}+P^{(\infty)}_{2}.
\end{equation}
Here $P^{(\infty)}_{1}$, $P^{(\infty)}_{2}$, $R_{1}$ and $R_{2}$ are defined in \eqref{4P1P2} and \eqref{5R1R2}.

From  \eqref{y(s)}, \eqref{H(s)}, \eqref{u(s)}, \eqref{5Psi}, \eqref{5Y} and \eqref{5Y1Y2}, we obtain the following asymptotics for $y$, $H$ and $u$ as $\tau=e^{\frac{\pi i}{4}}s\to-\infty$:
\begin{equation}\label{2y}
y(s)=-2(\Psi_{1})_{12}=-2(\Phi_{1})_{12}=2+\frac{2e^{s^{2}}}{\sqrt{\pi}s}+O(s^{-2}),
\end{equation}
\begin{equation}\label{2H}
	H(s)=(\Psi_{1})_{22}=-s(\Phi_{1})_{22}=-\frac{e^{s^{2}}}{\sqrt{\pi}}+O(s^{-1}),
\end{equation}
\begin{equation}\label{2u}
	u(s)=-2s-\frac{2e^{s^{2}}}{\sqrt{\pi}}+O(s^{-1}),
\end{equation}
as given in \eqref{th2u}, \eqref{th2y} and \eqref{th2H}.

\section*{Acknowledgements} 
We  thank the anonymous referees for their constructive suggestions.
The work of Shuai-Xia Xu was supported in part by the National Natural Science Foundation of China under grant numbers  12431008, 12371257 and 11971492, and by  Guangdong Basic and Applied Basic Research Foundation (Grant No. 2022B1515020063). Yu-Qiu Zhao was supported in part by the National Natural Science Foundation of China under grant numbers  11971489 and 12371077. 
\appendix

\begin{appendix}
\section{Confluent hypergeometric parametrix}\label{CHF}
As shown in \cite{IK2008}, the confluent hypergeometric parametrix $\Phi^{(CHF)}(\xi)=\Phi^{(CHF)}(\xi;\beta)$, with a parameter $\beta$, is a solution to the following RH problem. Some applications of this parametrix  in the studies of the asymptotics of the Fredholm determinants of  integrable  kernels with jump-type Fisher-Hartwig singularities  can be found in 
\cite{Charlier2021,DXZ2022} etc.. 
\begin{figure}
    \centering
    \includegraphics[width=0.7\linewidth]{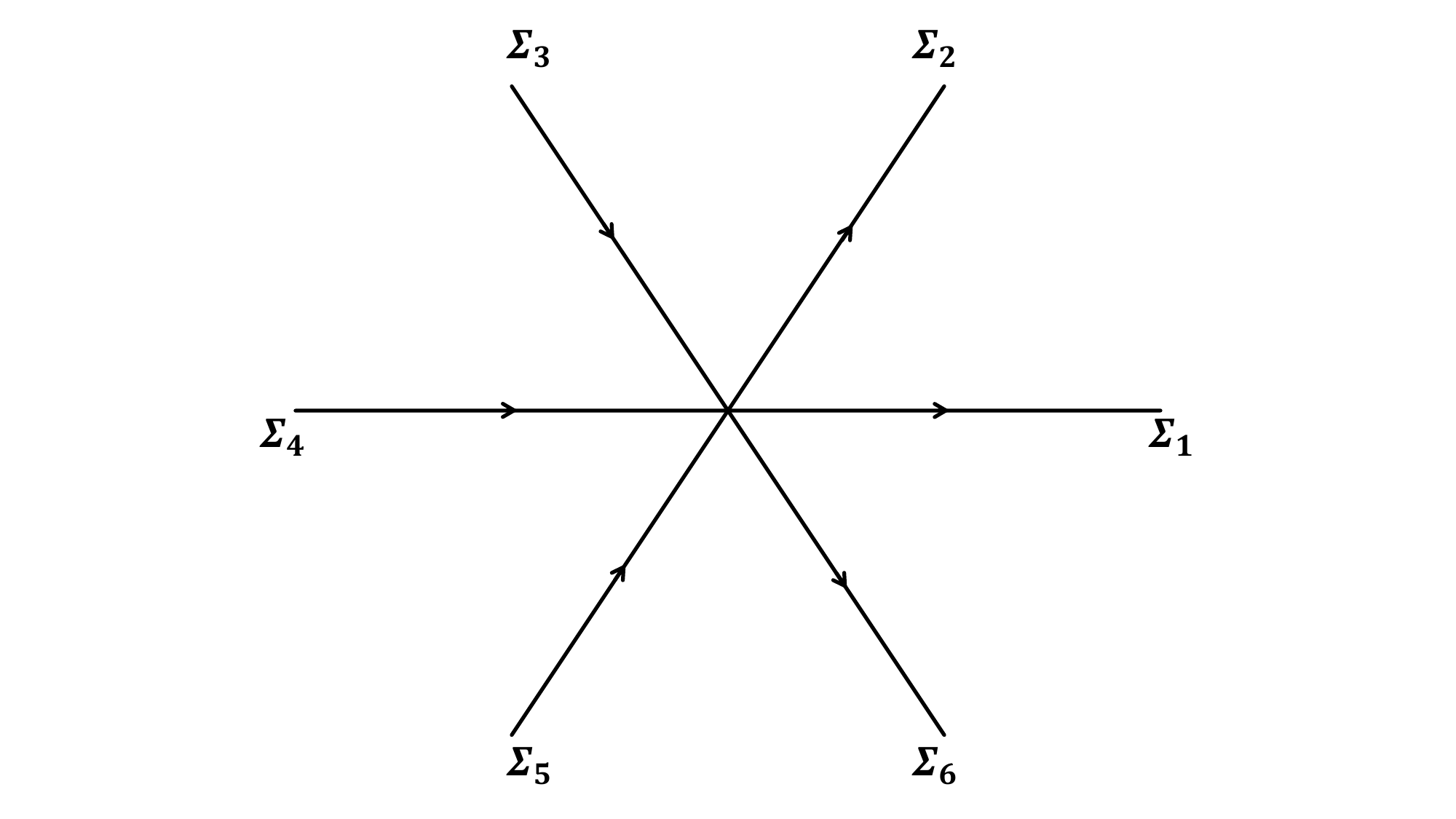}
    \caption{The jump contour for the RH problem for $\Phi^{(CHF)}$}
    \label{fCHF}
\end{figure}

\subsubsection*{RH problem for $\Phi^{(CHF)}$}

\begin{description}
	\item(1)
     $\Phi^{(CHF)}(\xi)$ is analytic in $\mathbb{C}\setminus{\cup^{6}_{i=1}\Sigma_{i}}$, where $\Sigma_{i}$, $i=1,\dots,6$, are shown in Fig. \ref{fCHF}.

\item(2)
 $\Phi^{(CHF)}_{+}(\xi)=\Phi^{(CHF)}_{-}(\xi)J_{i}(\xi)$,   $\xi\in\Sigma_{i}$, $i=1,\dots,6$, where
\begin{align*}
	J_{1}(\xi)=\begin{pmatrix}
		0 & e^{-\beta\pi i}\\-e^{\beta\pi i} & 0
	\end{pmatrix}, \quad
	J_{2}(\xi)=\begin{pmatrix}
		1&0\\e^{\beta\pi i} & 1
	\end{pmatrix},\quad
	J_{3}(\xi)=\begin{pmatrix}
			1&0\\e^{-\beta\pi i} & 1
	\end{pmatrix},\\\quad
	J_{4}(\xi)=\begin{pmatrix}
	0 & e^{\beta\pi i}\\-e^{-\beta\pi i} & 0
	\end{pmatrix}, \quad
	J_{5}(\xi)=\begin{pmatrix}
		1&0\\e^{-\beta\pi i} & 1
	\end{pmatrix},\quad
	J_{6}(\xi)=\begin{pmatrix}
	1&0\\e^{\beta\pi i} & 1
	\end{pmatrix}.
\end{align*}

\item(3)
 As $\xi\to\infty$,
\begin{align}
    \Phi^{(CHF)}(\xi)=\left(I+O(\xi^{-1})\right)\xi^{-\beta\sigma_{3}}e^{-\frac{i\xi}{2}\sigma_{3}}\left\{\begin{array}{ll}
    I, &\quad 0<\arg\xi<\pi,	\\
    \begin{pmatrix}
    	0& -e^{\beta\pi i}\\ e^{-\beta\pi i} &0
    \end{pmatrix}, &\quad \pi<\arg\xi<\frac{3}{2}\pi,	\\
    	\begin{pmatrix}
    		0&-e^{-\beta\pi i}\\e^{\beta\pi i}& 0
    	\end{pmatrix}, &\quad -\frac{\pi}{2}<\arg \xi<0.
    \end{array}\right.
\end{align}

\item(4)
As $\xi\to 0$, $\Phi^{(CHF)}(\xi)=O(\ln |\xi|)$.
\end{description}

For $\xi$ belonging to the region bounded by $\Sigma_{1}$ and $\Sigma_{2}$, the solution to the RH problem can be constructed by using the confluent hypergeometric function $\psi(a,b;\xi)$  \cite{IK2008}: 
\begin{equation}\label{CHF1}
	\Phi^{(CHF)}(\xi)=\begin{pmatrix}
	\psi(\beta,1,e^{\frac{\pi i}{2}}\xi)e^{\frac{\beta\pi i}{2}} &-\frac{\Gamma(1-\beta)}{\Gamma(\beta)}\psi(1-\beta,1,e^{-\frac{\pi i}{2}}\xi)e^{-\frac{\beta\pi i}{2}} \\
	-\frac{\Gamma(1+\beta)}{\Gamma(-\beta)}\psi(1+\beta,1,e^{\frac{\pi i}{2}}\xi)e^{\frac{3\beta\pi i}{2}} &
	\psi(-\beta,1,e^{-\frac{\pi i}{2}}\xi)
	\end{pmatrix}e^{-\frac{i\xi}{2}\sigma_{3}},
\end{equation}
where $\psi(a,b;\xi)$ is the unique solution of the Kummer's equation
\begin{equation}
	\xi\frac{d^{2}y}{d\xi^{2}}+(b-\xi)\frac{dy}{d\xi}-ay=0.
\end{equation}
If the parameter $b=1$, the expansions of the function $\psi(a,1;\xi)$, for $\arg \xi\in(-\frac{3}{2}\pi,\frac{3}{2}\pi)$, at infinty and zero are known to be 
\begin{equation}\label{CHF00}
	\psi(a,1;\xi)=\xi^{-a}\left[1-\frac{a^{2}}{\xi}+\frac{a^{2}(a-1)^{2}}{2\xi^{2}}+O(\xi^{-3})\right], \quad \xi\to\infty,
\end{equation}
\begin{equation}\label{CHF0}
	\psi(a,1;\xi)=-\frac{1}{\Gamma(a)}\left(\ln \xi+\frac{\Gamma^{'}(a)}{\Gamma(a)}+2\gamma_{E}\right)+O(\xi\ln \xi), \quad \xi\to 0,
\end{equation}
see \cite{Olver}, Chapter 13, where $\gamma_{E}$ is the Euler's constant.
From \eqref{CHF00} and \eqref{CHF0}, we obtain the asymptotics of $\Phi^{(CHF)}$ in \eqref{CHF1} near infinity and zero
\begin{equation}\label{CHFat00}
	\Phi^{(CHF)}(\xi)=\left(I+\frac{1}{\xi}\begin{pmatrix}
		i\beta^{2} & -i\frac{\Gamma(1-\beta)}{\Gamma(\beta)}e^{-\beta\pi i}\\
		i\frac{\Gamma(1+\beta)}{\Gamma(-\beta)}e^{\beta\pi i} &-i\beta^{2}
	\end{pmatrix}+O(\xi^{-2})\right)\xi^{-\beta \sigma_{3}}e^{-\frac{i\xi}{2}\sigma_{3}},\quad \xi\to\infty,
\end{equation}
\begin{equation}\label{CHFat0}
	\Phi^{(CHF)}(\xi)\begin{pmatrix}
		1 & 0\\  -e^{-\beta\pi i} &1
	\end{pmatrix}e^{-\frac{\beta \pi i}{2}\sigma_{3}}=\Upsilon_{0}\left(I+\Upsilon_{1}\xi+O(\xi^{2})\right)\begin{pmatrix}
	1 & -\frac{1-e^{2\beta\pi i}}{2\pi i}\ln\left(e^{-\frac{\pi i}{2}}\xi\right) \\ 0 & 1 
	\end{pmatrix} ,\quad \xi\to 0,
\end{equation}
where 
\begin{equation}
	\Upsilon_{0}=\begin{pmatrix}
		\Gamma(1-\beta)e^{-\beta\pi i} & \frac{1}{\Gamma(\beta)}\left(\frac{\Gamma^{'}(1-\beta)}{\Gamma(1-\beta)}+2\gamma_{E}\right) \\
		\Gamma(1+\beta) & -\frac{e^{\beta\pi i}}{\Gamma(-\beta)}\left(\frac{\Gamma^{'}(-\beta)}{\Gamma(-\beta)}+2\gamma_{E}\right)
	\end{pmatrix}.
\end{equation}

\end{appendix}

\end{document}